\documentclass[a4paper,11pt]{article}
\pdfoutput=1 

\usepackage{jheppub} 

\usepackage{float}

\usepackage{color}
\usepackage{xcolor}
\usepackage[utf8]{inputenc}
\usepackage[squaren]{SIunits}
\usepackage{amsmath}
\usepackage{amsfonts}
\usepackage{slashed}
\usepackage{graphicx}
\usepackage{subfigure}
\usepackage{rotating}
\usepackage{enumitem}  
\usepackage{pifont}
\usepackage{dsfont}

\usepackage{graphicx,amsmath,amssymb,epsfig,verbatim,mathrsfs,array,layout,textcomp,latexsym}
\usepackage{multirow}
\usepackage{makecell}

\allowdisplaybreaks[1]

\newcommand{\alphas}{\alpha_s}
\newcommand{\alphae}{\alpha_e}

\newcommand{\TeV}{\,\text{TeV}}
\newcommand{\GeV}{\,\text{GeV}}

\newcommand{\Cnine}{\mathcal{C}_{9,\mu}^{\phantom{\prime}}}
\newcommand{\Cten}{\mathcal{C}_{10,\mu}^{\phantom{\prime}}}
\newcommand{\Cninepr}{\mathcal{C}_{9,\mu}^{\prime}}
\newcommand{\Ctenpr}{\mathcal{C}_{10,\mu}^{\prime}}
\newcommand{\Cseven}{\mathcal{C}_{7,\mu}^{\phantom{\prime}}}
\newcommand{\Csevenpr}{\mathcal{C}_{7,\mu}^{\prime}}

\newcommand{\Cnineten}{\mathcal{C}_{(9,10),\mu}}
\newcommand{\Cninetenall}{\mathcal{C}_{(9,10),\mu}^{(\prime)}}
\newcommand{\Call}{\mathcal{C}_{(7,9,10),\mu}^{(\prime)}}

\title{\boldmath Interplay of dineutrino modes with semileptonic rare $B$--decays}

\author[a]{Rigo Bause,}
\author[a]{Hector Gisbert}
\author[a]{Marcel Golz}
\author[a]{and Gudrun Hiller}

\affiliation[a]{Fakult\"at Physik, TU Dortmund, Otto-Hahn-Str.\,4, D-44221 Dortmund, Germany}

\emailAdd{rigo.bause@tu-dortmund.de}
\emailAdd{hector.gisbert@tu-dortmund.de}
\emailAdd{marcel.golz@tu-dortmund.de}
\emailAdd{ghiller@physik.uni-dortmund.de}

\abstract{We present a systematic global analysis of dineutrino modes $b \to q \,\nu \bar \nu$, $q=d,s$, and charged dilepton $b \to q \,\ell^+ \ell^-$ transitions. 
We derive improved  or even entirely new limits on dineutrino branching ratios including decays $B^0 \to (K^0 , X_s)\, \nu \bar \nu$, $B_s \to \phi \,\nu \bar \nu$ and 
$B^0 \to (\pi^0, \rho^0)\, \nu \bar \nu$ from 
dineutrino modes which presently are best constrained: $B^+ \to (K^+,\pi^+, \rho^+) \,\nu \bar \nu$ and $B^0 \to K^{*0} \,\nu \bar \nu$. Using SMEFT we obtain new flavor constraints from the dineutrino modes, which are stronger than the corresponding ones  from charged dilepton rare $b$-decay or Drell-Yan data,
for $e \tau$ and $\tau \tau$ final states,  as well as for $\mu \tau$ ones in $b \to s$ processes.  The method also allows to put novel constraints on semileptonic 
four-fermion operators with top quarks. Implications for ditau modes $b \to s \, \tau^+ \tau^-$  and $b \to d \, \tau^+ \tau^-$ are worked out. Even stronger constraints are obtained in simplified BSM frameworks such as leptoquarks and $Z^\prime$-models.
Furthermore, the interplay between dineutrinos and charged dileptons 
allows for concrete, novel tests of lepton universality in rare $B$-decays. Performing a global fit to $b \to s \,\mu^+ \mu^-, \,s \gamma$ transitions  we find that lepton universality predicts the ratio of the $B^0 \to K^{*0} \,\nu \bar \nu$ to $B^0 \to K^0  \,\nu \bar \nu$ ($B^+ \to K^+  \,\nu \bar \nu$) branching fractions to be within 1.7 to 2.6 (1.6 to 2.4) at $1\,\sigma$, a region that includes the standard model, and that can be narrowed with improved charged dilepton data. There is sizable room outside this region where universality is broken and that can be probed with the Belle II experiment. Using results of a fit to $B^0 \to \mu^+ \mu^-$, $B^0_s\to \bar{K}^{\ast 0}\,\mu^+\mu^-$ and $ B^+ \to \pi^+\, \mu^+ \mu^-$ data we obtain an analogous relation for $|\Delta b|=|\Delta d|=1$ transitions: if lepton universality  holds the ratio of the $B^0 \to \rho^0 \, \nu \bar \nu$ to $B^0 \to \pi^0 \, \nu \bar \nu$ ($B^+ \to \pi^+ \, \nu \bar \nu$)  branching fractions is within 2.5 to 5.7 (1.2 to 2.6) at 1 $\sigma$. 
Putting upper limits on ${\cal{B}}(B_s \to \nu \bar \nu)$  at the level of $10^{-5}$ and ${\cal{B}}(B^0 \to \nu \bar \nu)$  below $10^{-6}$
would allow to control backgrounds from  (pseudo-)scalar operators such as those induced
by light right-handed neutrinos.}

\begin{document} 
\maketitle
\flushbottom

\section{Introduction}

Flavor-changing neutral current (FCNC) quark transitions provide promising avenues towards  new physics (NP)  due to their suppression within the standard model (SM)
by a weak loop,  the Glashow-Iliopoulos-Maiani (GIM) mechanism and  Cabibbo-Kobayashi-Maskawa (CKM) hierarchies.
NP effects could hence be large, and signal a breakdown of the SM.

Further  tests of  the SM  and its symmetries can be performed if leptons are involved. 
Rare $B$-decays into a pair of leptons $\ell^+\ell^-=e^+e^-,\mu^+\mu^-$ allow for clean tests of  lepton universality (LU),
a backbone of the $SU(3)_C \times SU(2)_L \times U(1)_Y$-SM with
the ratios $R_{H}=\mathcal{B}(B\to H \,\mu^+\mu^-)/\mathcal{B}(B\to H\,e^+e^-)$, $H=K,K^*,X_s, \dots$ \cite{Hiller:2003js}.
With  identical kinematic cuts for muons and electrons, deviations from the universality limit  $R_{H}=1$ are  induced by electron-muon mass splitting only and
are very small irrespective of hadronic uncertainties.

Interestingly, non-universality has recently been  evidenced by LHCb
$R_K= 0.846$ ${^{+0.042}_{-0.039}}{^{+0.013}_{-0.012}}$
at $3.1\,\sigma$~\cite{Aaij:2021vac}.  
A similar suppression of muons versus electrons has been observed in  $R_{K^*}$ at $\sim 2-3$ $\sigma$~\cite{Aaij:2017vbb}.
Further deviations from the SM in global fits, {\it e.g.,} recently~\cite{Alguero:2021anc,Kriewald:2021hfc,Geng:2021nhg},  which include data on angular distributions in $B^{0,+}\to K^{\ast\,0,+}\,\mu^+\mu^-$ decays, point to  semileptonic four-fermion operators $(\bar b_L\,\gamma_\mu\, s_L )(\bar\mu \,\gamma^\mu(\gamma_5)\,\mu )$ as minimal, joint 
solution to  these tensions. Although such operators are induced abundantly  by beyond the standard model (BSM) physics, only few models bring in the requisite LU violation. This singles out the importance of LU test observables such as  $R_H$ for model building, and demands further scrutiny.
FCNC quark processes into dineutrinos $q^\prime\to q \,\nu\, \bar \nu$ are ideal candidate modes to do so: Firstly,
they are subject to similar suppressions as  $q^\prime\to q\,\ell^+\,\ell^-$ transitions. 
Importantly,
the flavor of neutrinos is experimentally untagged, therefore a measurement of a dineutrino branching ratio
involves an incoherent sum of neutrino flavors $i,j=e,\mu,\tau$,
${\mathcal{B}(q^\prime\to q \,\nu\, \bar \nu)=\sum_{i,j}\mathcal{B}(q^\prime\to q \,\nu_i\, \bar \nu_j)}$.
This way, the dineutrino modes automatically include contributions from lepton universality violation, or lepton flavor  violation,
allowing for tests thereof~\cite{Bause:2020auq}.
      
In this work, we consider only left-handed (LH)  neutrinos such as those in the SM; 
we also discuss the impact of light right-handed (RH) neutrinos on our analysis,
as well as ways how to control them.
 
On the experimental side, dineutrino modes  require a clean environment such as an $e^+e^-$-facility to perform missing energy measurements. 
Presently only upper limits on $b\to q\,\nu\bar\nu$ branching ratios with $q=d,s$ from LEP~\cite{Adam:1996ts,Barate:2000rc}, Babar~\cite{Lees:2013kla} and Belle~\cite{Grygier:2017tzo,Lutz:2013ftz} exist. 
The most stringent upper limits  for $B\to K^{(\ast)}\,\nu\bar\nu$ decay modes exist for   $K^+$, $K^{\ast+}$ and $K^{\ast 0}$, and are
a factor of two to five above the SM predictions.
Belle II is expected  to observe all three decay modes with about 10$\,\text{ab}^{-1}$ (50$\,\text{ab}^{-1}$) of data leading to an accuracy on the branching ratio of $30\,\%$ ($10\,\%$), even if the NP contribution is subdominant compared to the SM one. 
Recent Belle II efforts can be found in Ref.~\cite{Dattola:2021cmw}.

Most of the current phenomenological studies for $b\to q\,\nu\bar\nu$ decays rely on specific extensions of the SM, see for instance Refs.~\cite{Bordone:2020lnb,Bordone:2019uzc,Gherardi:2020qhc,Crivellin:2019dwb,Sahoo:2015fla,Maji:2018gvz,Descotes-Genon:2020buf,Calibbi:2015kma,Buras:2014fpa,Altmannshofer:2009ma,Browder:2021hbl,Crivellin:2018yvo,BhupalDev:2021ipu,Altmannshofer:2020axr}. 
A key goal of this work is to exploit the $\text{SU}(2)_L$-link between charged dilepton and dineutrino couplings systematically within the standard model effective field theory
(SMEFT)  framework~\cite{Bause:2020auq}, and to work out  how dineutrino branching ratios  contribute  to deciphering the present flavor anomalies. 

This  paper is organized as follows: 
The effective field theory (EFT) framework is introduced in Sec.~\ref{sec:EFT}, discussing both high and low energy EFT descriptions of rare $B$ decays into charged dileptons and dineutrinos, and their relation.
In Sec.~\ref{sec:diffBr} we present  differential branching ratios and  SM branching ratios.
Phenomenological implications are presented in Sec.~\ref{sec:MI}: derived EFT limits on dineutrino branching ratios, bounds on dilepton couplings using the current upper limits on dineutrino modes. Test of LU with $b\to q\,\nu\bar\nu$ decays are presented in Sec.~\ref{sec:LU}, including effects from light RH neutrinos. 
We conclude in Sec.~\ref{sec:con}. 
Further details on renormalization group equation (RGE) effects, the differential branching ratios, form factors, global fits, and SM  and NP benchmark dineutrino decay distributions can be found in Appendices~\ref{app:RGEeffects}--\ref{app:diffBR}.

\section{Effective theory framework}\label{sec:EFT}

We give the weak effective theory framework for  $|\Delta b|=|\Delta q|=1$, $q=d,s$ transitions into dileptons and dineutrinos in Sec.~\ref{sec:wet}.
The SMEFT set-up is given in Sec.~\ref{sec:smeft}.

\subsection{Weak Effective Theory \label{sec:wet}}

Below the electroweak scale, $\mu<\mu_\text{EW}$, FCNC interactions between two quarks and two leptons, with flavors $\alpha,\beta$ and $i,j$, respectively, can be 
described by the following Hamiltonians for
dineutrinos
\begin{align}\label{eq:Heffnu}
    \mathcal{H}_{\text{eff}}^{\nu_i \bar \nu_j}\,=\,-\frac{4\,G_F}{\sqrt{2}}\,\frac{\alphae}{4\pi}\sum_{k}\mathcal{C}_k^{P_{\alpha\beta}ij}\,Q^{\alpha\beta ij}_k\,+\,\text{H.c.}~,
\end{align}
and for
 charged leptons, 
\begin{align}\label{eq:Heffell}
    \mathcal{H}_{\text{eff}}^{\ell_i^-\ell_j^+}\,=\,-\frac{4\,G_F}{\sqrt{2}}\,\frac{\alphae}{4\pi}\sum_{k}\mathcal{K}_k^{P_{\alpha\beta}ij}\,O^{\alpha\beta ij}_k\,+\,\text{H.c.}~.
\end{align}
The superscript $P=D$ ($P=U$) refers to the down-quark (up-quark) sector, {\it i.e.} $P_{\alpha\beta}=D_{13}$ ($U_{12}$) represents $b\to d$ ($c\to u$) transitions.
The fine structure (Fermi's) constant is denoted by $\alphae$ ($G_F$).
The low-energy dynamics of these FCNC transitions are described by dimension six operators  $Q^{\alpha\beta ij}_k$ and $O^{\alpha\beta ij}_k$. 
In absence of  light right-handed neutrinos, Eq.~\eqref{eq:Heffnu}  contains contributions from two operators only,
\begin{align}\label{eq:Qneutrinos}
Q_{L(R)}^{\alpha \beta i j} &= \left(\bar q_{L(R)}^{\,\alpha} \gamma_\mu q_{L(R)}^{\,\beta}\right)    \left(\bar \nu_{L}^{\,j} \gamma^\mu \nu_{L}^{\,i}\right) ~. 
\end{align}
The short-distance dynamics are encoded in the Wilson coefficients $\mathcal{K}_k^{P_{\alpha\beta}ij}=\mathcal{K}_{k,\text{SM}}^{P_{\alpha\beta}ij}+\mathcal{K}_{k,\text{NP}}^{P_{\alpha\beta}ij}$ 
and $\mathcal{C}_k^{P_{\alpha\beta}ij}=\mathcal{C}_{k,\text{SM}}^{P_{\alpha\beta}ij}+\mathcal{C}_{k,\text{NP}}^{P_{\alpha\beta}ij}$.
In the SM the Wilson coefficients $\mathcal{C}_{L}^{D_{\alpha\beta}ij}$ are lepton-flavor universal and 
can be written as 
\begin{align}\label{eq:SM}
    \mathcal{C}_{L,\,\rm SM}^{D_{\alpha\beta}ij}=\,V_{t\beta}V_{t\alpha}^*\,X_{\rm SM}\,\delta_{ij}~,
\end{align}
with $X_{\rm SM}=-\frac{2 \,X(x_t)}{\sin^2\theta_W}=-12.64 \pm 0.15$~\cite{Brod:2010hi,Brod:2021hsj}, where $X(x_t)$ is a loop function depending on
$x_t=\frac{m_t^2}{M_W^2}$~\cite{Misiak:1999yg,Buchalla:1998ba}. Here, $m_t$ ($M_W$) denotes the top ($W$-boson) mass
and $\theta_W$ the weak mixing angle.
The uncertainty of $X_\text{SM}$ is dominated by the one of the top mass.  Right-handed quark FCNCs
$ \mathcal{C}_{R,\,\rm SM}^{D_{\alpha\beta}ij}$ are suppressed relative to $ \mathcal{C}_{L,\,\rm SM}^{D_{\alpha\beta}ij}$ by light quark masses and neglected in this work.

The semileptonic four-fermion operators in \eqref{eq:Heffell}, which are relevant to the interplay with dineutrinos, read
\begin{align}\label{eq:Oleptons}
        O_{L(R)}^{\alpha\beta i j}&=  \left(\bar q_{L(R)}^{\,\alpha} \gamma_\mu q_{L(R)}^\beta\right) \left(\bar \ell_{L}^{\,j} \gamma^\mu \ell_{L}^{\,i}\right) ~.
\end{align}
Their Wilson coefficients are related to the ones customary to, for instance,  rare $b$-decay studies, {\it e.g.}, \cite{Kriewald:2021hfc},
\begin{align}
\begin{split}
    \mathcal{O}_{9\phantom{0}}^{ij} =& \left( \bar{s}_L\gamma_\mu b_L\right)\left( \bar{\ell}^j\gamma^\mu \ell^i \right)\,,\\ 
    \mathcal{O}_{10}^{ij} =& \left( \bar{s}_L\gamma_\mu b_L\right)\left( \bar{\ell}^j\gamma^\mu\gamma^5 \ell^i \right)\,, \\
     \mathcal{O}_{9\phantom{0}}^{ \prime ij } =& \left( \bar{s}_R\gamma_\mu b_R\right)\left( \bar{\ell}^j\gamma^\mu \ell^i \right)\,,\\ 
    \mathcal{O}_{10}^{\prime ij } =& \left( \bar{s}_R\gamma_\mu b_R\right)\left( \bar{\ell}^j \gamma^\mu\gamma^5 \ell^i \right)\,,
\end{split}
\end{align}
as
\begin{equation} \label{eq:C910}
\begin{split}
\mathcal{K}_L^{D_{23} ij} &=V_{tb} V_{ts}^* \, (\mathcal{C}_9^{ij}- \mathcal{C}_{10}^{ij} )\, , \\
\mathcal{K}_R^{D_{23}ij} &=V_{tb} V_{ts}^* \, ( \mathcal{C}_9^{\prime ij}- \mathcal{C}^{\prime ij}_{10}) \, ,
\end{split}
\end{equation}
with the CKM matrix $V$.
Further contributions to $b \to q \,\ell \ell^\prime$ transitions  arise from $SU(2)_L$-singlet leptons, via $\mathcal{C}_9^{ij}+ \mathcal{C}_{10}^{ij}$,  $\mathcal{C}_9^{\prime ij}+ \mathcal{C}_{10}^{\prime ij}$ and
dipole operators. All of these are taken into account in the global fits presented  in Sec.~\ref{sec:global} with details given in App.~\ref{app:globalfit}, however, do not matter when placing upper limits on flavorful couplings 
after matching onto  SMEFT.
We also neglect (pseudo-)scalar and tensor operators  except when considering light RH neutrinos in Sec.~\ref{sec:RH}.

\subsection{Standard Model Effective Field Theory \label{sec:smeft}}

Assuming the scale of NP to be sufficiently separated from the electroweak scale, $\mu_{\text{EW}}\lesssim \Lambda_{\text{NP}}$, 
allows to construct the SMEFT with the same dynamical matter fields (Higgs, fermions) as the SM, consistent with SM gauge symmetry $SU(3)_C\times SU(2)_L \times U(1)_Y$. 
This framework is suitable to link different sectors in flavor physics. Here we connect dineutrino and charged dilepton final states.

At leading order  in SMEFT, 
FCNC ${q_\beta\to q_\alpha\,(\ell_i^-\,\ell_j^+,\,\nu_{i}\,\bar \nu_{j})}$ transitions are governed by semileptonic four-fermion operators 
\begin{align}\label{eq:SMEFT}
    \mathcal{L}_{\text{SMEFT}}\supset    \mathcal{L}_{\text{SM}}+ \sum_i\,\frac{C_i}{v^2}\,\mathcal{O}_{C_i}~,
\end{align}
with
\begin{align}
\begin{split}
    \mathcal{O}_{C^{(1)}_{\ell q}} &=\bar Q \gamma_\mu Q \,\bar L \gamma^\mu L ~,\,\,\quad\quad \mathcal{O}_{C_{\ell u}}=  \bar U \gamma_\mu U \,\bar L \gamma^\mu L\,, \\
    \mathcal{O}_{C^{(3)}_{\ell q}}
    &=\bar Q \gamma_\mu  \tau^a Q \,\bar L \gamma^\mu \tau^a L ~,\,\,\mathcal{O}_{C_{\ell d}}=  \bar D \gamma_\mu D \,\bar L \gamma^\mu L ~,
\end{split}
\end{align}
and  $v=(\sqrt{2}\,G_F)^{-1/2} \simeq 246\,$GeV.
Here $\tau^a$ are Pauli-matrices, while $Q$ and $L$ denote quark and lepton $SU(2)_L$--doublets and $U(D)$ refer to
up-singlet (down-singlet) quarks,
where we have suppressed quark and lepton flavor indices for brevity.
Further dimension six operators,  notably penguins of type $\bar Q \gamma_\mu Q \,\phi^\dagger D^\mu \phi$, where $\phi$ denotes the Higgs and $D^\mu$
 the covariant derivative  are subject to constraints  \cite{Efrati:2015eaa,Brivio:2019ius}  and negligible for the purpose of this work.
 Operators with charged lepton singlets $E$, such as $\bar Q \gamma_\mu Q \, \bar E \gamma^\mu E$
are not connected to the dineutrino  processes. Note, in weak effective theory they break the relation $\mathcal{C}_9=-\mathcal{C}_{10}$, see the discussion after (\ref{eq:C910}).
By construction, all Wilson coefficients in SMEFT are induced by BSM physics.

Matching the SMEFT Lagrangian~\eqref{eq:SMEFT} onto Eq.~\eqref{eq:Heffnu}  and \eqref{eq:Heffell}  in the gauge basis, one finds in the down-sector,
\begin{align}  
\begin{split}
    C_{L}^{D}&={\frac{2\pi}{\alphae}}\left(C^{(1)}_{\ell q} - C^{(3)}_{\ell q}\right)  \,, \quad C_{R}^D={\frac{2\pi}{\alphae}}\,C_{\ell d} \, ,  \\
K_{L}^D&
={\frac{2\pi}{\alphae}}\left(C^{(1)}_{\ell q}  + C^{(3)}_{\ell q}  \right)\, , \quad
K_{R}^D
={\frac{2\pi}{\alphae}}\,C_{\ell d}  \, ,   
\end{split}
\end{align}
where analogous expressions for
 the up-sector are given in Refs.~\cite{Bause:2020auq,Bause:2020xzj}\footnote{In Ref.~\cite{Bause:2020xzj}, a factor $(2\pi)/\alphae$ is erroneously missing on the right-hand-side of Eq.~(46); numerical results are not affected.}.
Interestingly, there is a one-to-one map between the dineutrino and the dilepton Wilson coefficients  for right-handed quark currents, $C_R^D=K_R^D$. In contrast $C_L^D$ is not fixed in general by $K_L^D$ due to the different relative signs between $C^{(1)}_{\ell q}$ and $C^{(3)}_{\ell q}$, instead $C_L^D=K_L^U$ and $C_L^U=K_L^D$ in the gauge basis by $SU(2)_L$~\cite{Bause:2020auq}.

To express $C_k^D$ and $K_k^D$ in the mass basis, denoted by calligraphic  $\mathcal{C}_k^{D}$ and $\mathcal{K}_k^{D}$, it is necessary to perform a field rotation. 
Four different unitary rotations exist in the quark sector, corresponding to the left-handed $V_{u,d}$ and right-handed ones $U_{u,d}$, both for up- and down-type quarks.
In contrast, for leptons only two rotations are required, $V_\ell$ and $V_\nu$. 
Employing the rotations, the Wilson coefficients in the mass basis read 
\begin{align}\label{eq:massbasis}
\begin{split}
    \mathcal{C}_L^{D}\,&=\,V_\nu^\dagger\,V_d^\dagger\,C_L^{D}\,V_d\,V_\nu~,\,\,\mathcal{C}_R^{D}\,=\,V_\nu^\dagger\,U_d^\dagger\,C_R^{D}\,U_d\,V_\nu~,\\  \mathcal{K}_L^{D}\,&=\,V_\ell^\dagger\,V_d^\dagger\,K_L^{D}\,V_d\,V_\ell~,\,\,
    \mathcal{K}_R^{D}\,=\,V_\ell^\dagger\,U_d^\dagger\,K_R^{D}\,U_d\,V_\ell~.
\end{split}
\end{align}
With $C_R^D=K_R^D$, it follows that
\begin{align}\label{eq:WCsR}
\begin{split}
    \mathcal{C}_R^{D}\,=\,V_\nu^\dagger\,U_d^\dagger\,K_R^{D}\,U_d\,V_\nu\,=\,W^\dagger\,\mathcal{K}_R^D\,W~,   
\end{split} 
\end{align}
where $W=V_\ell^\dagger\,V_\nu$ is the Pontecorvo-Maki-Nakagawa-Sakata (PMNS) matrix.
For left-handed quark currents holds $C_L^U=K_L^D$ and $C_L^D=K_L^U$, hence
\begin{align}\label{eq:WCsL}
\mathcal{C}_L^{P_{\alpha\beta}}=W^\dagger\,L^P_{\alpha\beta}\,W~,
\end{align}
with
\begin{align}\label{eq:WCL}
\begin{split}
        L^{D}_{\alpha\beta}&=(V^\dagger\,\mathcal{K}_L^U\,V)_{\alpha\beta}~,  \\
        L^{U}_{\alpha\beta}&=(V \,\mathcal{K}_L^D\,V^\dagger)_{\alpha\beta}~,
\end{split}
\end{align}
where $V=V_u^\dagger\,V_d$ is the CKM-matrix. 
Expanding Eqs.~\eqref{eq:WCL} in the Wolfenstein parameter  $\lambda\approx0.2$, we obtain  for $b\to s$ transitions
\begin{align}\label{eq:bs}
    L_{23}^D\,=\,\mathcal{K}^{U_{23}}_L+\mathcal{O}(\lambda)~,
\end{align}  
 and for  $b\to d$ transitions:
\begin{align}\label{eq:bd}
    L_{13}^D\,=\,\mathcal{K}^{U_{13}}_L+\mathcal{O}(\lambda)~.
\end{align}  
Here we adopt these limits and neglect $\mathcal{O}(\lambda)$ corrections in Eqs.~\eqref{eq:bs} and \eqref{eq:bd}. Note that switching off mixing between the first two generations  causes CKM-corrections to be suppressed,
at $\mathcal{O}(\lambda^2)$ for $ L_{23}^D$ and $\mathcal{O}(\lambda^3)$ for $ L_{13}^D$. In addition, we omit renormalization group running effects generated when evolving Wilson coefficients from the NP scale $\Lambda_{\text{NP}}$ to $\mu_{\text{EW}}$. 
These effects represent a correction of less than $5\,\%$ for $\Lambda_{\text{NP}} \sim 10\,\TeV$ in Eqs.~\eqref{eq:WCsR} and \eqref{eq:WCsL}, see App.~\ref{app:RGEeffects} for details.
For $\mu<\mu_{\text{EW}}$, the vector operators given by Eqs.~\eqref{eq:Qneutrinos} and \eqref{eq:Oleptons} do not suffer from renormalization group effects since they are invariant under QCD-evolution. 

In the remainder of this work we employ the simpler notation for the NP couplings in the mass basis
\begin{align}
    \begin{split}
       \mathcal{K}_k^{tc{ij}}&=\mathcal{K}_{k,\,\text{NP}}^{U_{23}{ij}}, ~\mathcal{K}_k^{bs{ij}}\,=\mathcal{K}_{k,\,\text{NP}}^{D_{23}{ij}}, 
       ~ \mathcal{C}_k^{bsij}=\mathcal{C}_{k,\text{NP}}^{D_{23}ij} \, , \\
      \mathcal{K}_k^{tu{ij}}&=\mathcal{K}_{k,\,\text{NP}}^{U_{13}{ij}}\,,~ \mathcal{K}_k^{bd{ij}}=\mathcal{K}_{k,\,\text{NP}}^{D_{13}{ij}},
       ~ \mathcal{C}_k^{bdij}=\mathcal{C}_{k,\text{NP}}^{D_{13}ij} \, .
    \end{split}
\end{align}

\begin{table}[ht!]
 \centering
 \begin{tabular}{lcc}
  \hline
  \hline
  $B\to F_q $ & $A^{BF_q}_+$ & $A^{BF_q}_-$ \\
  &$[10^{-8}]$&$[10^{-8}]$ \\
  \hline
  $B^0\to K^0$ & $516\pm 68$ & $0$ \\
  $B^+\to K^+$ & $558 \pm 74$ & $0$ \\
  $B^0\to K^{*\,0}$ & $200\pm 29$ & $888 \pm 108$\\
  $B^+\to K^{*\,+}$ & $217 \pm 32$ & $961\pm 116$ \\
  $B^0_s\to \phi$ & $184\pm 9$ & $1110\pm 85$ \\
  $B^0\to X_s$ & $1834 \pm 193$ & $1834 \pm 193$ \\
  $B^+\to X_s$ & $1978 \pm 208$ & $1978 \pm 208$ \\ 
    && \\
  $B^0\to\pi^0$ & $154 \pm 16$ & $0$ \\
  $B^+\to\pi^+$ & $332\pm 34$ & $0$  \\
  $B^0\to\rho^{0}$ & $59 \pm 12$ & $573 \pm 233$ \\
  $B^+\to\rho^{+}$ & $126 \pm 26$ & $1236 \pm 502$\\ 
  $B^0_s\to K^0$ & $383 \pm 74$& $0$ \\
  $B^0_s\to K^{\ast 0}$ & $153\pm 9$ & $891\pm 86$ \\
  $B^0\to X_d$ & $1840 \pm 194$ & $1840 \pm 194$ \\
  $B^+\to X_d$ & $1985 \pm 209$ & $1985 \pm 209$ \\
  \hline
  \hline
  \end{tabular}
  \caption{Coefficients  $A^{BF_q}_\pm$  for $B \to F_q \,\nu \bar \nu$ as in Eq.~\eqref{eq:tapm}. 
  The uncertainties in exclusive transitions come from form factors.
  The latter induce correlations between $A_+^{BF_q}$ and $A_-^{BF_q}$ which have been taken into account in
  the SM branching ratios Tab.~\ref{tab:smPred}. 
  The uncertainty of inclusive modes is dominated by the $b$ quark mass in the 1S scheme, $m_b^{\text{1S}}=4.65\pm0.03\,\GeV$~\cite{Zyla:2020zbs}, in addition we have included $10\,\%$ of uncertainty to account for corrections of $\mathcal{O}(\Lambda^2/m_b^2)$~\cite{Altmannshofer:2009ma}. 
  As neither LCSR nor lattice results for $B_s^0 \to K^0$ are available the values of $A_\pm^{B_s^0 K^0}$ are obtained using $B \to K$ form factor input, see main text. }
    \label{tab:afactors}
\end{table}

\begin{table*}[t!]
\setlength{\tabcolsep}{6pt} 
\resizebox{\textwidth}{!}{ 
\renewcommand{\arraystretch}{1.2}
 \centering
 \begin{tabular}{lccccc}
  \hline
  \hline
   & SM, & SM, &Exp. limit& Derived& Belle II  \\
  $B\to F_q $&this work& literature &($90\,\%$ CL)& EFT limits & $5\,\text{ab}^{-1}$ ($50\,\text{ab}^{-1}$)\\
  & $[10^{-8}]$ & $[10^{-8}]$ & $[10^{-6}]$ &  $[10^{-6}]$ & $\%$ \\
  \hline
  $B^0\to K^0$ & $391 \pm 52$& $460\pm 50$~\cite{Kou:2018nap} & $26$~\cite{Grygier:2017tzo} & $15$ & --  \\  
  $B^+\to K^+$ & $423 \pm 56$ & $460\pm 50$~\cite{Kou:2018nap} & $16$~\cite{Lees:2013kla} & $16^a$ & $30\, (11)$~\cite{Kou:2018nap}\\
  $B^0\to K^{*\,0}$ & $824 \pm 99$ & $960\pm 90$~\cite{Kou:2018nap} & $18$~\cite{Grygier:2017tzo} & $18^a$ & $26 \,(9.6)$~\cite{Kou:2018nap} \\
  $B^+\to K^{*\,+}$ & $893 \pm 107$ & $960\pm 90$~\cite{Kou:2018nap} & $40$~\cite{Lutz:2013ftz} & $19$ & $25\, (9.3)$~\cite{Kou:2018nap} \\
  $B^0_s\to \phi$ & $981 \pm 69$ & $1400\pm 500$~\cite{Kim:2009mp} & $5400$~\cite{Adam:1996ts}& $23$ & --  \\
  $B^0\to X_s$ & $(28 \pm 3)\cdot 10^2$ & $(29 \pm 3)\cdot 10^2$~\cite{Buras:2014fpa} & $640$~\cite{Barate:2000rc} & $78$ & -- \\
  $B^+\to X_s$ & $(30 \pm 3)\cdot 10^2$ & $(29 \pm 3)\cdot 10^2$~\cite{Buras:2014fpa} & $640$~\cite{Barate:2000rc} & $84$ & -- \\
  && & &\\
  $B^0\to\pi^0$ & $5.4 \pm 0.6$ & $7.3\pm 0.7$~\cite{Du:2015tda} & $9$~\cite{Grygier:2017tzo} & $6$ &  --  \\
  $B^+\to\pi^+$ & $12 \pm 1$ & $14\pm 1$~\cite{Du:2015tda}& $14$~\cite{Grygier:2017tzo} & $14^a$ & --\\
  \multirow{2}{*}{$B^0\to\rho^{0}$} &  $22 \pm 8$ &   \multirow{2}{*}{$20 \pm 10$~\cite{Kim:2009mp}}& \multirow{2}{*}{$40$~\cite{Grygier:2017tzo}} & \multirow{2}{*}{$14$} & \multirow{2}{*}{--}  \\
  & $16 \pm 2^{\,\dagger}$ & & & \\
  \multirow{2}{*}{$B^+\to\rho^{+}$} & 
  $48 \pm 18$ & \multirow{2}{*}{$42 \pm 20$~\cite{Kim:2009mp}} &  \multirow{2}{*}{$30$~\cite{Grygier:2017tzo}} &   \multirow{2}{*}{$30^a$} &   \multirow{2}{*}{--}\\
  & $34 \pm 4^{\,\dagger}$ & & & \\
  $B^0_s\to K^0$ & $13 \pm 3$ & $27 \pm 16$~\cite{Kim:2009mp} & -- & $26$ & --  \\
  $B^0_s\to K^{\ast 0}$ & $36 \pm 3$ & -- & -- & $24$ & --  \\  
  $B^0\to X_d$ & $(1.3 \pm 0.1)\cdot 10^2$ & $(1.7 \pm 0.5)\cdot 10^2$~\cite{Kim:2009mp} & -- & $114$ & --\\
  $B^+\to X_d$ & $(1.4 \pm 0.1)\cdot 10^2$ & $(1.7 \pm 0.5)\cdot 10^2$~\cite{Kim:2009mp} & -- & $123$ & -- \\
  \hline
  \hline
  \end{tabular}}
  \caption{SM predictions for dineutrino modes (this work, second column) as well as SM predictions available in the literature (third column). 
  Current experimental limits at $90\,\%$ CL are displayed in the fourth column. 
  Derived EFT limits using Eqs.~\eqref{eq:MIbs} and \eqref{eq:MIbd} are displayed in the fifth column, while
  Belle II sensitivities for $5\,\text{ab}^{-1}$ ($50\,\text{ab}^{-1}$) from Ref.~\cite{Kou:2018nap} are displayed in the last column. 
  ${}^a$Input.
  ${ }^{\dagger}$Normalized to $\mathcal{B}(B\to\rho \,\ell\,\nu_\ell)_{\text{exp}}$, see App.~\ref{sec:BtorhoFFfit} for details. {Differences between our SM predictions and the literature  are due to updated CKM values and form factor improvements.}
  }
  \label{tab:smPred}
\end{table*}

\section{Dineutrino Branching ratios}\label{sec:diffBr}

In this section we present a unified description of $|\Delta b|=|\Delta q|=1$, $q=d,s$ dineutrino modes in terms of Wilson coefficients as in
(\ref{eq:xpm}). The impatient reader may jump to the parameterization of differential branching ratios (\ref{eq:dBR}) with  model-independent, decay mode specific  coefficients $a_\pm(q^2)$.
Here, $q^2$ denotes the invariant mass-squared of the dineutrinos.
The $q^2$-differential branching ratio is related to the final hadron's energy $E$-distribution in the $B$ rest frame as 
$\text{d}  {\cal{B}}/\text{d}\, q^2= 1/(2 m_B) \text{d} {\cal{B}} /\text{d} E$.
Integrated over the full  $q^2$-regions one obtains the coefficients $A_\pm^{BF_q}$ (\ref{eq:tapm}), presented in Tab.~\ref{tab:afactors}.
The SM dineutrino branching ratios are compiled in Tab.~\ref{tab:smPred}. See the following for details on decay specifics, form factors and backgrounds 
(\ref{eq:bgd}), (\ref{eq:bgd2}), or go directly
to the phenomenological implications in Sec.~\ref{sec:MI}.

The differential branching ratio of a $B$ meson decaying into a hadronic state $F_q$ with quark content $q=d,\,s$ and dineutrinos can be written as
\begin{align}\label{eq:dBR}
    \frac{\text{d}\mathcal{B}(B\to F_q\,\nu \bar \nu)}{\text{d}\, q^2} = a_+^{B F_q}(q^2)\, x_{{bq}}^+ +a_-^{B F_q}(q^2)\, x_{{bq}}^-~,
\end{align}
where only two combinations of Wilson coefficients enter
\begin{align}\label{eq:xpm}
    x_{{bq}}^\pm=  \sum_{i,j}  \big\vert{\mathcal{C}_{L,\,\rm SM}^{D_{\alpha 3}ij}+}\mathcal{C}_L^{bq ij}\pm\mathcal{C}_R^{bq ij}\big\vert^2 ~, 
\end{align}
where $\alpha=1$ for $q=d$ and $\alpha=2$ for $q=s$.
The $q^2$--dependence of $a_\pm^{B F_q}$ for different decay modes can be extracted from~\cite{Melikhov:1998ug,Colangelo:1996ay,Kim:2009mp,Altmannshofer:2009ma}, and is presented in App.~\ref{app:diffBR}. 
Information on $B\to P$ and $B\to V$ form factors is provided via supplemented files in Refs.~\cite{Gubernari:2018wyi,Straub:2015ica}, further details are provided in App.~\ref{app:formfactors}.
The authors of Refs.~\cite{Gubernari:2018wyi,Straub:2015ica} perform a fit including information from light-cone sum rules (LCSRs) at low--$q^2$ and lattice QCD for large-$q^2$, with the exception of $B_s^0 \to K^0$ and $B\to\rho$. 
For the latter, we perform a fit combining data from LCSR at low-$q^2$ from Ref.~\cite{Gubernari:2018wyi} and the available lattice QCD data from the SPQcdR~\cite{Abada:2002ie} and UKQCD~\cite{Bowler:2004zb} collaborations, 
see App.~\ref{sec:BtorhoFFfit} for details. The $B_s^0 \to K^0$ form factors are presently not available from LCSR or lattice computations, and we follow Ref.~\cite{Kim:2009mp} and use the $B^0 \to K^0$  form factor together with an estimate of flavor breaking,
\begin{align}
    f_+^{B_s^0 K^0}(q^2) = f_+^{B^0 K^0}(q^2)\,\frac{V^{B_s^0 K^{\ast 0}}(q^2)}{V^{B^0 K^{\ast 0}}(q^2)}\,.
\end{align}
Plugging the SM coefficient Eq.~\eqref{eq:SM} into the master formula Eq.~\eqref{eq:dBR}  we obtain the SM differential branching ratios with their uncertainties for the different modes, cf. black shaded regions in Fig.~\ref{fig:plotdiffBP} in App.~\ref{app:diffBR}.

Integrating the differential branching ratios given in Eq.~\eqref{eq:dBR}, one finds
\begin{align}\label{eq:BR}
    \mathcal{B}(B\to F_q\,\nu \bar \nu)= A_+^{B F_q}\,x_{bq}^+\,+ A_-^{B F_q}\,x_{bq}^-\,,
\end{align}
where 
\begin{align}\label{eq:tapm}
    A_\pm^{B F_q}&=\int_{q^2_{\text{min}}}^{q^2_{\text{max}}}\,\text{d}q^2\,a_\pm^{B F_q}(q^2)\,.
\end{align}
Here $q^2_{\text{max}}=(m_{B}-m_{F_q})^2$ for the exclusive modes and $q^2_{\text{max}}=(m_b-m_q)^2$ for inclusive modes, while $q^2_{\text{min}}=0$ in all modes. $m_{F_q}$ ($m_B$) denotes the mass of the hadronic final state ($B$ meson). 
In Tab.~\ref{tab:afactors}, we provide the central values of $A^{B F_q}_\pm$ with their symmetrized uncertainties. 

The  values of $A^{B F_q}_\pm$, for decays to pseudoscalars $A_-^{BP}=0$ and for vectors $A_+^{BV} \ll A_-^{BV}$, highlight the complementarity between the different decays modes as a result of Lorentz invariance and parity conservation in the strong interactions.

Using the values of $A_\pm^{B F_q}$ in Tab.~\ref{tab:afactors}, together with Eqs.~\eqref{eq:SM}, \eqref{eq:xpm} and \eqref{eq:BR}, and $|V_{tb}\,V_{ts}^*|=0.0397$, $|V_{tb} V_{td}^*|=0.0085$~\cite{Zyla:2020zbs},  we obtain the SM branching ratios.
Central values with their respective uncertainties from form factors are presented in the second column of Tab.~\ref{tab:smPred}. The third column of Tab.~\ref{tab:smPred} collects the SM branching ratios available in the literature, which are in good agreement with our predictions,  with  differences  due to updated CKM values and improved results of form factors.
The fourth column provides the current experimental limits at $90\,\%$ CL, while the last column displays the available Belle II sensitivities for $5\,\text{ab}^{-1}$ ($50\,\text{ab}^{-1}$)~\cite{Kou:2018nap}.

Resonant backgrounds in charged meson decays through $\tau$-leptons, $B^+ \to \tau^+ (\to F_q^+\bar \nu_\tau) \,\nu_\tau$ lead to the same final state as the search channels $B^+ \to  F_q^+ \,\bar \nu \nu$. 
The interference between the long- and short-distance contribution is negligible~\cite{Kamenik:2009kc}. The resonant  branching ratios can be written as~\cite{Du:2015tda}
\begin{align}\label{eq:br-long}
    \mathcal{B}(B^+\to F_q^+\, \bar \nu_\tau \nu_\tau)_{\text{LD}} &=\frac{G_F^4|V_{ub}V_{uq}^*|^2f_{B^+}^2f_{F_q^+}^2}{128\,\pi^2\, m_{B^+}^3\Gamma_\tau\Gamma_{B^+}}\times m_\tau(m_{B^+}^2-m_\tau^2)^2(m_{F_q^+}^2-m_\tau^2)^2~,  \end{align}
where $\Gamma_{\tau,\,B^+}$ are the decay widths of the $\tau$ and the $B^+$--meson, while $f_{B^+}$ and $f_{F^+_q}$ refer to the decay constants of 
the $B^+$ and $F_q^+$ mesons, respectively. The branching ratio in Eq.~\eqref{eq:br-long} is suppressed with respect to the short-distance contribution by two additional powers of $G_F$, however, since $\Gamma_\tau\sim \mathcal{O}(G_F^2)$, this suppression is cancelled. 
In addition, the long-distance contribution contains an enhancement with respect to the short-distance contribution triggered by the large mass of $\tau$, which yields
\begin{align} \label{eq:bgd}
    \mathcal{B}(B^+\to K^+\, \bar \nu_\tau \nu_\tau)_{\text{LD}}\sim 5 \cdot 10^{-7} ~, \\
  \mathcal{B}(B^+\to \pi^+\, \bar \nu_\tau \nu_\tau)_{\text{LD}}\sim 8 \cdot 10^{-6}~,
 \label{eq:bgd2}
\end{align}
in agreement with Ref.~\cite{Du:2015tda}. 
In rare charm dineutrino modes the analogous $\tau$-background can be avoided by appropriate cuts~\cite{Bause:2020xzj}, while in $b\to s\, \bar{\nu}\nu$ and $b\to d\, \bar{\nu}\nu$ it is irreducible and corresponds to an additional uncertainty of $\sim 10\%$ on the SM value in $b\to s\,\bar{\nu}\nu$. 
In contrast for $b\to d\,\bar{\nu}\nu$ the background yields branching ratios almost two orders of magnitude above the SM expectation. 
Since only experimental upper limits exist, we consider the full-$q^2$ region for the short distance contribution but remark that $\tau$-backgrounds will become relevant if a future measurement in this type of modes becomes available.

\section{Phenomenological implications}\label{sec:MI}

In this section, we study $b\to q\,\nu\bar\nu$ transitions and their interplay with $b\to q\,\ell^+\ell^-$ transitions in the context of the EFT framework presented in Sec.~\ref{sec:EFT}. Specifically, in Sec.~\ref{sec:derived} we work out derived limits on dineutrino modes that follow from the strongest limits on $b \to q \,\nu \bar \nu$ transitions and the 2-parameter EFT framework \eqref{eq:xpm}, \eqref{eq:BR}.
In Sec.~\ref{sec:boundcoupling} we employ SMEFT to obtain constraints from dineutrino data on charged dilepton modes.
Implications depending on lepton flavor patterns are discussed. They  turn out to be most interesting for modes into taus.
We present improved and new limits on $b \to q\, \tau^+ \tau^-$ transitions in Sec.~\ref{sec:tau}.
The impact of lepton-specific $b \to q \,\ell \ell^{(\prime)}$ data on dineutrino modes is analyzed in the next section, Sec.\ref{sec:LU}.

\subsection{Derived EFT limits \label{sec:derived}}

The different sensitivities to Wilson coefficients in $x^\pm_{bq}$ in the modes $B\to P\,\nu\bar\nu$, $B\to V\,\nu\bar\nu$, and $B\to X_q\,\nu\bar\nu$ decays 
can be exploited via Eq.~\eqref{eq:BR}, together with the current experimental limits of $B\to F_q\,\nu\bar\nu$ decays provided in Tab.~\ref{tab:smPred}. 
We extract  the following bounds on $x^\pm_{bq}$,  
\begin{align}\label{eq:MIbs}
x^+_{bs} \lesssim 2.9~, \quad x^-_{bs}+ 0.2\, x^+_{bs} \lesssim 2.0~,
\end{align}
from $B^+\to K^+\,\nu\bar\nu$ and $B^0\to K^{*0}\,\nu\bar\nu$, while limits on $x^\pm_{bd}$ are fixed by $B^+\to \pi^+\,\nu\bar\nu$ and $B^+\to \rho^{+}\,\nu\bar\nu$,
\begin{align}\label{eq:MIbd}
x^+_{bd} \lesssim 4.2~, \quad x^-_{bd}+ 0.1\, x^+_{bd} \lesssim 2.4~,
\end{align}
which are  of the same order but  weaker than~\eqref{eq:MIbs}. We derive indirect limits on branching ratios of other dineutrino modes that hold within our EFT framework.
The  limits obtained in this way are displayed in the fifth column of Tab~\ref{tab:smPred}. 
A violation of these limits would be a sign of NP carried by missing information in the EFT description, 
{\it i.e.,} light BSM particles. 

\begin{table*}
    \centering
\resizebox{\textwidth}{!}{   
\setlength{\tabcolsep}{7pt} 
\renewcommand{\arraystretch}{1.2}
    \begin{tabular}{c|c|cccccc}
    \hline
    \hline
     Data & $|\kappa_{A}^{q_1q_2\ell\ell^\prime}|$ & $ee$ & $\mu \mu $ & $\tau \tau$  & $e \mu$ & $e \tau$ & $\mu \tau$\\
     \hline
      &$| \kappa_{R}^{bd\ell\ell^\prime}|$ &  $210$ & $210$ & $210$ & $210$ & $210$ & $210$ \\
       Rare $B$ decays to & $ \kappa_{L}^{tu\ell\ell^\prime}$ &  $[-197,223]$ & $[-197,223]$ & $[-197,223]$ & $210$ & $210$ & $210$ \\
    \cline{2-8} 
    Dineutrinos & $|\kappa_{R}^{bs\ell\ell^\prime} |$ &  $35$ & $35$ & $35$ & $32$ & $32$ & $32$ \\
      & $\kappa_{L}^{tc\ell\ell^\prime}$ &  $[-22,47]$ & $[-22,47]$ & $[-22,47]$ & $32$ & $32$ & $32$ \\
    \hline
    \hline
     &$ \kappa_{R}^{bd\ell\ell^\prime}$ &  $\sim10$ & $[-4,4]$ & $\sim 2500$ & $\sim20$ & $\sim280$ & $\sim200$ \\
   Rare $B$ decays to  &$ \kappa_{L}^{bd\ell\ell^\prime}$ &  $\sim10$ & $[-8,2]$ & $\sim 2500$ & $\sim20$ & $\sim280$ & $\sim200$ \\
    \cline{2-8}
    Charged dileptons  &$ \kappa_{R}^{bs\ell\ell^\prime}$ &  $\mathcal{O}(1)$ & $[0.2,0.8]$ & $\sim 800$ & $\sim2$ & $\sim 50$ & $\sim 60$ \\
    &$ \kappa_{L}^{bs\ell\ell^\prime}$ &  $\mathcal{O}(1)$ & $[-1.6,-1.1]$ & $\sim 800$ & $\sim2$ & $\sim 50$ & $\sim 60$ \\
    \hline
    \hline
     Drell-Yan  & $| \kappa_{L,R}^{bd\ell\ell^\prime}|$ &  ${583}$ & ${314}$ & ${1122}$ & ${260}$ & ${800}$ & ${866}$ \\
    \cline{2-8} 
       & $|\kappa_{L,R}^{bs\ell\ell^\prime} |$ & ${331}$ & ${178}$ & ${637}$ & ${142}$ & ${486}$ & ${529}$ \\
    \hline
    \hline
    $t + \ell$   & $ \kappa_{L}^{tt\ell\ell^\prime}$ &  $[-196,243]$ & $[-196,243]$ & $-$ & $-$ & $-$ & $-$ \\
    \hline
    \hline
    \end{tabular}
    }
    \caption{Upper limits on $bd,\,bs,\,tu,\,tc$ charged lepton couplings $\kappa_{A}^{q_1q_2\ell\ell^\prime}$. The first row displays results   \eqref{eq:cLFCkappabs}-\eqref{eq:cLFVkappabd} from dineutrino modes worked out in Sec.~\ref{sec:boundcoupling}. 
    The second row gives constraints  from charged dilepton modes from this work, see Sec.~\ref{sec:boundcoupling} and Eqs.~\eqref{eq:KRglobalfit1sigma} and \eqref{eq:globalfitkappaRbd} for the  global $b\to q\,\mu^+\mu^-$ fit results.
    The last two rows display upper limits extracted from high--$p_T$ data, that is, Drell-Yan processes~\cite{Fuentes-Martin:2020lea,Angelescu:2020uug}, and top quark production plus leptons (admixture of electrons and muons)~\cite{Sirunyan:2020tqm}. 
    The LFV-bounds from Drell-Yan are quoted as charge-averaged, $\sqrt{ |\kappa^{\ell^+ \ell^{\prime-}}|^2 + |\kappa^{\ell^- \ell^{\prime +}}|^2}/\sqrt{2}$, whereas the other bounds are for a single coupling.}
    \label{tab:limitsonK}
\end{table*}

\subsection{Charged dilepton couplings bounded by dineutrino modes}\label{sec:boundcoupling}

The $SU(2)_L$--links provided by Eqs.~\eqref{eq:WCsR} and \eqref{eq:WCsL} allows us to connect flavor-summed branching ratios of dineutrino modes with Wilson coefficients of dilepton transitions. 
This idea was presented in Ref.~\cite{Bause:2020auq}, and phenomenologically studied for $c\to u\,\nu\bar\nu$ transitions in Ref.~\cite{Bause:2020xzj}. 
Applying this link to $b\to q\,\nu\bar\nu$ transitions, the quantities $x_{bq}^\pm$ read
\begin{align}\label{eq:xpmgen}
\begin{split}
       x_{bs}^\pm = \sum_{i,j}  \big\vert{\mathcal{C}_{L,\,\rm SM}^{D_{23}ij}}+\mathcal{K}_{L}^{{tc} ij}\pm\mathcal{K}_{R}^{bs ij}\big\vert^2 ~,\\
       x_{bd}^\pm = \sum_{i,j}  \big\vert {\mathcal{C}_{L,\,\rm SM}^{D_{13}ij}}+\mathcal{K}_{L}^{{tu} ij}\pm\mathcal{K}_{R}^{bd ij}\big\vert^2 ~,
\end{split}
\end{align}
where the sum runs over charged lepton flavors $i,\,j=e,\,\mu,\,\tau$. 
In the following we employ
\begin{align}
\begin{split}
\kappa_{L,R}^{{bq} ij} = 
\mathcal{K}_{L,R}^{bq ij} \cdot  \left( V_{tb}V_{tq}^\ast \right)^{-1}\,~,   \\
\kappa_{L,R}^{{tc} ij} = 
\mathcal{K}_{L,R}^{tc ij} \cdot  \left( V_{tb}V_{ts}^\ast \right)^{-1}\,~,   \\
\kappa_{L,R}^{{tu} ij} = 
\mathcal{K}_{L,R}^{tu ij} \cdot  \left( V_{tb}V_{td}^\ast \right)^{-1}\,~,   
\end{split}
\end{align}
where the dependence of the CKM matrix elements has been factorized for better comparison of $b\to s$ and $b \to d$ transitions.

Using Eqs.~\eqref{eq:MIbs} and \eqref{eq:MIbd}, one obtains\footnote{Since $x^\pm_{bq}\geq 0$, we conservatively considered $x^-_{bq} \,+\,0.1(0.2)\,x^+_{bq}$ as $x^-_{bq}$ in Eqs.~\eqref{eq:MIbs} and \eqref{eq:MIbd} to obtain Eqs.~\eqref{eq:genboundbs} and \eqref{eq:genboundbd}.}
\begin{align}\label{eq:genboundbs}
\begin{split}
       \sum_{i,j}  \big\vert X_{\text{SM}}\,\delta_{ij}+\kappa_{L}^{{tc} ij}+\kappa_{R}^{{bs} ij}\big\vert^2\lesssim 1.8\cdot 10^{3} ~,\\
       \sum_{i,j}  \big\vert X_{\text{SM}}\,\delta_{ij}+\kappa_{L}^{{tc} ij}-\kappa_{R}^{{bs} ij}\big\vert^2\lesssim 1.3\cdot 10^{3} ~,
\end{split}
\end{align}
for $b\to s\,\nu\bar\nu$ transitions, and
\begin{align}\label{eq:genboundbd}
\begin{split}
       \sum_{i,j}  \big\vert X_{\text{SM}}\,\delta_{ij}+\kappa_{L}^{tu ij}+\kappa_{R}^{{bd} ij}\big\vert^2\lesssim 5.8\cdot 10^{4} ~,\\
       \sum_{i,j}  \big\vert X_{\text{SM}}\,\delta_{ij}+\kappa_{L}^{{tu} ij}-\kappa_{R}^{{bd} ij}\big\vert^2\lesssim 3.3\cdot 10^{4}~,
\end{split}
\end{align}
for $b\to d\,\nu\bar\nu$ transitions.

Eqs.~\eqref{eq:genboundbs} and \eqref{eq:genboundbd} allow to set bounds on $\kappa_{L}^{tc ij}$, $\kappa_{L}^{tu ij}$ and $\kappa_{R}^{{bq} ij}$ depending on lepton 
flavor. We first discuss lepton universality, followed by charged Lepton Flavor Conservation (cLFC) and then the general case.

If LU holds, that is,  $\kappa_{A}^{{q_1q_2} ij}\propto \delta_{ij}$,  the double-sums in Eqs.~\eqref{eq:genboundbs} and \eqref{eq:genboundbd}
collapse to an overall factor of 3.
Assuming  real-valued  couplings $\kappa_{A}^{q_1q_2\ell \ell}\in \mathds{R}$  one obtains
\begin{align}\label{eq:LUkappabs}
    |\kappa_{R}^{bs\ell\ell}|\lesssim 23~,\quad
    -10\lesssim\kappa_{L}^{{tc} \ell\ell}\lesssim 35~,
\end{align}
and
\begin{align}\label{eq:LUkappabd}
   |\kappa_{R}^{bd\ell\ell}|\lesssim 122~,\quad
    -109\lesssim\kappa_{L}^{tu\ell\ell}\lesssim 134~, 
\end{align}
for $b\to s$ and $b\to d$ transitions, respectively. 

If cLFC holds, the double-sums in Eqs.~\eqref{eq:genboundbs} and \eqref{eq:genboundbd} only run over diagonal charged lepton flavor indices. Resulting bounds are weaker than the LU ones in Eqs.~\eqref{eq:LUkappabs} and \eqref{eq:LUkappabd}, since we only consider one of the BSM couplings entering the sums at a time, whereas the $X_\text{SM}$ term contributes for each generation:
$ 2  \,\big\vert X_{\text{SM}} \big\vert^2+  \big\vert X_{\text{SM}} +\kappa_{L}^{{tc} \ell \ell }+\kappa_{R}^{{bs} \ell \ell}\big\vert^2\lesssim 1.8\cdot 10^{3}$ for the example of
the first equation of \eqref{eq:genboundbs}.
Assuming real-valued couplings, we obtain
\begin{align}\label{eq:cLFCkappabs}
    |\kappa_{R}^{bs\ell\ell}|\lesssim 35~,\quad
    -22\lesssim\kappa_{L}^{tc \ell\ell}\lesssim 47~,
\end{align}
and
\begin{align}\label{eq:cLFCkappabd}
    |\kappa_{R}^{bd\ell\ell}|\lesssim 210\,~,\quad
    -197\lesssim\kappa_{L}^{tu \ell\ell}\lesssim 223~,
\end{align}
for $b\to s$ and $b\to d$ transitions, respectively.

In general also lepton flavor violating (LFV) couplings $\kappa_{A}^{q_1q_2 \ell{\ell^\prime}}$ with $\ell \neq \ell^\prime$, that is $\ell\ell^\prime= e\mu,\, \mu \tau,\, e \tau$ and permutations, appear. We consider one LFV-coupling at a time, without SM-interference, which gives the constraint
$ 3 \, \big\vert X_{\text{SM}} \big\vert^2+  \big\vert \kappa_{L}^{{tc} \ell \ell^\prime }+\kappa_{R}^{{bs} \ell \ell^\prime }\big\vert^2\lesssim 1.8\cdot 10^{3}$ for the example of
the first equation of \eqref{eq:genboundbs}.
We obtain
\begin{align}\label{eq:cLFVkappabs}
    |\kappa_{R}^{bs\ell\ell^\prime}|\lesssim 32~,\quad
    |\kappa_{L}^{{tc} \ell\ell^\prime}|\lesssim 32~,
\end{align}
and
\begin{align}\label{eq:cLFVkappabd}
    |\kappa_{R}^{bd\ell\ell^\prime}|\lesssim 210~,\quad
    |\kappa_{L}^{tu \ell\ell^\prime}|\lesssim 210~,
\end{align}
for $b\to s$ and $b\to d$ transitions, respectively. 
Constraints on the lepton flavor diagonal couplings as in \eqref{eq:cLFCkappabs}, \eqref{eq:cLFCkappabd} continue to hold in the general case.

In Tab.~\ref{tab:limitsonK} we compile the  limits presented in \eqref{eq:cLFCkappabs}-\eqref{eq:cLFVkappabd} from dineutrino data, together with limits 
from decays to dileptons, 
using Drell-Yan~\cite{Fuentes-Martin:2020lea,Angelescu:2020uug},~\footnote{In previous works~\cite{Bause:2020auq,Bause:2020xzj}, 
the LFV bounds on $|\kappa^{\ell^+ \ell^{\prime-}}|$, $\ell\neq\ell^\prime$, are not normalized by $1/\sqrt{2}$, i.e. presented flavor-summed and not averaged.
} and top quark production plus lepton~\cite{Sirunyan:2020tqm} data. 

The limits on $b$-FCNCs with dimuons, $\kappa_{L,R}^{bd\mu\mu}$ and $\kappa_{L,R}^{bs\mu\mu}$, are the strongest. They have been extracted 
from global fits to $b\to q\,\mu^+\mu^-$ data, presented in Sec.~\ref{sec:global}.

We work out bounds on all other couplings $\kappa_{A}^{bq \ell \ell^\prime}$, $A=L,R$, from (semi)leptonic rare $B$-decays using \textit{flavio} \cite{Straub:2018kue}, assuming one  coupling at a time, $|\mathcal{C}_9|=|\mathcal{C}_{10}|=\kappa_L/2$ or $|\mathcal{C}_9^\prime|=|\mathcal{C}_{10}^\prime|=\kappa_R/2$, see Eq.~\eqref{eq:C910}.
The strongest non-$\mu \mu$ limits stem from the current experimental upper bounds on the branching ratios of $B^+\to\pi^+ e^+e^-$, $B^0\to\tau^+\tau^-$, $B^0_s\to\tau^+\tau^-$, $B\to K^{(*)}\mu^\pm e^\mp$, $B^+\to\pi^+\mu^\pm e^\mp$, $B^+\to K^+\tau^\pm e^\mp$, $B^0\to \tau^\pm e^\mp$, $B_s^0\to \mu^\pm\tau^\mp$ and $B^0\to \tau^\pm\mu^\mp$, given in Ref.~\cite{Zyla:2020zbs}. 
For $b s e e$ couplings, we note that, while in principle doable, a global $b\to s e^+ e^-$ fit is beyond the scope of this work.

For $b \to d$ transitions, we observe that dineutrino constraints are by a factor of 1.3 and 12 stronger than limits from charged dilepton modes into $e \tau$ and $\tau \tau$, respectively.
For $b \to s$ transitions,  constraints from dineutrinos modes improve  limits from charged dilepton data by a factor of 2, 2 and 23
in $e \tau$, $\mu \tau$ and $\tau \tau$ final states, respectively.

In addition, Tab.~\ref{tab:limitsonK} shows that dineutrino bounds on $\kappa_R^{bs\ell\ell^\prime}$ couplings are a factor 4 or more stronger (depending on the coupling) than Drell-Yan data. For $\kappa_R^{bd\ell\ell^\prime}$, our limits on $\ell\ell^\prime=\mu\mu, e\mu$ couplings are slightly better than Drell-Yan data, while for the rest our limits are a factor 3 stronger or more, again depending on the coupling.

We also obtain constraints from dineutrino modes on left-handed  couplings  with top-quarks on  $t\to u$ and $t\to c$ FCNCs.
We compare them to the constraints from a recent LHC analysis with tops and  dielectrons and dimuons by the CMS experiment~\cite{Sirunyan:2020tqm}. 
We find that the limits from dineutrino modes on $\kappa_{L}^{tc\ell \ell}$, $\ell=e, \mu$ are stronger than the ones with ditops by roughly a factor 5, whereas the ones on 
$\kappa_{L}^{tu\ell\ell}$ are comparable. Assuming a top-philic flavor pattern the ditop coupling induces FCNC ones (in the down mass basis),
{\it e.g.,} \cite{Bissmann:2020mfi},
$\kappa_{L}^{tc\ell \ell} \sim \kappa_{L}^{tt \ell \ell} \cdot (V_{tb} V_{ts}^*)$ and $\kappa_{L}^{tu\ell\ell} \sim \kappa_{L}^{tt\ell\ell} \cdot  (V_{tb} V_{td}^*)$.
Under these assumptions we obtain the bounds 
$\kappa_{L}^{tc\ell \ell} \sim [-8,10]$, $\kappa_{L}^{tu\ell\ell} \sim [-2 ,2]$, stronger than the dineutrino ones. On the other hand, the constraints from dineutrino data are 
available and of similar size for all lepton flavors $\ell \ell^\prime$, whereas the collider limits from \cite{Sirunyan:2020tqm} are limited to $ee$ and $\mu \mu$.

\subsection{Improved limits on $b\to q\,\tau^+\tau^-$ decays \label{sec:tau}}

In the previous section, we have shown that dineutrino data establish the most stringent bounds on $\kappa_R^{bq\tau\tau}$ couplings, followed by the Drell-Yan data where also $\kappa_L^{bq\tau\tau}$ is constrained. Using the complementarity between both approaches, that is, bounds on $\kappa_L^{bq\tau\tau}$ from Drell-Yan data, and bounds on $\kappa_R^{bq\tau\tau}$ from dineutrino data, this allows us to improve the current experimental upper limits on branching ratios of $b\to q\,\tau^+\tau^-$ decays~\cite{Zyla:2020zbs,Belle:2021ndr} at $95\%$\,CL ($90\%$\,CL for $B^+\to K^+\tau^+\tau^-$ and $B^0\to K^{*0}\tau^+\tau^-$)
\begin{align}
\begin{split}
    \mathcal{B}(B^0\to\tau^+\tau^-)_\text{exp} &<2.1\times 10^{-3}~,\\
    \mathcal{B}(B^0_s\to\tau^+\tau^-)_\text{exp} &<6.8\times 10^{-3}~,\\
    \mathcal{B}(B^+\to K^+\tau^+\tau^-)_\text{exp} &<2.25\times 10^{-3}~,\\
    \mathcal{B}(B^0\to K^{*0}\tau^+\tau^-)_\text{exp} &<2.0\times 10^{-3}~,
\end{split}
\end{align}
or even obtain novel ones. 

Our indirect limits on branching ratios of $b\to q\,\tau^+\tau^-$ decays are obtained using the limits on $\kappa_R^{bq\tau\tau}$ given by Eqs.~\eqref{eq:cLFCkappabs} and \eqref{eq:cLFCkappabd}, while limits on $\kappa_L^{bq\tau\tau}$ from Drell-Yan data from Tab.~\ref{tab:limitsonK}. Using \textit{flavio}~\cite{Straub:2018kue}, neglecting effects from scalar and tensor operators, and considering two couplings at a time  with $\kappa_L\sim 2\,\mathcal{C}_9\sim 2\,\mathcal{C}_{10}$ and $\kappa_R\sim 2\,\mathcal{C}_9 ^\prime\sim 2\,\mathcal{C}_{10}^\prime$,\footnote{We avoided the possibility of large cancellations by varying signs in  Eqs.~\eqref{eq:cLFCkappabs} 
and \eqref{eq:cLFCkappabd}.} we find the following upper limits for $b\to s\,\tau^+\tau^-$ transitions
\begin{equation}
\label{eq:clfctaulimits}
\begin{split}
    \mathcal{B}(B_s\to \tau^+\tau^-)\lesssim 5.0\cdot10^{-3}\,,\\
    \mathcal{B}(B^0\to K^0 \,\tau^+\tau^-)^{[15,\,22]}\lesssim 7.8\cdot10^{-4}\,,\\
    \mathcal{B}(B^+\to K^+ \,\tau^+\tau^-)^{[15,\,22]}\lesssim 8.4\cdot10^{-4}\,,\\
    \mathcal{B}(B^0\to K^{\ast 0} \,\tau^+\tau^-)^{[15,\,19]}\lesssim 7.4\cdot10^{-4}\,,\\
    \mathcal{B}(B^+\to K^{\ast +} \,\tau^+\tau^-)^{[15,\,19]}\lesssim 8.1\cdot10^{-4}\,,\\
    \mathcal{B}(B_s\to \phi \,\tau^+\tau^-)^{[15,\,18.8]}\lesssim 6.8\cdot10^{-4}\,. \\
\end{split}
\end{equation}
These are well above their respective SM predictions
\begin{equation}
\begin{split}
     \mathcal{B}(B_s\to \tau^+\tau^-)_{\text{SM}}= (7.78\pm 0.31)\cdot10^{-7}\,,\\
    \mathcal{B}(B^0\to K^0 \tau^+\tau^-)^{[15,\,22]}_{\text{SM}}= (1.17\pm 0.12)\cdot10^{-7}\,,\\
    \mathcal{B}(B^+\to K^+ \tau^+\tau^-)^{[15,\,22]}_{\text{SM}}= (1.26\pm 0.14)\cdot10^{-7}\,,\\
    \mathcal{B}(B^0\to K^{\star 0} \tau^+\tau^-)^{[15,\,19]}_{\text{SM}}= (0.97\pm 0.10)\cdot10^{-7}\,,\\
    \mathcal{B}(B^+\to K^{\star +} \tau^+\tau^-)^{[15,\,19]}_{\text{SM}}= (1.05\pm 0.11)\cdot10^{-7}\,,\\
    \mathcal{B}(B_s\to \phi\, \tau^+\tau^-)^{[15,\,18.8]}_{\text{SM}}=(0.90\pm0.07)\cdot10^{-7}\,,
\end{split}
\end{equation}
consistent with \cite{Capdevila:2017iqn}, 
where the superscript indicates the $q^2$--range in GeV$^2$ for the dilepton invariant mass squared. The broad bins above $15\, \mbox{GeV}^2$ remove the
$\psi(2S)$ resonance and support  the use of the operator product expansion in $1/Q, Q=(m_b, \sqrt{q^2})$~\cite{Grinstein:2004vb}. 

Following the same procedure for $b\to d\, \tau^+ \tau^-$ transitions, we obtain the upper limits
\begin{align}
\label{eq:taulimits-bd}
\begin{split}
    \mathcal{B}(B^0\to\tau^+\tau^-)&\lesssim 6.0\cdot10^{-4}~,\, \\
    \mathcal{B}(B^0\to \pi^0 \tau^+ \tau^-)^{[15,22]}&\lesssim {2.5}\cdot10^{-5}~,\,\\
    \mathcal{B}(B^+\to \pi^+ \tau^+ \tau^-)^{[15,22]}&\lesssim {5.3}\cdot10^{-5}~,\, 
\end{split}
\end{align}
several orders above their respective SM predictions
\begin{equation}
\begin{split}
     \mathcal{B}(B^0\to \tau^+\tau^-)_{\text{SM}}= (2.39 \pm 0.24) \cdot 10^{-8}\,,\\
    \mathcal{B}(B^0\to \pi^0 \tau^+ \tau^-)_{\text{SM}}^{[15,22]}=({0.20}\pm 0.02)\cdot 10^{-8} \,,\\
    \mathcal{B}(B^+\to \pi^+ \tau^+ \tau^-)_{\text{SM}}^{[15,22]}=({0.44}\pm0.05)\cdot 10^{-8} \,.
\end{split}
\end{equation}

Belle II with $5\,\text{ab}^{-1}$ ($50\,\text{ab}^{-1}$) is expected to place following (projected) upper limits on the branching ratios~\cite{Kou:2018nap}
\begin{align}
    \begin{split}
        \mathcal{B}(B_s\to\tau^+\tau^-)_{\text{proj}}&<\,8.1\,(-)\cdot 10^{-5}~,\\
        \mathcal{B}(B^+\to K^+\,\tau^+\tau^-)_{\text{proj}}&<\,6.5\,(2.0)\cdot 10^{-5}~,\\
        \mathcal{B}(B^0\to\tau^+\tau^-)_{\text{proj}}&<\,30\,(9.6)\cdot 10^{-5}~,
    \end{split}
\end{align}
which cover the regions \eqref{eq:clfctaulimits}, \eqref{eq:taulimits-bd}.
We stress that the latter are based on the general  bounds given by Eqs.~\eqref{eq:cLFCkappabs} and \eqref{eq:cLFCkappabd}, and allow to constrain
models of new physics.

\section{Testing universality with $\boldsymbol{b\to q\,\nu\bar\nu}$  \label{sec:LU}}

In the previous sections we have exploited the $\text{SU}(2)_L$--link, given by Eqs.~\eqref{eq:WCsR} and \eqref{eq:WCsL}, using the current experimental upper limits on dineutrino branching ratios from Tab.~\ref{tab:smPred} to extract bounds on flavor specific charged dilepton couplings, $\kappa_{R}^{bs\ell\ell^\prime}$ and $\kappa_{R}^{bd\ell\ell^\prime}$. 
Since this link is \emph{bidirectional}, 
we can also explore the implications of charged dilepton data on dineutrino modes. 
To do so we use global fits to dimuon data as the strongest available bounds. The numerical results of the $|\Delta b|=|\Delta s|=1$ and $|\Delta b|=|\Delta d|=1$ fits have been already presented in Tab.~\ref{tab:limitsonK}.

\begin{figure*}[ht!]
    \centering
    \includegraphics[width=7cm,height=5.5cm]{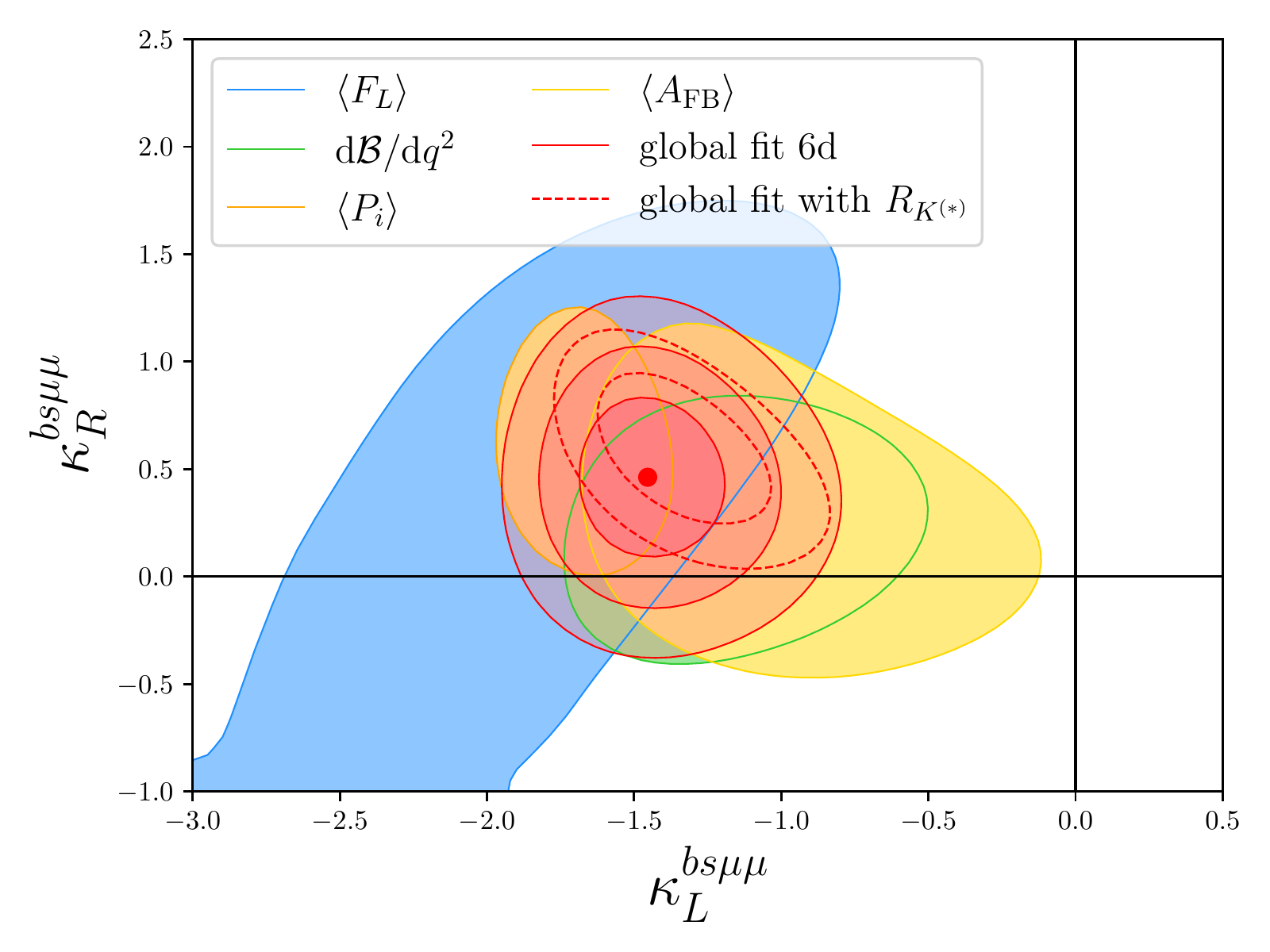}
    \includegraphics[width=7cm,height=5.5cm]{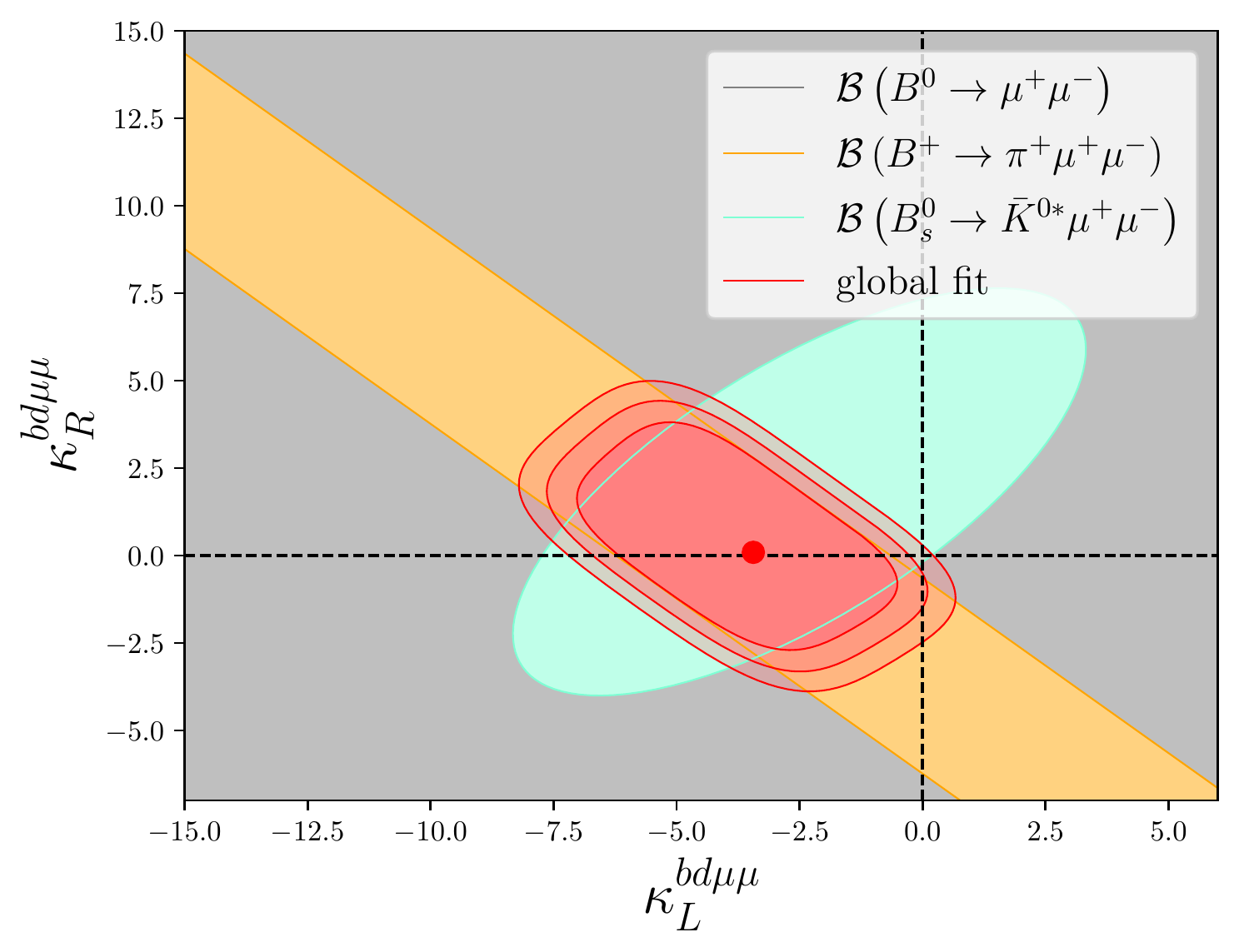}   
    \caption{
    Global fits to rare $B$-decay data on $|\Delta b|=|\Delta s|=1$ (left panel) and on $|\Delta b|=|\Delta d|=1$  transitions (right panel).
        \textbf{ Left plot:} $\kappa_{L}^{bs\mu\mu}$--$\kappa_{R}^{bs\mu\mu}$ plane with its best fit values (red point), and the $1,~2,$ and $3\,\sigma$ contours (red shaded areas). 
         $1\,\sigma$ contours for different sets of observables are shown in blue for $\langle F_L\rangle$, green for $\langle\text{d}\,\mathcal{B}/\text{d}q^2\rangle$, orange for $\langle P_i\rangle $, and yellow  for $\langle A_{\text{FB}}\rangle $. 
         Red dashed contours show the impact of $R_{K^{(*)}}$ data, when included in the global fit.
 \textbf{ Right plot:}  $1,~2,$ and $3\,\sigma$ fit contours (red shaded areas) in the $\kappa_{L}^{bd\mu\mu}$--$\kappa_{R}^{bd\mu\mu}$ plane and their best fit values (red point). 
    The impact of $\mathcal{B}(B^+\to\pi^+\,\mu^+\mu^-)$ and $\mathcal{B}(B^0_s\to \bar{K}^{\ast 0}\,\mu^+\mu^-)$ can be read off from their $1\sigma$ contours, orange and celeste, respectively.  The $B^0 \to \mu^+ \mu^-$ limit is included in the global fit, but of smaller impact (grey area, which fills the whole plot region). The $|\Delta b|=|\Delta d|=1$ fit results are adapted from \cite{Bause:inprep}.
    }
    \label{fig:globalfit}
\end{figure*}

\subsection{Global fits \label{sec:global}}

\subsubsection{ $|\Delta b|=|\Delta s|=1$ }

Using the available experimental information
on $b\to s\,\mu^+\mu^-$ data (excluding $R_{K^{(*)}}$), we perform a global fit with \textit{flavio} \cite{Straub:2018kue} of the semileptonic Wilson coefficients $\Call$. Results are given in Tab.~\ref{tab:FitValues_noRk}.
The six-dimensional fit  yields the following $1\,\sigma$ fit values for the NP coupling $\kappa_{L,R}^{bs\mu\mu}$, see also (\ref{eq:C910}),
\begin{align}\label{eq:KRglobalfit1sigma}
\begin{split}
\kappa_{L}^{bs\mu\mu} &= \Cnine - \Cten =-1.45 \pm 0.29~,\\
 \kappa_{R}^{bs\mu\mu} &= \Cninepr - \Ctenpr =\phantom{-} 0.46 \pm 0.26~. 
\end{split}
\end{align}
Eq.~\eqref{eq:KRglobalfit1sigma} exhibits a clear tension between $b\to s \,\mu^+\mu^-$ data and the SM, which can  be described
by  a \emph{pull} from the SM, $\text{pull}_\text{SM}$, in units of standard deviations $\sigma$.
This fit gives $\text{pull}_\text{SM}=4.6\,\sigma$, with a goodness of fit $\chi^2/\text{dof}=0.91$.

In the  left plot of Fig.~\ref{fig:globalfit} we show the $1,~2,$ and $3\,\sigma$ fit contours (red shaded areas) in the $\kappa_{L}^{bs\mu\mu}$--$\kappa_{R}^{bs\mu\mu}$ plane and its best fit values (red point).  
The $1\,\sigma$ regions for different sets of observables are shown in blue for $\langle F_L\rangle$, green for $\langle\text{d}\,\mathcal{B}/\text{d}q^2\rangle$, orange for $\langle P_i\rangle $, and yellow  for $\langle A_{\text{FB}}\rangle$. 
Red dashed lines show the impact of $R_{K^{(*)}}$ data  when included in the global fit.
The bounds provided by Eq.~\eqref{eq:KRglobalfit1sigma} 
are a factor 20 stronger than the universality limit  extracted from dineutrino data~\eqref{eq:LUkappabs}. 
Further information and additional global fits including $R_{K^{(*)}}$ data can be found in App.~\ref{app:globalfit}.

The reason for keeping the universality ratios $R_{K^{(*)}}$ out of the global fit  is that they can be affected by NP in muon, but also in electron couplings. While presently
the consistency of the fits gives no reason to include electron effects, they cannot be excluded and need to be studied with electron-specific measurements and fits.

\subsubsection{ $|\Delta b|=|\Delta d|=1$ }

In $b\to d\,\mu^+\mu^-$ transitions, information from global fits is currently only available in Refs.~\cite{Du:2015tda,Ali:2013zfa,Rusov:2019ixr}, and is mainly based on the current experimental information on $B^+\to\pi^+\,\mu^+\mu^-$. 
However, further information can be obtained from the recent update on  $\mathcal{B}(B^0\to\mu^+\mu^-)=(0.56\pm0.7)\cdot 10^{-10}$~\cite{Altmannshofer:2021qrr} at $95$\% CL, where the quoted value includes the recent result from LHCb~\cite{LHCb:2021awg,LHCb:2021vsc}, 
in addition to the first evidence for $\mathcal{B}(B_s^0\to \bar{K}^{*\,0} \mu^+\mu^-)=$ $(2.9\pm 1.1)\cdot 10^{-8}$ \cite{Aaij:2018jhg}. 

We employ results of Ref.~\cite{Bause:inprep} on a four-dimensional fit to the aforementioned modes and data from $B^+\to\pi^+\mu^+\mu^-$ to obtain
constraints on $\kappa_{L}^{bd\mu\mu}$ and $\kappa_{R}^{bd\mu\mu}$. 
The main difference with the results in $b\to s$ is that in $ b \to d$ we obtain two solutions for the four-dimensional fit\footnote{In contrast to the $b\to s\,\mu^+\mu^-$ global fit, for $b\to d\,\mu^+\mu^-$ we do not consider contributions from dipole couplings $\mathcal{C}_7^{(\prime)}$.}, in this work we consider the solution with the smallest $\chi^2/\text{dof}=0.28$ that gives
\begin{align}\label{eq:globalfitkappaRbd}
\begin{split}
    \kappa_{L}^{bd\mu\mu}&=-3 \pm 5\,,\\
    \kappa_{R}^{bd\mu\mu}&= \phantom{-}0 \pm 4\,,
\end{split}
\end{align}
with $\text{pull}_\text{SM}=1.92\,\sigma$. Eq.~\eqref{eq:globalfitkappaRbd} is a factor 40 stronger than the limit in Eq.~\eqref{eq:LUkappabd}. 
In the  right plot of Fig.~\ref{fig:globalfit} we display the $1,~2,$ and $3\,\sigma$ fit contours (red shaded areas) in the $\kappa_{L}^{bs\mu\mu}$--$\kappa_{R}^{bs\mu\mu}$ plane and its best fit values (red point). 
The impact of $\mathcal{B}(B^+\to\pi^+\,\mu^+\mu^-)$  and $\mathcal{B}(B^0_s\to \bar{K}^{* 0}\,\mu^+\mu^-)$  in the global fit can be read off from its $1\sigma$ contours, orange and celeste, respectively. 
The $B^0 \to \mu^+ \mu^-$ limit is presently of lesser importance (grey area, which covers  the whole plot region).

Future measurements of $b \to d\, \,\mu^+ \mu^-$ modes are necessary to improve the fit and exclude one of the possible two solutions. 
For details of the $b\to d\,\mu^+\mu^-$ global fit we refer to Ref.~\cite{Bause:inprep}.

\subsection{Universality tests  with $b\to q\,\nu\bar\nu$, $q=d,s$}

Particularizing Eq.~\eqref{eq:BR} to the LU limit via Eq.~\eqref{eq:xpmgen}, the branching ratios for $B\to V\,\nu\bar\nu$ and $B\to P\,\nu\bar\nu$ decays
assuming lepton universality are obtained as
\begin{align}\label{eq:LUBtoV}
    &\mathcal{B}(B\to V\,\nu\bar\nu)_{\text{LU}}\,=\,A_+^{BV}\,x_{bq,\text{LU}}^+ +\,A_-^{BV}\,x_{bq,\text{LU}}^-~,\nonumber\\
    &\mathcal{B}(B\to P\,\nu\bar\nu)_{\text{LU}}\,=\,A_+^{BP}\,x_{bq,\text{LU}}^+~,
\end{align}
respectively, with
\begin{align}\label{eq:xpmLU}
    x_{bq,\text{LU}}^\pm\,=\,3\,\left|V_{tb}V_{tq}^*\right|^2\,\left(X_{\text{SM}}+\kappa_{L}^{tq^\prime\ell\ell}\pm\kappa_{R}^{bq\ell\ell}\right)^2~,
\end{align}
with $q^\prime=u,\,(c)$ for $q=d,\,(s)$, respectively.
Solving $\mathcal{B}(B\to P\,\nu\bar\nu)_{\text{LU}}$ given by Eq.~\eqref{eq:LUBtoV}, we find two solutions
\begin{align}\label{eq:LUsol}
   \kappa_{L}^{tq^\prime\ell\ell}=-X_{\text{SM}}-\kappa_{R}^{bq\ell\ell}\pm\sqrt{ \frac{\mathcal{B}(B\to P\,\nu\bar\nu)_{\text{LU}}}{3\,\left|V_{tb}V_{tq}^*\right|^2 A_+^{BP} }}~.
\end{align}

One may plug Eq.~\eqref{eq:LUsol} into Eq.~\eqref{eq:LUBtoV} which yields a correlation between two branching ratios assuming LU,
\begin{align}\label{eq:luregion}
    &\mathcal{B}(B\to V\,\nu\bar\nu)_{\text{LU}}\,=\,\frac{A_+^{BV}}{A_+^{BP}}\,\mathcal{B}(B\to P\,\nu\bar\nu)_{\text{LU}}\\
    &\,+\,3\,A_{-}^{BV}\,\left|V_{tb}V_{tq}^*\right|^2\left(\sqrt{ \frac{\mathcal{B}(B\to P\,\nu\bar\nu)_{\text{LU}}}{3\,\left|V_{tb}V_{tq}^*\right|^2 A_+^{BP} }}\mp 2\,\kappa_{R}^{bq\ell\ell}\right)^2~.\nonumber
\end{align}
Information on $A_\pm^{BF_q}$ is provided in Tab.~\ref{tab:afactors}, and the most stringent limits on $\kappa_R^{bq\ell\ell}$ are given for $\ell\ell=\mu\mu$ by Eqs.~\eqref{eq:KRglobalfit1sigma} and \eqref{eq:globalfitkappaRbd}.
Performing a Taylor expansion up to $\mathcal{O}(\kappa_{R}^{bq\mu\mu})$ we observe that Eq.~\eqref{eq:luregion} results in
\begin{align}\label{eq:LUratio_series}
 \frac{\mathcal{B}(B\to V\,\nu\bar\nu)_{\text{LU}}}{\mathcal{B}(B\to P\,\nu\bar\nu)_{\text{LU}}} \,&= \,\frac{A_+^{BV}+A_-^{BV}}{A_+^{BP}}\\
 \pm\, 4\sqrt{3}\,\left|V_{tb}V_{tq}^*\right|&\,\frac{A_-^{BV}}{\sqrt{A_+^{BP}\cdot\mathcal{B}(B\to P\,\nu\bar\nu)_{\text{LU}}}}\,\kappa_{R}^{bq\mu\mu}\,, \nonumber
\end{align}    
where for $\mathcal{B}(B\to P\,\nu\bar\nu)_{\text{LU}}\gtrsim \mathcal{O}(10^{-6})$ (SM-like or larger) and $\kappa_{R}^{bq\mu\mu}\lesssim \mathcal{O}(1)$ (as in Eqs.~\eqref{eq:KRglobalfit1sigma} and \eqref{eq:globalfitkappaRbd}), the first term in (\ref{eq:LUratio_series}) becomes an excellent approximation for
the ratio of branching fractions into vectors and pseudoscalars given that universality holds.
Importantly,  it is otherwise independent of new physics with uncertainties  fully dominated by form factor ones.  In the subsequent analysis we use the full expression.

\begin{figure*}[ht!]
    \centering
    \includegraphics[width=7cm,height=5.5cm]{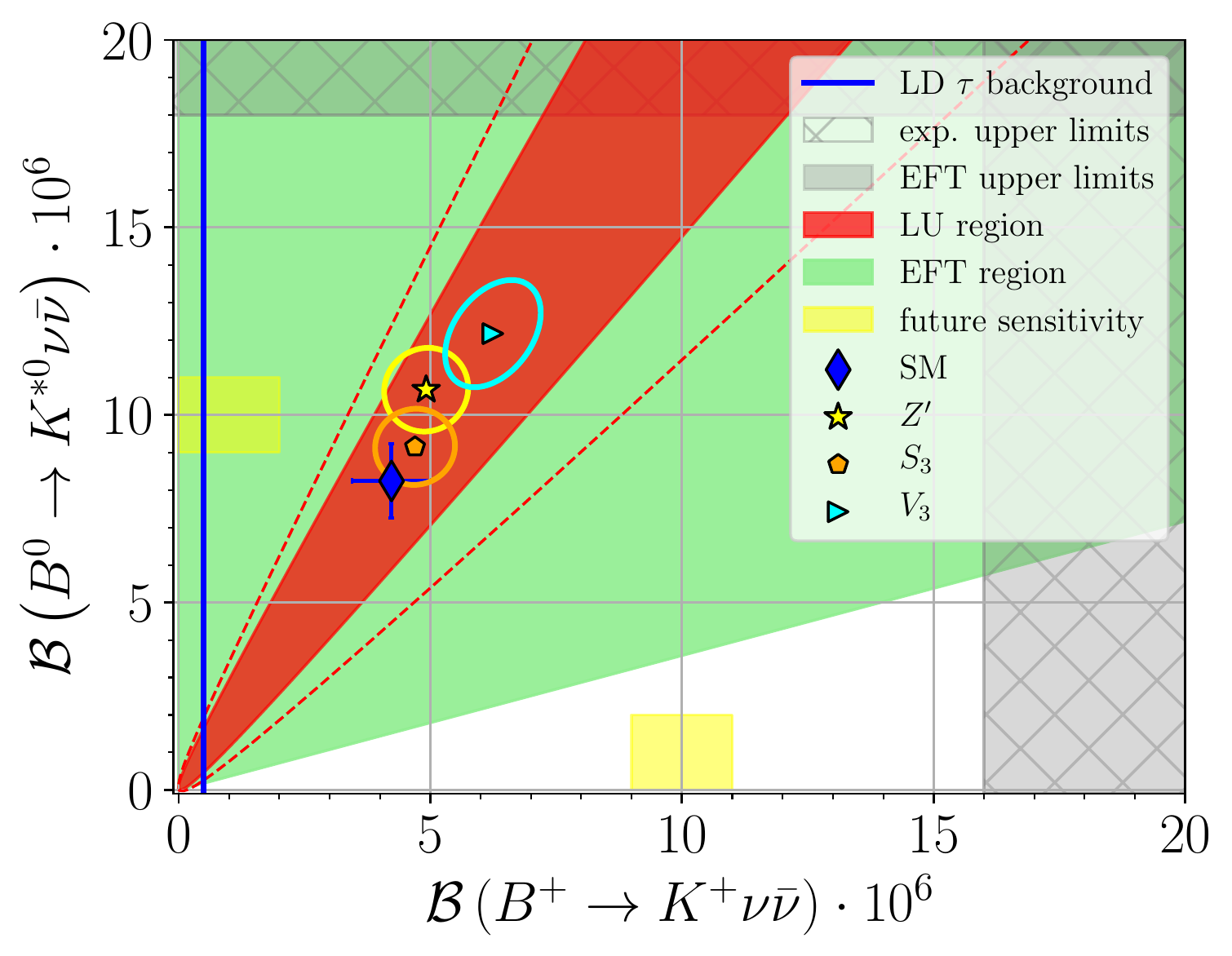}
   \includegraphics[width=7cm,height=5.5cm]{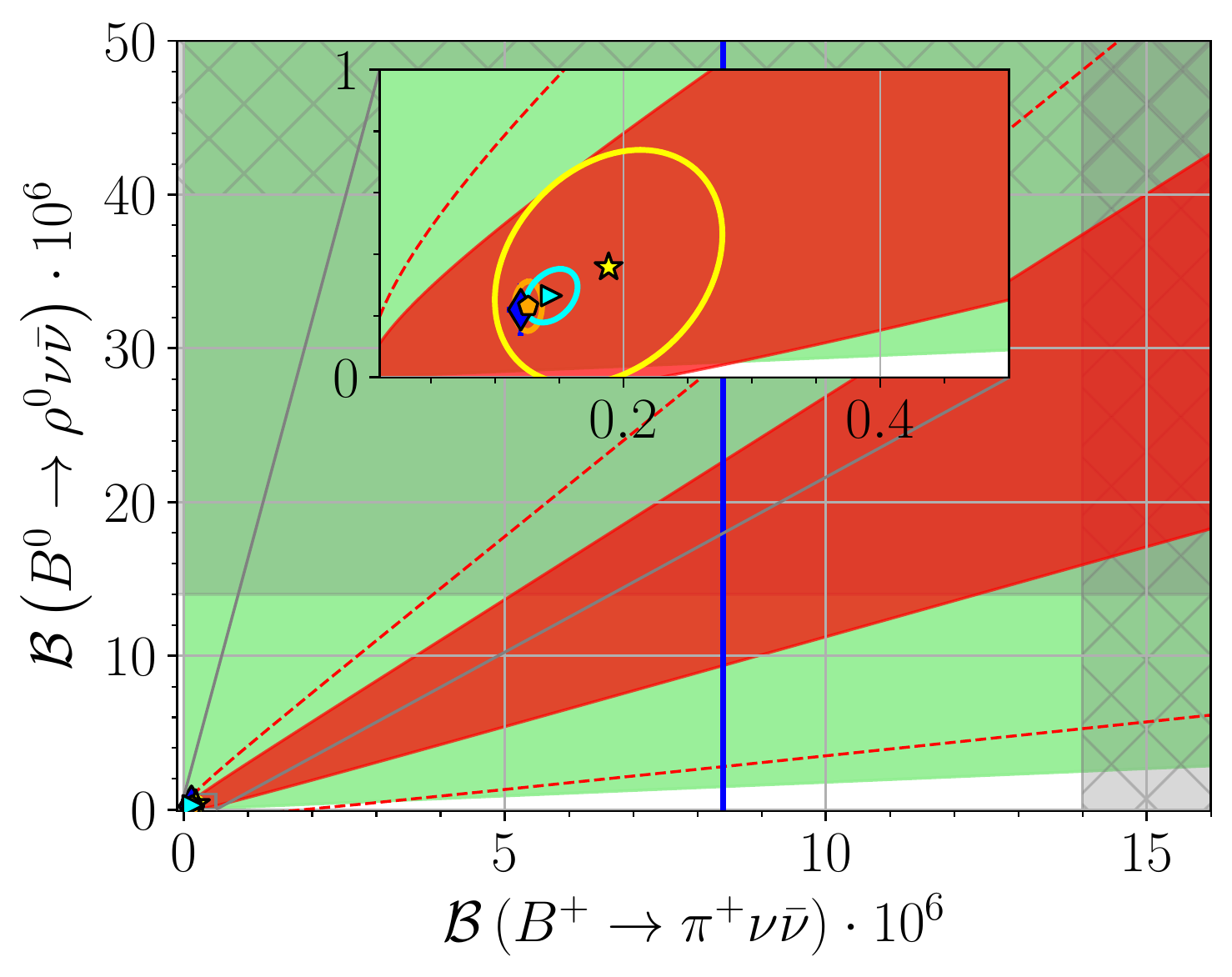}
    \includegraphics[width=7cm,height=5.5cm]{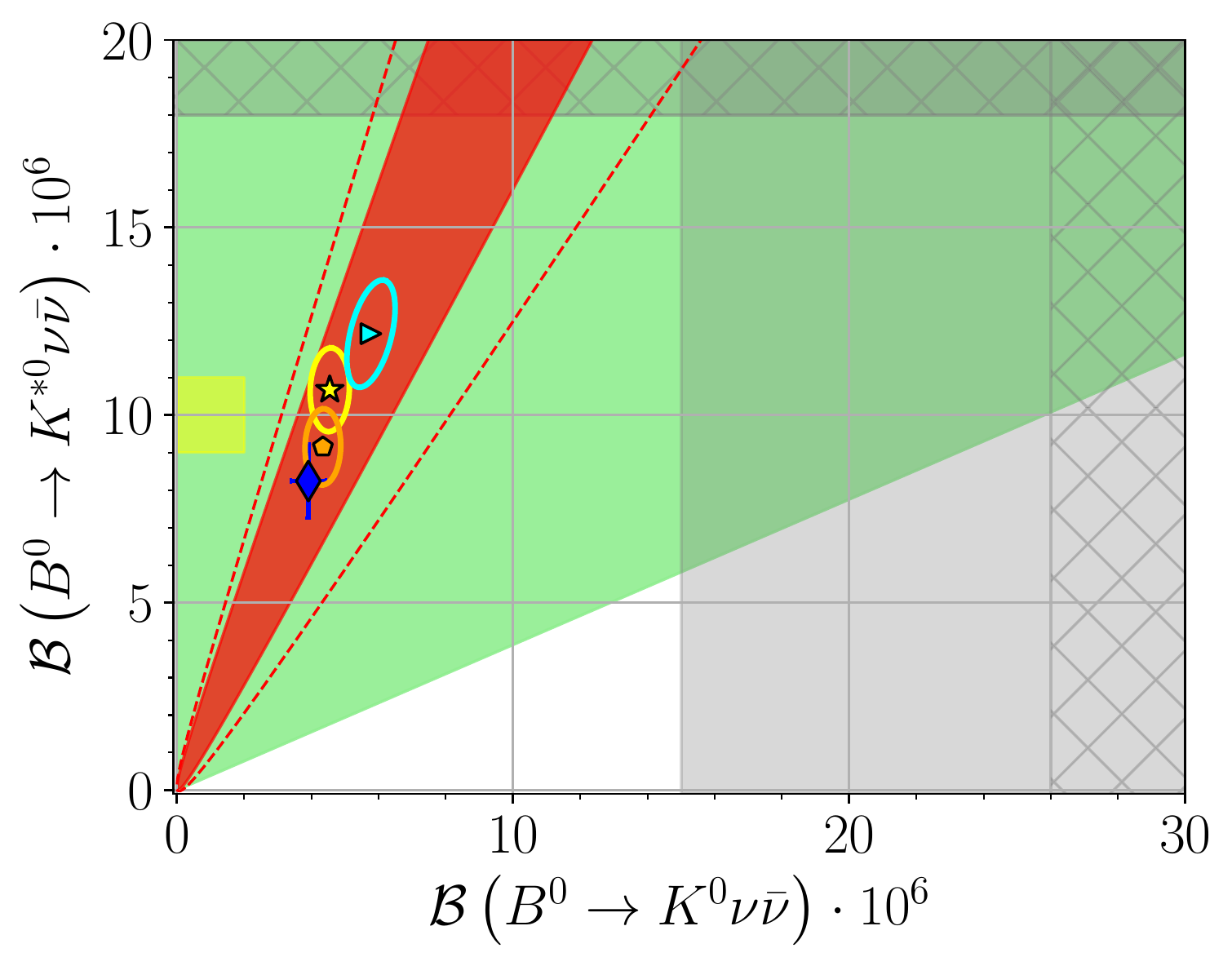}
    \includegraphics[width=7cm,height=5.5cm]{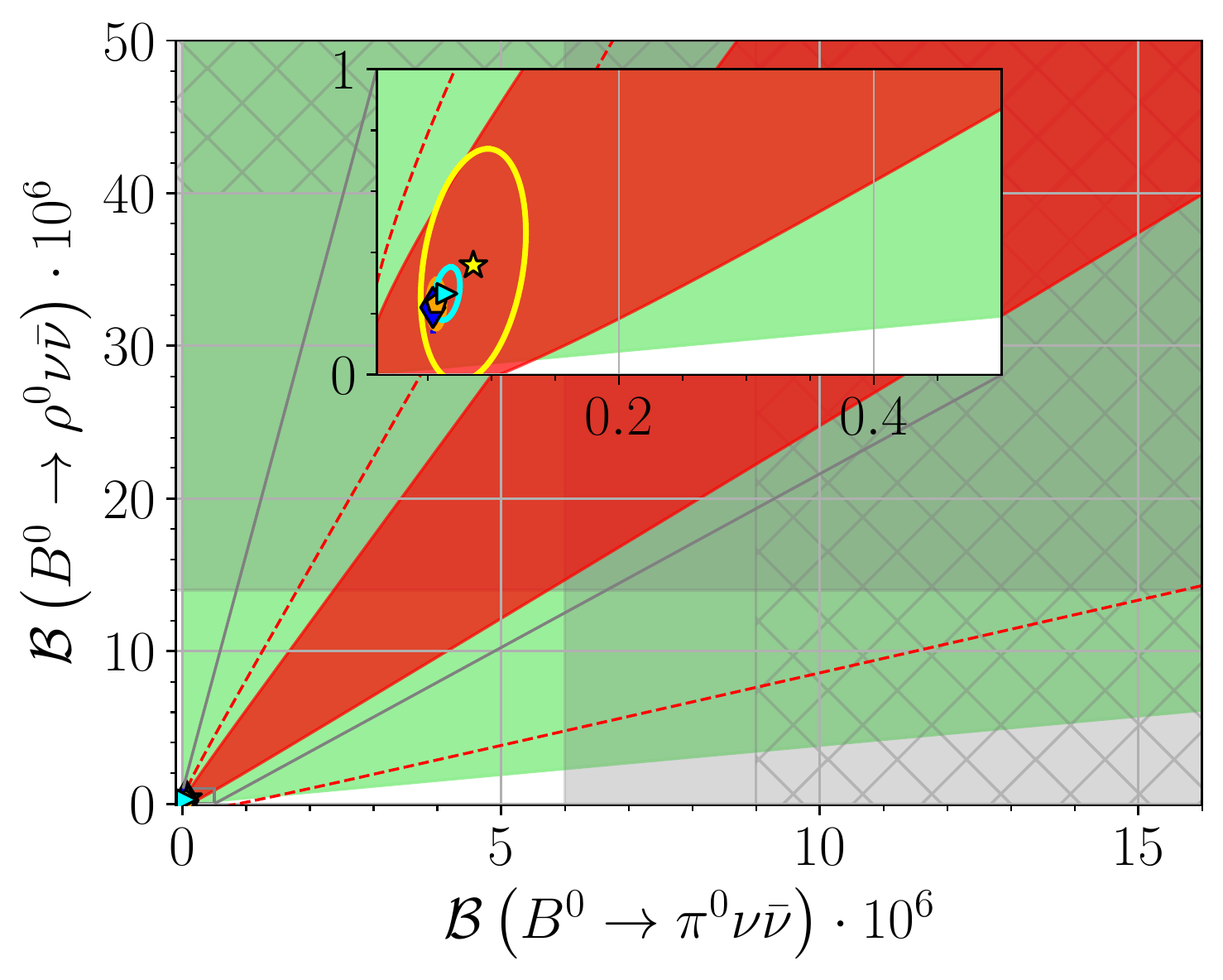}
   \caption{The figure is split into a left panel for $b\to s$ and a right panel for $b\to d$ transitions. 
     \textbf{Upper left:} $\mathcal{B}(B^0 \to K^{*0}\,\nu\bar{\nu})$ versus $\mathcal{B}(B^+ \to K^+\,\nu\bar{\nu})$.
    SM predictions (blue diamond) with their uncertainties (blue bars) from Tab.~\ref{tab:smPred}, where we have included the resonant $\tau$-background in the charged mode as an additional uncertainty (cf. Sec.~\ref{sec:diffBr}).The region to the left of the solid blue line is governed by  pure resonant contributions (\ref{eq:bgd}), (\ref{eq:bgd2}). 
    Dark red region (dashed red lines) represent the LU region given by Eq.~\eqref{eq:luregion} where $\kappa_{R}^{bs\mu\mu}$ and $A_\pm$ have been scanned within their $1\,\sigma$ ($2\,\sigma$) uncertainties.
    The light green region represents the validity of our EFT framework, Eqs.~\eqref{eq:MIbs} and \eqref{eq:MIbd}. Assuming Eq.~\eqref{eq:BSMleftlink}, we provide specific LU BSM benchmarks, which result in best fit values (markers) and $1\,\sigma$ regions (ellipses) for $Z^\prime$ (red star), LQ representations $S_3$ (pink pentagon) and $V_3$ (celeste triangle) from $b \to s \,\mu^+ \mu^-$ global fits. Hatched gray bands correspond to current experimental $90\%$ CL upper limits in Tab.~\ref{tab:smPred}.
    The widths of the yellow boxes illustrate the projected experimental sensitivity  of Belle II with $50\,\text{ab}^{-1}$ in Tab.~\ref{tab:smPred}. \textbf{Lower left:} Similar to upper left plot, but for $\mathcal{B}(B^0 \to K^{*0}\,\nu\bar{\nu})$ versus $\mathcal{B}(B^0 \to K^0\,\nu\bar{\nu})$.
     \textbf{Upper right:} $\mathcal{B}(B^0 \to \rho^0\nu\bar{\nu})$ versus $\mathcal{B}(B^+ \to \pi^+\nu\bar{\nu})$ with labeling similar to upper left plot.
    The plot includes a zoom into the region around the SM expectation.
    The $\tau$-background (solid blue line) is not included as an uncertainty in $\mathcal{B^+ \to \pi^+}\,\nu\bar{\nu}$ as it dominates the SM prediction by two orders of magnitude. \textbf{Lower right:} Similar to upper right plot but for $\mathcal{B}(B^0 \to \rho^0\,\nu\bar{\nu})$ versus $\mathcal{B}(B^0 \to \pi^0\,\nu\bar{\nu})$. 
    }
    \label{fig:plotKstarversusK}
\end{figure*}

In the upper (lower) left plot of Fig.~\ref{fig:plotKstarversusK}, we display the correlation between $\mathcal{B}(B^0 \to K^{*0}\nu\bar{\nu})$ and $\mathcal{B}(B^+ \to K^+\nu\bar{\nu})$ ($\mathcal{B}(B^0 \to K^0\nu\bar{\nu})$) using Eq.~\eqref{eq:luregion}, where the value of $\kappa_{R}^{bs\mu\mu}$ is given by Eq.~\eqref{eq:KRglobalfit1sigma}. 
Scanning $\kappa_{R}^{bs\mu\mu}$ and the form factors within their $1\,\sigma$ regions, we obtain the dark red region which represents the LU region. The dashed red lines indicate the $2\,\sigma$ contour. Two measurements of branching ratios outside this region will represent a violation of LU, while a measurement inside this region does not necessarily imply LU conservation. 
The SM predictions from Tab.~\ref{tab:smPred} are depicted as a blue diamond with their uncertainties (blue bars). The light green region represents the validity of our EFT framework, previously given by Eqs.~\eqref{eq:MIbs} and \eqref{eq:MIbd}. 
The hatched bands correspond to current experimental $90\,\%$ CL upper limits in Tab.~\ref{tab:smPred}. 
The gray bands represent our derived EFT limits from Tab~\ref{tab:smPred}. 
A measurement between gray and hatched area would infer a clear hint for BSM physics not covered by our EFT framework. 
The widths of the yellow boxes illustrate the projected experimental sensitivity ($10\,\%$ at the chosen point) of Belle II with $50\,\text{ab}^{-1}$.

Interestingly, we observe that a measurement of $B^0\to K^0\,\nu\bar\nu$ in the range of $\sim (13-15)\cdot 10^{-6}$ would represent a clear sign of LU violation, independent of $B^0 \to K^{*0}\nu\bar{\nu}$. 
Similar conclusions can be inferred for other modes, again looking at the $\mathcal{B}(B\to V\,\nu\bar\nu)$--$\mathcal{B}(B\to P\,\nu\bar\nu)$ plane, as can be observed in Fig.~\ref{fig:plotdiffBP}.

Correlations between $b\to d\,\nu\bar\nu$ decays are shown in the right plots of Fig.~\ref{fig:plotKstarversusK}. Here, we project the $\mathcal{B}(B^0 \to\rho^0\nu\bar{\nu})$ --$\mathcal{B}(B^+ \to \pi^+\nu\bar{\nu})$ plane (upper right plot) and the $\mathcal{B}(B^0 \to\rho^0\nu\bar{\nu})$ --$\mathcal{B}(B^0 \to \pi^0\nu\bar{\nu})$ plane (lower right plot) using the $1\,\sigma$ fitted values of $\kappa_{R,\,\text{NP}}^{bd\mu\mu}$ according to Eq.~\eqref{eq:globalfitkappaRbd}.
For the plots we use the $B\to \rho$ form factors fitted to LCSR and lattice data, see App.~\ref{sec:BtorhoFFfit}.

Using information on $e e$, $\mu\mu$ and $\tau\tau$ couplings, a test of cLFC would also be possible, following a similar procedure as in the LU case~\eqref{eq:luregion}. However, scanning $e e$, $\mu\mu$ and $\tau\tau$ couplings within its allowed ranges provided by Tab.~\ref{tab:limitsonK}, we observe that the current limits on $\tau\tau$ couplings are so weak that the resulting range covers the whole green region in Fig.~\ref{fig:plotKstarversusK}.
Note that in the region to the left of the solid blue lines sensitivity to NP is lost as these correspond to
branching ratios of a $B^+$ annihilating via $\tau^+ \nu \to P^+ \nu \bar \nu$  \eqref{eq:bgd}, \eqref{eq:bgd2}.
The lower plots which show correlations between neutral $B$-decays, on the other hand, are not affected.

\subsection{BSM tree-level mediators}    

In this section we explore the implications of specific BSM extensions with the following generic alignment $C^{(3)}_{\ell q}=\alpha\,C^{(1)}_{\ell q}$, and therefore
\begin{align}\label{eq:BSMleftlink}
    K_L^D=\gamma\, C_L^D\,=\,(1+\alpha)\,{\left(\frac{2\pi}{\alphae}\right)}\,C^{(1)}_{\ell q},~~\gamma=\frac{1+\alpha}{1-\alpha}~.
\end{align}

\begin{table}[ht!]
    \centering
    \resizebox{\textwidth}{!}{
    \begin{tabular}{c|c|cc|cc|cc|c}
    \hline
    \hline
    model & {\small$(SU(3)_C,SU(2)_L,Y)$} & $\alpha$ & $\gamma$ & $\kappa_{L}^{bs\mu\mu}$ & $\kappa_{R}^{bs\mu\mu}$ & $\kappa_{L}^{bd\mu\mu}$ & $\kappa_{R}^{bd\mu\mu}$ & $C^D_R$\\
    \hline
    $Z^\prime$ & $(1,1,0)$ & $\phantom{-}0$ & $1$ & $-1.45 \pm 0.29$ & $0.46\pm 0.26$ & $-3 \pm 5$ & $0 \pm 4$ & $\neq 0$ \\
    \hline
    $S_3$ & $(3,3,-\frac{1}{3})$ & $\phantom{-}\frac{1}{3}$ & $2$ & \multirow{2}{2.1cm}{$-1.36 \pm 0.32$} & \multirow{2}{0.2cm}{$0$} & \multirow{2}{1.4cm}{$0.6 \pm 0.4$} & \multirow{2}{0.2cm}{$0$} & $0$ \\
    $V_3$ & $(3,3,-\frac{2}{3})$& $-\frac{1}{3}$ & $\frac{1}{2}$ & & & &  & $0$ \\
    \hline
    \hline
    \end{tabular}}
    \caption{Values for $\alpha$ and $\gamma$~\eqref{eq:BSMleftlink} for different BSM tree-level mediators~\cite{Hiller:2016kry}. In the second column the representation of the corresponding mediator under the SM gauge group is shown. 
    The values of $\kappa_{L(R)}^{bs\mu\mu}$ within $1\sigma$ uncertainties are from $b\to s\,\mu^+\mu^-$ global fit results, see App.~\ref{app:globalfit} for details, whereas the $\kappa_{L(R)}^{bd\mu\mu}$ 
    values are provided by a fit of $b \to d\,\mu^+\mu^-$ observables, see main text and Ref.~\cite{Bause:inprep} for details.
    The last column displays which dineutrino Wilson coefficient is not generated by the model. 
    }
    \label{tab:BKvv_para}
\end{table}

Eq.~\eqref{eq:BSMleftlink} allows us to predict all LU branching ratios using only information from $b\to s\,\mu^+\mu^-$ and $b\to d\,\mu^+\mu^-$ global fit results, at the price of giving up the model-independent framework analyzed in the previous section.

The BSM extensions listed in Tab.~\ref{tab:BKvv_para} generate non-zero Wilson coefficients $K_L^D$ and $C_L^D$ allowing to connect dineutrino modes to charged dilepton data as in Eq.~\eqref{eq:BSMleftlink}, e.g.,~\cite{Hiller:2016kry}.

The third column of Tab.~\ref{tab:BKvv_para} displays the values of $\alpha$ and $\gamma$ for different BSM models as defined in Eq.~\eqref{eq:BSMleftlink}. 
The fourth and fifth columns provide the central values with their $1\,\sigma$ uncertainties (including correlations) of $\kappa_{L}^{bs\mu\mu}$ and $\kappa_{R}^{bs\mu\mu}$ extracted from $b\to s\,\mu^+\mu^-$ and $b\to d\,\mu^+\mu^-$ global fit results, respectively. 
For $Z^\prime$ models, we use the corresponding 6-dimensional results in ~\eqref{eq:KRglobalfit1sigma}, while for the leptoquark representations, where $C_R=0$, we employ results from a 1-dimensional fit assuming $\Cnine=-\Cten$. 
Further details on these global fits can be found in App.~\ref{app:globalfit}. 
Using Eqs.~\eqref{eq:BR},~\eqref{eq:xpmgen}, and \eqref{eq:BSMleftlink} together with the values of Tab.~\ref{tab:BKvv_para}, we obtain LU BSM $1\,\sigma$ best fit branching ratio predictions that are listed in Tab.~\ref{tab:tree_level_BM},
where correlations between $\kappa_{L}^{bs\mu\mu}$ and $\kappa_{R}^{bs\mu\mu}$ have been included.

Fig.~\ref{fig:plotKstarversusK} shows the best fit branching ratio predictions for the three tree-level mediator benchmarks (markers) with their $1\,\sigma$ uncertainties (ellipses) derived from Tab.~\ref{tab:BKvv_para}. 

For the computation of the ellipses (and similar for the LU region) we separate the branching ratio contributions into those coming only from the SM, new physics and their interference terms.
The error propagation is then handled by taking only the central values of $A_\pm^{BF_q}$ for the pure NP contribution, whereas the SM contribution is given in Tab.~\ref{tab:smPred} and includes correlations between the $A_\pm^{BF_q}$ factors.
To avoid doubling counting of uncertainties in the interference terms, we scale $A \sim \sqrt{A^\text{cen}\cdot A^\text{unc}}$, where $A^\text{cen}$ and $A^\text{unc}$ refer to the central value and the value including uncertainties of the corresponding $A_\pm^{BF_q}$, respectively. Therefore, form factor uncertainties are only included once per term.

Fig.~\ref{fig:plotKstarversusK} shows that future data from Belle II on dineutrino modes combined with the new test presented in this work 
allows to probe and  potentially exclude concrete new physics models, such as leptoquarks $S_3$, $V_3$ and flavorful $Z^\prime$-extensions that play a role in
explaining the $b \to s \ell^+ \ell^-$-anomalies.

\begin{table}[h!]
 \centering
\setlength{\tabcolsep}{5pt} 
\renewcommand{\arraystretch}{1.1}
\begin{tabular}{c|c|ccc}
\hline
\hline
& LU max & \multicolumn{3}{c}{LU benchmark} \\
$\mathcal{B}\left(B \to F_q\,\nu\bar{\nu}\right)$ & & $Z^\prime$ & $S_3$ & $V_3$ \\
& $[10^{-6}]$ & $[10^{-6}]$ & $[10^{-6}]$ & $[10^{-6}]$ \\
\hline 
$B^0 \to K^0$ & $8.5$ & $4.5 \pm 0.6$ & $4.3 \pm 0.5$ & $5.8 \pm 0.7$ \\
$B^+ \to K^+$ & $9.2$ & $4.9 \pm 0.8$ & $4.7 \pm 0.8$ & $6.2 \pm 0.9$ \\
$B^0 \to K^{\ast 0}$ & $18^a$ & $10.7 \pm 1.1$ & $9.2 \pm 1.0$ & $12.2 \pm 1.4$ \\
$B^+ \to K^{\ast +}$ & $19$ & $11.6 \pm 1.2$ & $9.9 \pm 1.1$ & $13.2 \pm 1.5$ \\
& & & \\
$B^0 \to \pi^0$ & $3.9$ & $0.09 \pm 0.04$ & $0.057 \pm 0.006$ & $0.065 \pm 0.010$ \\
$B^+ \to \pi^+$ & $8.3$ & $0.19 \pm 0.09$ & $0.126 \pm 0.011$ & $0.144 \pm 0.020$ \\
$B^0 \to \rho^0$ & $14$ & $0.4 \pm 0.4$ & $0.23 \pm 0.08$ & $0.27 \pm 0.09$ \\
$B^+ \to \rho^+$ & $30^a$ & $0.8 \pm 0.8$ & $0.50 \pm 0.18$ & $0.58 \pm 0.19$ \\
\hline
\hline
\end{tabular}
\caption{Maximally allowed lepton universal branching ratio by the LU limits~\eqref{eq:LUkappabs},~\eqref{eq:LUkappabd} as well as predictions in a $Z^\prime$ model as well as two leptoquark representations, $S_3$ and $V_3$, for selected $b\to s$ and $b \to d$ modes. ${}^a$Input.}
\label{tab:tree_level_BM}
\end{table}

\subsection{Including light right-handed neutrinos}\label{sec:RH}

Light RH neutrinos induce additional dimension six dineutrino operators in Eq.~\eqref{eq:Heffnu}, such as (pseudo-) scalar, (axial-) vector and (pseudo-)tensor operators. 
These operators can spoil the model-independent results presented in the previous sections. 
In this section we study their impact considering scalar and pseudoscalar contributions from RH neutrinos triggered by the following operators
\begin{equation}\label{eq:opsplusOSP}
\begin{split}
Q_{S (P)}^{\alpha\beta i j}  &= (\bar q_{L}^{\alpha} q_{R}^{\beta}) \,(\bar \nu^{j}\,(\gamma_5)\,\nu^{i}) \,,\\
Q_{S (P)}^{ \prime \alpha\beta i j}  &= (\bar q_{R}^{\alpha} q_{L}^{\beta}) \,(\bar \nu^{j}\,(\gamma_5)\,\nu^{i})\,. 
\end{split}
\end{equation}

It is convenient to define the following combination of Wilson coefficients
\begin{align}\label{eq:yD}
\begin{split}
    y_{D_{\alpha\beta}}=
    \sum_{i,j}\bigg(
    \vert \mathcal{C}_S^{D_{\alpha\beta}ij}-\mathcal{C}_{S}^{\prime D_{\alpha\beta}ij} \vert^2
    +\vert \mathcal{C}_P^{D_{\alpha\beta}ij}-\mathcal{C}_{P}^{\prime D_{\alpha\beta}ij} \vert^2\bigg)~.
\end{split}
\end{align}
This particular combination enters  the branching ratio of $B^0\to \nu \bar \nu$ decays, 
\begin{align}\label{eq:brSP}
\mathcal{B}(B^0 \to \nu \bar \nu)=\frac{G_{\text{F}}^2\, \alphae^2\, m_B^5\, f_B^2\, \tau_B}{64\,\pi^3\, m_b^2}\,y_{D_{\alpha 3}}~, 
\end{align}
where contributions from vector and axial-vector operators
are helicity suppressed by two powers of the neutrino mass, and negligible. $\tau_B$ refers to the lifetime of the $B$-meson.
Tensor operators do not contribute to $B^0\to\nu \bar \nu$ decays. 

Therefore, only scalar and pseudoscalar operators as in $y_{D_{\alpha 3}}$ are constrained by $B^0\to\nu \bar \nu$. 
The mode $B^0\to\nu\bar\nu$ is experimentally constrained as~\cite{Zyla:2020zbs}
\begin{align}\label{eq:B0invExp}
\mathcal{B}\left(B^0 \to \nu \bar \nu\right)_{\text{exp}} < 2.4 \cdot 10^{-5}
\end{align}
at $90\,\%$ CL, while $B_s^0\to\nu\bar\nu$ remains currently unconstrained and only projections for Belle with $0.12\,\text{ab}^{-1}$ (Belle II with $0.5\,\text{ab}^{-1}$) exist~\cite{Kou:2018nap},
\begin{align}\label{eq:BsinvExp}
\begin{split}
\mathcal{B}\left(B^0_s \to \nu \bar \nu\right)_{\text{proj}}  < 9.7\,(1.1) \cdot 10^{-5}~.
\end{split}
\end{align}
From Eq.~\eqref{eq:yD} and \eqref{eq:B0invExp} we obtain the limit
\begin{align}\label{eq:boundSPbd}
y_{D_{13}}\lesssim 0.3~, 
\end{align}
while for $b \to s$ transitions we use the projected limits given by Eq.~\eqref{eq:BsinvExp}, which yields
\begin{align}\label{eq:boundSPbs}
y_{D_{23}} \lesssim 0.79\,(0.09)~.
\end{align}

When considering either $\mathcal{C}^{ij}_{P,S}=0$ or $\mathcal{C}^{\prime ij}_{P,S}=0$  to avoid cancellations between the two, the branching ratio of $B\to P\,\nu\bar\nu$ decays which, unlike $B^0 \to \nu\bar \nu$, 
depends on the sum of $\mathcal{C}^{ij}_{P,S}$ and $\mathcal{C}^{\prime ij}_{P,S}$, 
can be written as
\begin{align}\label{eq:BRSP}
    \mathcal{B}(B\to P\,\nu\bar\nu)_{S,P}\,=\,A_0^{BP}\,y_{D_{\alpha 3}} ~,
\end{align}
with
\begin{align}
    A_0^{BP}=\int_{q^2_{\text{min}}}^{q^2_{\text{max}}}\,\text{d}q^2\,a_0^{BP}(q^2)~,
\end{align}
and
\begin{align}
    a_0^{BP}(q^2)=\frac{\tau_B\,G_{\text{F}}^2\, \alphae^2\, {\lambda(m_B^2,m_P^2,q^2)}^{\frac{1}{2}}}{1024\,\pi^5\, m_B^3\,c_P^2}
   \times  \frac{q^2}{m_b^2}\,(m_B^2-m_P^2)^2\, ({f_0^{BP}(q^2)})^2~,
\end{align}
where $q^2_{\text{min}}$ and $q^2_{\text{max}}$ denote the kinematic limits of $B\to P\,\nu\bar\nu$, see Sec.~\ref{sec:diffBr}. For further clarifications of $c_P,\,\lambda$ we refer to App.~\ref{app:diffBR_a}.

We provide the impact exemplarily on $B \to P\,\nu\bar\nu$ decays since there is no specific enhancement or suppression in semileptonic decays for $S,P$--operators.
We obtain the following upper limits  based on \eqref{eq:boundSPbd}, and projected limits from  \eqref{eq:boundSPbs}, respectively, as 
\begin{align}\label{eq:SP}
    \begin{split}
        \mathcal{B}\left( B^{0,+}  \to \pi^{0,+} \,\nu\bar{\nu}\right)_{S,P} &\lesssim 1.2 \cdot 10^{-7}~,\\
        \mathcal{B}\left( B^0  \to K^0\, \nu\bar{\nu}\right)_{S,P}^{\text{proj}} &\lesssim 11.4\,(1.3)  \cdot 10^{-7}~,\\
        \mathcal{B}\left( B^+  \to K^+\, \nu\bar{\nu}\right)_{S,P}^{\text{proj}} &\lesssim 12.3\,(1.4)  \cdot 10^{-7}~. 
    \end{split}
\end{align}
Comparing to the SM predictions in Tab.~\ref{tab:smPred} we learn that  (pseudo-)scalar contributions in $b \to d$  transitions
can amount to an $\mathcal{O}(100\%)$ correction. An  improved experimental limit of $\mathcal{B}(B^0\to\nu\bar\nu)$ at the level of $\sim 5\cdot10^{-7}$ or smaller
would suffice to bring the correction to the SM at the percent-level.
The projected reach in the decay $B_s\to\nu\bar\nu$ from Eq.~\eqref{eq:BsinvExp} constrains $S,P$- contributions to $b\to s$ transitions to be less than a $\mathcal{O}(30\%)$ (Belle with $0.12\,\text{ab}^{-1}$), and  a $\mathcal{O}(3\%)$ (Belle II with $0.5\,\text{ab}^{-1}$) correction to the SM branching ratios. In the latter case,
 (pseudo-)scalar contributions would not be observable in $b\to s$ dineutrino modes such as $B \to K \,\nu \bar \nu$ within uncertainties.

\section{Conclusions \label{sec:con}}

We present a comprehensive, global analysis of FCNC $b$-dineutrino modes, and the interplay with charged dilepton $b \to q \,\ell^+ \ell^-$ transitions.
The study is timely for several reasons:\\
{\it i)} Belle II is expected to improve knowledge on several dineutrino modes in the nearer future~\cite{Kou:2018nap}.
{\it ii)} Information on semileptonic 4-fermion operators is improving from rare decay studies at flavor factories LHCb  and Belle II, as well as
Drell-Yan studies at the LHC.
{\it iii)} Correlations and synergies across sectors provide a useful and informative  path in  the present situation without direct observations of BSM physics at colliders, 
in particular given the  hints for new physics in rare $B$-decays, aka the $B$-anomalies, see \cite{Bissmann:2020mfi} for a recent
study connecting top and beauty observables in SMEFT.
{\it iv)} The first evidence for electron-muon universality violation by LHCb~\cite{Aaij:2021vac} makes further analyses and cross checks of this phenomenon vital.
In particular, dineutrino studies can provide independent  tests of lepton universality,  
as has been pointed out recently \cite{Bause:2020auq,Bause:2020xzj}, and shed light on the hints for lepton non-universality.

The main results of this study are the following:

First, exploiting correlations within the weak effective theory we derive improved  or even entirely new limits on dineutrino branching ratios 
presented in Tab.~\ref{tab:smPred},
including inclusive and exclusive decays $B^0 \to (K^0 , X_s)\, \nu \bar \nu$, $B_s \to \phi\, \nu \bar \nu$ and 
$B^0 \to (\pi^0, \rho^0)\, \nu \bar \nu$. These follow from  upper limits on Wilson coefficients imposed by those dineutrino modes which presently are best constrained: 
$B^+ \to K^+ \,\nu \bar \nu$ and $B^0 \to K^{*0} \,\nu \bar \nu$ in $b \to s$ FCNCs and $B^+ \to (\pi^+, \rho^+) \,\nu \bar \nu$  for $b\to d$ ones.
Any improvement on these modes, which is expected from Belle II, impacts upper limits on the other modes.

Secondly, using SMEFT we obtain new flavor constraints from the dineutrino modes, which are stronger than the corresponding ones  from charged dilepton rare $b$-decay or Drell-Yan data,
for $e \tau$ and $\tau \tau$ final states,  as well as for $\mu \tau$ ones in $b \to s$ processes, see Tab.~\ref{tab:limitsonK}.
Improved upper limits on branching ratios of  $b \to s \, \tau^+ \tau^-$  and $b \to d \, \tau^+ \tau^-$ transitions are obtained in Eqs.~\eqref{eq:clfctaulimits}, \eqref{eq:taulimits-bd}.
Even stronger constraints are obtained in simplified BSM frameworks such as leptoquarks and $Z^\prime$-models, see Tab.~\ref{tab:tree_level_BM}.
Interestingly, also constraints on left-handed couplings for top quarks with charm or up and leptons, 
$tc\ell \ell^\prime$ and  $tu\ell \ell^\prime$, are obtained.
These are quite unique, as top-couplings cannot be obtained from Drell-Yan production.
The $tc\ell \ell$ bounds are stronger than existing limits on left-handed couplings with ditops and dielectrons/dimuons~\cite{Sirunyan:2020tqm},
while our $tu\ell \ell$ ones are comparable.
We also stress that dineutrino data constrains \textit{all} dilepton final states, including LFV ones.

Furthermore, we also perform a global fit to the semileptonic Wilson coefficients for $|\Delta b|=|\Delta s|=1$ transitions,  shown in Fig.~\ref{fig:globalfit} (left panel)
and employ findings from a fit to $|\Delta b|=|\Delta d|=1$ transitions \cite{Bause:inprep} (right panel). This
enables a relation between $B \to$ vector and $B \to$ pseudoscalar dineutrinos branching ratios, displayed in 
Fig.~\ref{fig:plotKstarversusK}, that allows to test lepton universality. This is a key result of this work.
For $b \to s$ transitions we identify the $1\sigma\,(2\sigma)$ regions, 
\begin{align}
\begin{split}
\frac{\mathcal{B}(B^0 \to K^{* 0} \nu \bar \nu)}{\mathcal{B}(B^0 \to K^0 \nu \bar \nu)}&=1.7 \ldots 2.6  \quad  (1.3 \ldots 2.9) ~,  \\
\frac{\mathcal{B}(B^0 \to K^{* 0} \nu \bar \nu)}{\mathcal{B}(B^+ \to K^+ \nu \bar \nu)}&=1.6 \ldots 2.4 \quad  (1.2 \ldots 2.7) ~,  
\end{split}
\end{align}
shown as red cones. 
Outside of them lepton flavor universality is broken.

Corresponding ranges for $b \to d$ transitions have larger uncertainties due to $B \to \rho$ form factors.
We thus quote $1\sigma\,(2\sigma)$ results based on a fit   and in addition using $B \to \rho\, \ell \nu$ data (``norm'') assuming the latter to be SM-dominated, see App.~\ref{sec:BtorhoFFfit} for details, as
\begin{align}
\begin{split}
    \frac{\mathcal{B}(B^0 \to \rho^0 \nu \bar \nu)}{\mathcal{B}(B^0 \to \pi^0 \nu \bar \nu)}&=2.5 \ldots 5.7 \quad (1.0 \ldots 7.3) ~, \\
\frac{\mathcal{B}(B^0 \to \rho^0 \nu \bar \nu)}{\mathcal{B}(B^+ \to \pi^+ \nu \bar \nu)}&=1.2 \ldots 2.6 \quad (0.4 \ldots 3.4) ~,    
\end{split} \\
\begin{split}\label{eq:BR_ratio_rhonorm}
    \frac{\mathcal{B}(B^0 \to \rho^0 \nu \bar \nu)\vert_\text{norm}}{\mathcal{B}(B^0 \to \pi^0 \nu \bar \nu)}&= 2.6 \ldots 3.3 \quad (2.4 \ldots 3.4) ~, \\
\frac{\mathcal{B}(B^0 \to \rho^0 \nu \bar \nu)\vert_\text{norm}}{\mathcal{B}(B^+ \to \pi^+ \nu \bar \nu)} &= 1.2 \ldots 1.5 \quad (1.1 \ldots 1.6) ~,    
\end{split}
\end{align}

Outside of them  lepton universality is broken. 

Both $b \to s$ and $b \to d$ universality tests with dineutrino modes can be sharpened by improving constraints on semi-muonic four-fermion operators.
In addition, improving the  knowledge on form factors, in particular $B\to \rho$ ones,  would be desirable.
We also remark that  light right-handed neutrinos, which are outside of our framework, could be controlled by bounding ${\cal{B}}(B_s \to \nu \bar \nu)$ at the level of 
Belle II sensitivities, $\sim 10^{-5}$.  An improvement of the present  limit on ${\cal{B}}(B^0 \to \nu \bar \nu)$ by a factor $\sim 50$ would exclude such contributions to $b\to d$ branching ratios at the few percent level.

We look forward to more global analyses of dineutrino and charged dilepton modes 
together to fully exploit flavorful synergies.

\bigskip 

{\bf Note added: }
While we were finishing this paper, a preprint~\cite{He:2021yoz} appeared  in which also correlations between dineutrino branching ratios and $B_s\to\tau\tau$ decays  in $Z^\prime$ and Leptoquark models are discussed.

\acknowledgments
 
We would like to thank Stefan Bi\ss{}mann, Jonathan Kriewald, Ana Pe\~{n}uelas and Emmanuel Stamou for useful discussions. This work is supported by the \textit{Studienstiftung des Deutschen Volkes} (MG) and the \textit{ Bundesministerium f\"ur Bildung und Forschung} -- BMBF (HG).


\appendix

\section{RGE effects from $\boldsymbol{\Lambda_{\text{NP}}}$ to $\boldsymbol{\mu_{\text{EW}}}$}\label{app:RGEeffects}

In this appendix we explore effects from the renormalization group equation (RGE) on Eqs.~\eqref{eq:massbasis}. 
These corrections can be accounted by the following Hamiltonian
\begin{align}
    \mathcal{H}_{\text{eff}}\,=\,\mathcal{H}^{(0)}\,+\,\,\delta\mathcal{H}~,
\end{align}
where $\mathcal{H}^{(0)}= \mathcal{H}_{\text{eff}}^{\ell^-\ell^+}+\mathcal{H}_{\text{eff}}^{\nu \bar \nu}$ represents the leading order contribution, see Eqs.~\eqref{eq:Heffell} and \eqref{eq:Heffnu}. 
Their Wilson coefficients in the mass basis read
\begin{align}\label{eq:Clo}
    \begin{split}
        (\mathcal{C}_{L}^U)_{prst}^{(0)}&=\lambda_{\gamma s t \lambda}^{V_u}\,\lambda_{\alpha p r \beta}^{V_\nu}\left(\frac{2\pi}{\alphae}\right)\left(C^{(1)}_{\underset{\alpha\beta\gamma\lambda}{\ell q}}+C^{(3)}_{\underset{\alpha\beta\gamma\lambda}{\ell q}}\right)~,\\
        (\mathcal{C}_{L}^D)_{prst}^{(0)}&=\lambda_{\gamma s t \lambda}^{V_d}\,\lambda_{\alpha p r \beta}^{V_\nu}\left(\frac{2\pi}{\alphae}\right)\left(C^{(1)}_{\underset{\alpha\beta\gamma\lambda}{\ell q}}-C^{(3)}_{\underset{\alpha\beta\gamma\lambda}{\ell q}}\right)~,\\
        (\mathcal{C}_{R}^U)_{prst}^{(0)}&=\lambda_{\gamma s t \lambda}^{U_u}\,\lambda_{\alpha p r \beta}^{V_\nu}\,\left(\frac{2\pi}{\alphae}\right)C_{\underset{\alpha\beta\gamma\lambda}{\ell u}}~,\\
        (\mathcal{C}_{R}^D)_{prst}^{(0)}&=\lambda_{\gamma s t \lambda}^{U_d}\,\lambda_{\alpha p r \beta}^{V_\nu}\,\left(\frac{2\pi}{\alphae}\right)C_{\underset{\alpha\beta\gamma\lambda}{\ell d}}~,
    \end{split}
\end{align}
for dineutrino modes, while
\begin{align}\label{eq:Klo}
    \begin{split}
        (\mathcal{K}_{L}^U)_{prst}^{(0)}&=\lambda_{\gamma s t \lambda}^{V_u}\,\lambda_{\alpha p r \beta}^{V_\ell}\left(\frac{2\pi}{\alphae}\right)\left(C^{(1)}_{\underset{\alpha\beta\gamma\lambda}{\ell q}}-C^{(3)}_{\underset{\alpha\beta\gamma\lambda}{\ell q}}\right)~,\\
        (\mathcal{K}_{L}^D)_{prst}^{(0)}&=\lambda_{\gamma s t \lambda}^{V_d}\,\lambda_{\alpha p r \beta}^{V_\ell}\left(\frac{2\pi}{\alphae}\right)\left(C^{(1)}_{\underset{\alpha\beta\gamma\lambda}{\ell q}}+C^{(3)}_{\underset{\alpha\beta\gamma\lambda}{\ell q}}\right)~,\\
        (\mathcal{K}_{R}^U)_{prst}^{(0)}&=\lambda_{\gamma s t \lambda}^{U_u}\,\lambda_{\alpha p r \beta}^{V_\ell}\,\left(\frac{2\pi}{\alphae}\right)C_{\underset{\alpha\beta\gamma\lambda}{\ell u}}~,\\
        (\mathcal{K}_{R}^D)_{prst}^{(0)}&=\lambda_{\gamma s t \lambda}^{U_d}\,\lambda_{\alpha p r \beta}^{V_\ell}\,\left(\frac{2\pi}{\alphae}\right)C_{\underset{\alpha\beta\gamma\lambda}{\ell d}}~,
    \end{split}
\end{align}
for charged leptons, where $\lambda_{\gamma s t \lambda}^{X}\,=\,(X)^\dagger_{\gamma s}\,X_{t\lambda}$~. 
In contrast to Eqs.~\eqref{eq:WCsL} and \eqref{eq:WCsR}, we keep the flavor indices in Eqs.~\eqref{eq:Clo} and \eqref{eq:Klo}.

The piece $\delta\mathcal{H}$ accounts for RGE corrections from gauge~\cite{Alonso:2013hga}, Yukawa~\cite{Jenkins:2013wua}, and QED~\cite{Jenkins:2017dyc} coupling dependencies. 
These corrections contain the same operator basis as $\mathcal{H}^{(0)}$, therefore these effects can be parametrized as
\begin{align}\label{eq:rge_corr}
    C_i=C^{(0)}_i+\left(\frac{2\pi}{\alphae}\right)\frac{    L}{(4\pi)^2}\, \xi_{C_i}~,
\end{align}
where $L=\text{log}(\Lambda_{\text{NP}}/\mu_{\text{EW}})$ and $C_i$ is a Wilson coefficient, {\it i.e.} $\mathcal{C}_{L}^U$, $\mathcal{C}_{L}^D$, etc.\,. 
The values of $\xi_{C_i}$ contain non-trivial combinations of Wilson coefficients $C_j$, with $j$ not necessary equal to $i$. 
Using Refs.~\cite{Alonso:2013hga,Jenkins:2013wua,Jenkins:2017dyc}, and solving their RGEs in the leading log approximation, we find the values of $\xi_i$ displayed in Tab.~\ref{tab:rge_corr}.

\begingroup
\renewcommand{\arraystretch}{1.7}
\begin{table*}[tp]  
\centering
\resizebox{\textwidth}{!}{
\begin{tabular}{|c||c|c|c|c|}
        \hline
        \hline  
        & \makecell{rotation \\ SMEFT $\to$ WET} & gauge & Yukawa & QED \\
        \hline
         & & 
        $(g_1^2-3\,g_2^2)\big(C^{(1)}_{\underset{\alpha\beta\gamma\lambda}{\ell q}}+C^{(3)}_{\underset{\alpha\beta\gamma\lambda}{\ell q}}\big)$ &
        $\lambda^{U_u}_{\gamma^{\prime\prime}\gamma^{\prime}\lambda^{\prime}\lambda^{\prime\prime}}\,[Y_u^\dagger]_{\gamma^\prime\gamma}[Y_u]_{\lambda\lambda^\prime}\,C_{\underset{\alpha\beta\gamma^{\prime\prime}\lambda^{\prime\prime}}{\ell u}}$ 
        &  \\
        $(\mathcal{C}_{L}^U)_{prst}$ & $\lambda_{\gamma s t \lambda}^{V_u}\,\lambda_{\alpha p r \beta}^{V_\nu}$ & $-\frac{2}{3}\bigg[g_1^2\, C^{(1)}_{\underset{ww\gamma\lambda}{\ell q}}+g_2^2\, C^{(3)}_{\underset{ww\gamma\lambda}{\ell q}}\bigg]\delta_{\alpha\beta}$ & $-\lambda^{V_u}_{\gamma^{\prime\prime}\gamma ^\prime\lambda\lambda^{\prime\prime}}\,\frac{1}{2}[Y_u^\dagger Y_u]_{\gamma\gamma^\prime}\,\big(C^{(1)}_{\underset{\alpha\beta\gamma^{\prime\prime}\lambda^{\prime\prime}}{\ell q}}+C^{(3)}_{\underset{\alpha\beta\gamma^{\prime\prime}\lambda^{\prime\prime}}{\ell q}}\big)$ & $--$ \\
        & & & $-\,\lambda^{V_u}_{\gamma^{\prime\prime}\gamma\lambda^\prime\lambda^{\prime\prime}}\,\frac{1}{2}[Y_u^\dagger Y_u]_{\lambda^\prime\lambda}\,\big(C^{(1)}_{\underset{\alpha\beta\gamma^{\prime\prime}\lambda^{\prime\prime}}{\ell q}}+C^{(3)}_{\underset{\alpha\beta\gamma^{\prime\prime}\lambda^{\prime\prime}}{\ell q}}\big)$ & \\
        \hline
        & & $(g_1^2+3\,g_2^2)\,C^{(1)}_{\underset{\alpha\beta\gamma\lambda}{\ell q}}-(g_1^2+15\,g_2^2)\,C^{(3)}_{\underset{\alpha\beta\gamma\lambda}{\ell q}}$ &
        $\lambda^{U_u}_{\gamma^{\prime\prime}\gamma^\prime\lambda^\prime\lambda^{\prime\prime}}[Y_u^\dagger]_{\gamma^\prime\gamma}[Y_u]_{\lambda\lambda^\prime}\,C_{\underset{\alpha\beta\gamma^{\prime\prime}\lambda^{\prime\prime}}{\ell u}}$ & \\
        $(\mathcal{C}_L^D)_{prst}$ & $\lambda_{\gamma s t \lambda}^{V_d}\,\lambda_{\alpha p r \beta}^{V_\nu}$ & $-\frac{2}{3}\bigg[g_1^2\, C^{(1)}_{\underset{ww\gamma\lambda}{\ell q}} -g_2^2\, C^{(3)}_{\underset{ww\gamma\lambda}{\ell q}}\bigg]\delta_{\alpha\beta}$ & $-\lambda^{V_u}_{\gamma^{\prime\prime}\gamma^\prime\lambda\lambda^{\prime\prime}}\,\frac{1}{2}[Y_u^\dagger Y_u]_{\gamma\gamma^\prime}\,\big(C^{(1)}_{\underset{\alpha\beta\gamma^{\prime\prime}\lambda^{\prime\prime}}{\ell q}}-C^{(3)}_{\underset{\alpha\beta\gamma^{\prime\prime}\lambda^{\prime\prime}}{\ell q}}\big)$ & $--$ \\
        & & & $-\lambda^{V_u}_{\gamma^{\prime\prime}\gamma\lambda^\prime\lambda^{\prime\prime}}\,\frac{1}{2}[Y_u^\dagger Y_u]_{\lambda^\prime\lambda}\,\big(C^{(1)}_{\underset{\alpha\beta\gamma^{\prime\prime}\lambda^{\prime\prime}}{\ell q}}-C^{(3)}_{\underset{\alpha\beta\gamma^{\prime\prime}\lambda^{\prime\prime}}{\ell q}}\big)$ & \\
        \hline
        & & & $2\,\lambda^{V_u}_{\gamma^{\prime\prime}\gamma^\prime\lambda^\prime\lambda^{\prime\prime}}\,[Y_u^\dagger]_{\gamma^\prime\gamma}[Y_u]_{\lambda\lambda^\prime}\,C^{(1)}_{\underset{\alpha\beta\gamma^{\prime\prime}\lambda^{\prime\prime}}{\ell q}}$ &  \\
        $(\mathcal{C}_{R}^U)_{prst}$ & $\lambda_{\gamma s t \lambda}^{U_u}\,\lambda_{\alpha p r \beta}^{V_\nu}$ & $-\frac{2}{3}\,g_1^2\bigg[C_{\underset{ww\gamma\lambda}{\ell u}}\delta_{\alpha\beta}+ 6\,  C_{\underset{\alpha\beta\gamma\lambda}{\ell u}}\bigg]$ & 
        $-\lambda^{U_u}_{\gamma^{\prime\prime}\gamma^\prime\lambda\lambda^{\prime\prime}}\,[Y_u Y_u^\dagger]_{\gamma\gamma^\prime}\,C_{\underset{\alpha\beta\gamma^{\prime\prime}\lambda^{\prime\prime}}{\ell u}}$ & $--$ \\
        & & & $-\,\lambda^{U_u}_{\gamma^{\prime\prime}\gamma\lambda^\prime\lambda^{\prime\prime}}\,[Y_u Y_u^\dagger]_{\lambda^\prime\lambda}\,C_{\underset{\alpha\beta\gamma^{\prime\prime}\lambda^{\prime\prime}}{\ell u}}$ & \\
        \hline
        & & & & \\
        $(\mathcal{C}_{R}^D)_{prst}$ & $\lambda_{\gamma s t \lambda}^{U_d}\,\lambda_{\alpha p r \beta}^{V_\nu}$ & $-\frac{2}{3}\,g_1^2\bigg[C_{\underset{ww\gamma\lambda}{\ell d}}\delta_{\alpha\beta} -3\,  C_{\underset{\alpha\beta\gamma\lambda}{\ell d}}\bigg]$ & $--$ & $--$ \\
        & & & & \\      
        \hline
        \hline
        & & $(g_1^2+3\,g_2^2)\,C^{(1)}_{\underset{\alpha\beta\gamma\lambda}{\ell q}}-(g_1^2+15\,g_2^2)\,C^{(3)}_{\underset{\alpha\beta\gamma\lambda}{\ell q}}$ & $\lambda^{U_u}_{\gamma^{\prime\prime}\gamma^\prime\lambda^\prime\lambda^{\prime\prime}}[Y_u^\dagger]_{\gamma^\prime\gamma}[Y_u]_{\lambda\lambda^\prime}\,C_{\underset{\alpha\beta\gamma^{\prime\prime}\lambda^{\prime\prime}}{\ell u}}$ 
        & $8\,e^2\,\big(C^{(1)}_{\underset{\alpha\beta\gamma\lambda}{\ell q}}-C^{(3)}_{\underset{\alpha\beta\gamma\lambda}{\ell q}}\big)$ \\        
        $(\mathcal{K}_{L}^U)_{prst}$ &  $\lambda_{\gamma s t \lambda}^{V_u}\,\lambda_{\alpha p r \beta}^{V_\ell}$ &         $-\frac{2}{3}\bigg[g_1^2\, C^{(1)}_{\underset{ww\gamma\lambda}{\ell q}}-g_2^2\, C^{(3)}_{\underset{ww\gamma\lambda}{\ell q}}\bigg]\delta_{\alpha\beta}$ & 
        $-\lambda^{V_u}_{\gamma^{\prime\prime}\gamma^\prime\lambda\lambda^{\prime\prime}}\,\frac{1}{2}[Y_u^\dagger Y_u]_{\gamma\gamma^\prime}\,\big(C^{(1)}_{\underset{\alpha\beta\gamma^{\prime\prime}\lambda^{\prime\prime}}{\ell q}}-C^{(3)}_{\underset{\alpha\beta\gamma^{\prime\prime}\lambda^{\prime\prime}}{\ell q}}\big)$
        & $-\frac{4}{3}\,e^2\,\big(C^{(1)}_{\underset{ww\gamma\lambda}{\ell q}}-C^{(3)}_{\underset{ww\gamma\lambda}{\ell q}}\big)\,\delta_{\alpha\beta}$ \\
        & & & $-\lambda^{V_u}_{\gamma^{\prime\prime}\gamma\lambda^\prime\lambda^{\prime\prime}}\,\frac{1}{2}[Y_u^\dagger Y_u]_{\lambda^\prime\lambda}\,\big(C^{(1)}_{\underset{\alpha\beta\gamma^{\prime\prime}\lambda^{\prime\prime}}{\ell q}}-C^{(3)}_{\underset{\alpha\beta\gamma^{\prime\prime}\lambda^{\prime\prime}}{\ell q}}\big)$&  \\
        \hline
        & &
        $(g_1^2-3\,g_2^2)\big(C^{(1)}_{\underset{\alpha\beta\gamma\lambda}{\ell q}}+C^{(3)}_{\underset{\alpha\beta\gamma\lambda}{\ell q}}\big)$ 
        & $\lambda^{U_u}_{\gamma^{\prime\prime}\gamma^{\prime}\lambda^{\prime}\lambda^{\prime\prime}}\,[Y_u^\dagger]_{\gamma^\prime\gamma}[Y_u]_{\lambda\lambda^\prime}\,C_{\underset{\alpha\beta\gamma^{\prime\prime}\lambda^{\prime\prime}}{\ell u}}$
        & $-4\,e^2\,\big(C^{(1)}_{\underset{\alpha\beta\gamma\lambda}{\ell q}}+C^{(3)}_{\underset{\alpha\beta\gamma\lambda}{\ell q}}\big)$ \\
        $(\mathcal{K}_{L}^D)_{prst}$ & $\lambda_{\gamma s t \lambda}^{V_d}\,\lambda_{\alpha p r \beta}^{V_\ell}$ &      $-\frac{2}{3}\bigg[g_1^2\, C^{(1)}_{\underset{ww\gamma\lambda}{\ell q}}+g_2^2\, C^{(3)}_{\underset{ww\gamma\lambda}{\ell q}}\bigg]\delta_{\alpha\beta}$ &
        $-\lambda^{V_u}_{\gamma^{\prime\prime}\gamma^\prime\lambda\lambda^{\prime\prime}}\,\frac{1}{2}[Y_u^\dagger Y_u]_{\gamma\gamma^\prime}\,(C^{(1)}_{\underset{\alpha\beta\gamma^{\prime\prime}\lambda^{\prime\prime}}{\ell q}}+C^{(3)}_{\underset{\alpha\beta\gamma^{\prime\prime}\lambda^{\prime\prime}}{\ell q}})$
        & $-\frac{4}{3}\,e^2\,\big(C^{(1)}_{\underset{ww\gamma\lambda}{\ell q}}+C^{(3)}_{\underset{ww\gamma\lambda}{\ell q}}\big)\,\delta_{\alpha\beta}$ \\
        & & &       
        $-\lambda^{V_u}_{\gamma^{\prime\prime}\gamma\lambda^\prime\lambda^{\prime\prime}}\,\frac{1}{2}[Y_u^\dagger Y_u]_{\lambda^\prime\lambda}\,(C^{(1)}_{\underset{\alpha\beta\gamma^{\prime\prime}\lambda^{\prime\prime}}{\ell q}}+C^{(3)}_{\underset{\alpha\beta\gamma^{\prime\prime}\lambda^{\prime\prime}}{\ell q}})$  & \\
        \hline
        & & &
        $2\,\lambda^{V_u}_{\gamma^{\prime\prime}\gamma^\prime\lambda^\prime\lambda^{\prime\prime}}\,[Y_u^\dagger]_{\gamma^\prime\gamma}[Y_u]_{\lambda\lambda^\prime}\,C^{(1)}_{\underset{\alpha\beta\gamma^{\prime\prime}\lambda^{\prime\prime}}{\ell q}}$ & $-8\,e^2\,C_{\underset{\alpha\beta\gamma\lambda}{\ell u}}$ \\
        $(\mathcal{K}_{R}^U)_{prst}$ & $\lambda_{\gamma s t \lambda}^{U_u}\,\lambda_{\alpha p r \beta}^{V_\ell}$ & 
        $-\frac{2}{3}\,g_1^2\bigg[C_{\underset{ww\gamma\lambda}{\ell u}}\delta_{\alpha\beta}+ 6\,  C_{\underset{\alpha\beta\gamma\lambda}{\ell u}}\bigg]$ &
        $-\lambda^{U_u}_{\gamma^{\prime\prime}\gamma^\prime\lambda\lambda^{\prime\prime}}\,[Y_u Y_u^\dagger]_{\gamma\gamma^\prime}\,C_{\underset{\alpha\beta\gamma^{\prime\prime}\lambda^{\prime\prime}}{\ell u}}$
        & $-\frac{4}{3}\,e^2\,C_{\underset{ww\gamma\lambda}{\ell u}}\,\delta_{\alpha\beta}$ \\
        & & & $-\lambda^{U_u}_{\gamma^{\prime\prime}\gamma\lambda^\prime\lambda^{\prime\prime}}\,[Y_u Y_u^\dagger]_{\lambda^\prime\lambda}\,C_{\underset{\alpha\beta\gamma^{\prime\prime}\lambda^{\prime\prime}}{\ell u}}$ & \\ 
        \hline
        & & & &
        $4\,e^2\,C_{\underset{\alpha\beta\gamma\lambda}{\ell d}}$ \\
        $(\mathcal{K}_{R}^D)_{prst}$ & $\lambda_{\gamma s t \lambda}^{U_d}\,\lambda_{\alpha p r \beta}^{V_\ell}$ &
        $-\frac{2}{3}\,g_1^2\bigg[C_{\underset{ww\gamma\lambda}{\ell u}}\delta_{\alpha\beta} -3\, C_{\underset{\alpha\beta\gamma\lambda}{\ell u}}\bigg]$ & 
        $--$ & $-\frac{4}{3}\,e^2\,C_{\underset{ww\gamma\lambda}{\ell d}}\,\delta_{\alpha\beta}$ \\
        & & & & \\
        \hline
        \hline
\end{tabular}
}
\caption{Coefficients $\xi_{C_i}$ as Eq.~\eqref{eq:rge_corr} separated by RGE corrections from gauge~\cite{Alonso:2013hga}, Yukawa~\cite{Jenkins:2013wua}, and QED~\cite{Jenkins:2017dyc} coupling dependencies.}
\label{tab:rge_corr}
\end{table*}
\endgroup

We observe that, if we do not consider the global prefactor due to rotations, several terms share identical corrections except for QED corrections (due to different values of electric quark charges $q_u\neq q_d$). 
Using the coefficients $\xi_{C_i}$ from Tab.~\ref{tab:rge_corr}, we find (omitting flavor indices)
\begin{align} 
    V_u\,V_\nu\,\mathcal{C}_L^U\,V_\nu^\dagger\, V_u^\dagger-V_d\, V_\ell\,\mathcal{K}_L^D\, V_\ell^\dagger\, V_d^\dagger\,=\,\left(\frac{2\pi}{\alphae}\right)\delta^U_L~,\\
    V_d\,V_\nu\,\mathcal{C}_L^D\,V_\nu^\dagger\, V_d^\dagger-V_u\, V_\ell\,\mathcal{K}_L^U\, V_\ell^\dagger\, V_u^\dagger\,=\,\left(\frac{2\pi}{\alphae}\right)\delta^D_L~,\\
    U_u\,V_\nu\,\mathcal{C}_R^U\,V_\nu^\dagger\, U_u^\dagger-U_u\, V_\ell\,\mathcal{K}_R^U\, V_\ell^\dagger\, U_u^\dagger\,=\,\left(\frac{2\pi}{\alphae}\right)\delta^U_R~,\\
    U_d\,V_\nu\,\mathcal{C}_R^D\,V_\nu^\dagger\, U_d^\dagger-U_d\, V_\ell\,\mathcal{K}_R^D\, V_\ell^\dagger\, U_d^\dagger\,=\,\left(\frac{2\pi}{\alphae}\right)\delta^D_R~,
\end{align}
where
\begin{align}
    (\delta^U_L)_{\alpha\beta\gamma\lambda}&=\bigg(\frac{1}{3}\,(C^{(1)}_{\underset{ww\gamma\lambda}{\ell q}}+C^{(3)}_{\underset{ww\gamma\lambda}{\ell q}})\,\delta_{\alpha\beta}\,+\,(C^{(1)}_{\underset{\alpha\beta\gamma\lambda}{\ell q}}+C^{(3)}_{\underset{\alpha\beta\gamma\lambda}{\ell q}})\bigg)\,\frac{\alphae}{\pi}\,L~,\\
    (\delta^D_L)_{\alpha\beta\gamma\lambda}&=\bigg(\frac{1}{3}\,(C^{(1)}_{\underset{ww\gamma\lambda}{\ell q}}-C^{(3)}_{\underset{ww\gamma\lambda}{\ell q}})\,\delta_{\alpha\beta}\, -2\,(C^{(1)}_{\underset{\alpha\beta\gamma\lambda}{\ell q}}-C^{(3)}_{\underset{\alpha\beta\gamma\lambda}{\ell q}})\bigg)\,\frac{\alphae}{\pi}\,L~,\\
    (\delta^U_R)_{\alpha\beta\gamma\lambda}&=\bigg(\frac{1}{3}\,C_{\underset{ww\gamma\lambda}{\ell u}}\,\delta_{\alpha\beta} +2\,C_{\underset{\alpha\beta\gamma\lambda}{\ell u}}\bigg)\,\frac{\alphae}{\pi}\,L~,\\
    (\delta^D_R)_{\alpha\beta\gamma\lambda}&=\bigg(\frac{1}{3}\,C_{\underset{ww\gamma\lambda}{\ell d}}\,\delta_{\alpha\beta} -\,C_{\underset{\alpha\beta\gamma\lambda}{\ell d}}\bigg)\,\frac{\alphae}{\pi}\,L~.
\end{align}
Assuming that Wilson coefficients are of similar size (${\sim C}$) and interfere constructively, we obtain for ${\mu_{\text{EW}}\sim 80\,\GeV}$ and $\Lambda_{\text{NP}}\sim 10\,\TeV$,
\begin{align}
\begin{split}\label{eq:RGE1}
    \delta_L^U&\sim \delta_L^D\sim\,2\,\delta_R^U\sim\,2\,\delta_R^D\sim 2\,C\,\frac{\alphae}{\pi}\, L\sim \,0.02\,C~,
\end{split}
\end{align}
for $\alpha=\beta$, and
\begin{align}
\begin{split}\label{eq:RGE2}
    \delta_L^U&\sim\,\frac{1}{2}\,\delta_L^D\,\sim\, \delta_R^U\,\sim\, 2\,\delta_R^D\,\sim\, 2\,C\,\frac{\alphae}{\pi}\, L\sim \,0.02\,C~,
\end{split}
\end{align}
for $\alpha\neq\beta$.

As already mentioned in the main text, Eqs.~\eqref{eq:RGE1} and \eqref{eq:RGE2} represent a correction of less than $5\,\%$ for $\Lambda_{\text{NP}} \sim 10\,\TeV$ in Eqs.~\eqref{eq:WCsR} and \eqref{eq:WCsL}.

\section{Differential branching ratios}\label{app:diffBR_a}

In this appendix we present the $q^2$--dependent functions $a_\pm^{B F_q}$ for the exclusive transitions,
$B\to P \,\nu_i\bar\nu_j$ and $B\to V \,\nu_i\bar\nu_j$ 
where $P$ and $V$ are pseudoscalar ($P=K,\pi$) and vector ($V=K^*,\rho,\phi$) particles, respectively, 
and for the inclusive modes, $B\to X_{q}\,\nu_i\bar\nu_j$ with $q=s,d$.

\subsection{$B\to P\,\nu_i\bar\nu_j$}\label{sec:BtoP}

The $B\to P\,\nu_i\bar\nu_j$ mode, where $B=B^0,\,B^+$ and $P=\pi^0,\,\pi^+,\,K^0,\,K^+$, respectively, is described by only one form factor,  $f_+^{B P}$. 
The $a_+^{B P}$--function of the differential branching ratio is given by~\cite{Melikhov:1998ug,Colangelo:1996ay,Kim:2009mp} 
\begin{align}\label{eq:BPvv_apm}
a_+^{B P}(q^2)= \frac{G_F^2\,\alphae^2\,\tau_B\,(\lambda_{BP}(q^2))^{3/2}\left(f^{BP}_+(q^2)\right)^2}{3072\,\pi^5\,m_B^3\,c_P^2}~, 
\end{align}
while $a_-^{B P}(q^2)=0$. 
Here, $\tau_B$ denotes the lifetime of the $B$ meson.
The parameter $c_P$ accounts for the flavor content of the pseudoscalar particles, 
in particular $c_{\pi^0}=\sqrt{2}$ and $c_{\pi^+,K^{0},K^{+}}=1$. 
The function $\lambda_{BP}(q^2)$ is the usual Källén function $\lambda(m_B^2,m_P^2,q^2)$ with $\lambda(a,b,c)=a^2+b^2+c^2-2\,(a\,b+a\,c+b\,c)$. 
Notice that Eq.~\eqref{eq:BPvv_apm} is equivalent to the one provided in the literature, {\it e.g.} \cite{Kim:2009mp}, when the sum over the neutrino flavors is performed. Information about $f^{BP}_+(q^2)$ is provided in App.~\ref{app:formfactors}.

\subsection{$B\to V\,\nu_i\bar\nu_j$}\label{sec:BtoV}

In contrast to $B\to P\,\nu_i\bar\nu_j$, the differential distribution of $B\to V\,\nu_i\bar\nu_j$ is enriched with three form factors. 
The functions $a_\pm^{B V}$ associated with $B\to V\,\nu_i\bar\nu_j$ transitions can be written as~\cite{Altmannshofer:2009ma,Melikhov:1998ug,Colangelo:1996ay}
\begin{align}
    &a_+^{B V}(q^2)=\frac{G_F^2\,\alphae^2\,\tau_B\,(\lambda_{BV}(q^2))^{3/2}}{3072\,\pi^5\,m_B^5\,c_V^2}\,\frac{2\,q^2\,(V(q^2))^2}{\left(1+\frac{m_V}{m_B}\right)^2}~, \label{eq:apBV}\\
        &a_-^{B V}(q^2)=\frac{G_F^2\,\alphae^2\,\tau_B\,(\lambda_{BV}(q^2))^{1/2}}{1536\,\pi^5\,m_B\,c_V^2} \label{eq:amBV}\\
    &\times\left[32\,m_V^2\,(A_{12}(q^2))^2+\left(1+\frac{m_V}{m_B}\right)^2 q^2 (A_1(q^2))^2\right]~,\nonumber
\end{align}
with $\lambda_{BV}(q^2)=\lambda(m_B^2,m_V^2,q^2)$. 
The parameter $c_P$ accounts for the flavor content of the vector particles, in particular $c_{\rho^0}=\sqrt{2}$ and $c_{\rho^+,K^{*\,0},K^{*\,+},\phi}=1$.
Information about $V(q^2)$, $A_1(q^2)$, and $A_{12}(q^2)$ is provided in App.~ \ref{app:formfactors}.

\subsection{$B\to X_{d,s}\,\nu_i\bar\nu_j$}\label{sec:BtoX}

The functions $a_\pm^{B X_q}$ associated with $B\to X_{q}\,\nu_i\bar\nu_j$ with $q=d,s$ transitions are given by~\cite{Altmannshofer:2009ma}
\begin{align}
a_\pm^{B X_q}(q^2)&=\frac{G_F^2\,\alphae^2\,\tau_B\,\kappa(0)}{3072\,\pi^5\,m_b^3}\sqrt{\lambda(m_b^2,m_q^2,q^2)}\label{eq:inclusiveB}\\ 
&\times\left[\lambda(m_b^2,m_q^2,q^2)+3\, q^2\left(m_b^2+m_q^2-q^2\right)\right]  ~,\nonumber
\end{align}
where
\begin{align}
    \kappa(0)=1\,+\,\frac{\alphas(m_b)}{\pi}\left[\frac{25}{6}-\frac{2}{3}\,\pi^2\right]\approx0.83\,,
\end{align}
includes QCD corrections to the $b\to q\,
\nu\bar{\nu}$ matrix element due to virtual and bremsstrahlung
contributions
\cite{Bobeth:2001jm}.

\section{Form factors}\label{app:formfactors}

Here, we provide detailed information on the form factors. 
In general any form factor, denoted by $\mathcal{F}$, can be parametrized as~\cite{Straub:2015ica}
\begin{align}\label{eq:formfactor}
    \mathcal{F}(q^2)=\frac{1}{1-\frac{q^2}{m_{R_{\mathcal{F}}}^2}}\sum_{k=0}^2\alpha_k^{(\mathcal{F})}\left[z(q^2)-z(0)\right]^2~,
\end{align}
where 
\begin{align}
    z(q^2)=\frac{\sqrt{t_+-q^2}-\sqrt{t_+-t_0}}{\sqrt{t_+-q^2}+\sqrt{t_+-t_0}}~,
\end{align}
with $t_\pm=(m_B\pm m_{P,V})^2$ and $t_0=t_+\big(1-\sqrt{1-t_-/t_+}\big)$. 
Here, $m_{R_{\mathcal{F}}}$ represents the mass of sub-threshold resonances compatible with the quantum numbers of the form factor $\mathcal{F}$. 
The values of $m_{R_{\mathcal{F}}}$ can be found in Refs.~\cite{Straub:2015ica,Gubernari:2018wyi}.  

\subsection{$B\to P,V$}

We use the latest form factors results from Ref.~\cite{Gubernari:2018wyi,Straub:2015ica}, where a fit of LCSR and lattice data is performed. 
Central values of $\alpha_k^{(\mathcal{F})}$ as well as uncertainties and correlations for each form factor $\mathcal{F}$, can be found in supplemented files of these references. 
For almost all modes we employ these fit results, with the exception of the $B \to \rho$ mode, where the previous fit was performed using only LCSR data at low-$q^2$. 
In the following section, we employ the latest LCSR results and perform a fit with the available lattice data.

\subsection{$B \to \rho$}\label{sec:BtorhoFFfit}

We perform a fit of three $B \to \rho$ form factors $V,\,A_1$ and $A_2$ following a similar procedure as in Refs.~\cite{Albertus:2014xwa,Flynn:2008zr}.
The form factor $A_{12}$, which is used in our parametrization in Eq.~\eqref{eq:amBV}, is obtained via the relation
\begin{align}\label{eq:ff_A12}
    A_{12}= \frac{(m_B+m_V)^2(m_B^2-m_V^2-q^2)A_1-\lambda_{BV}A_2}{16\,m_B\,m_V^2\,(m_B+m_V)}\,.
\end{align}
For low $q^2$, we use LCSR data from Ref.~\cite{Gubernari:2018wyi}, while for high $q^2$ we use the available data from the SPQcdR~\cite{Abada:2002ie} and UKQCD~\cite{Bowler:2004zb} collaborations. 

\begin{figure}[h!]
    \centering
    \includegraphics[width=7cm,height=5.5cm]{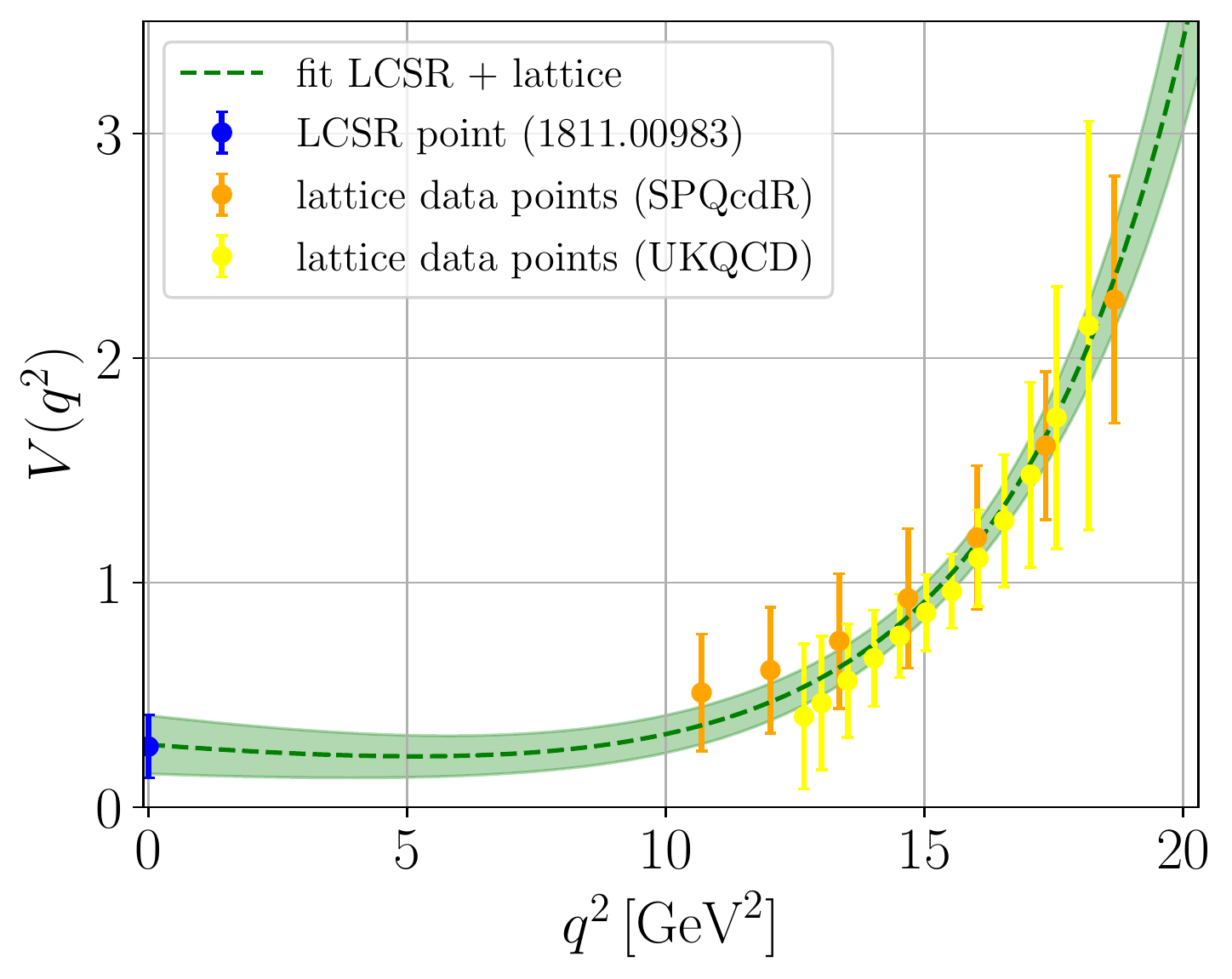}
    \includegraphics[width=7cm,height=5.5cm]{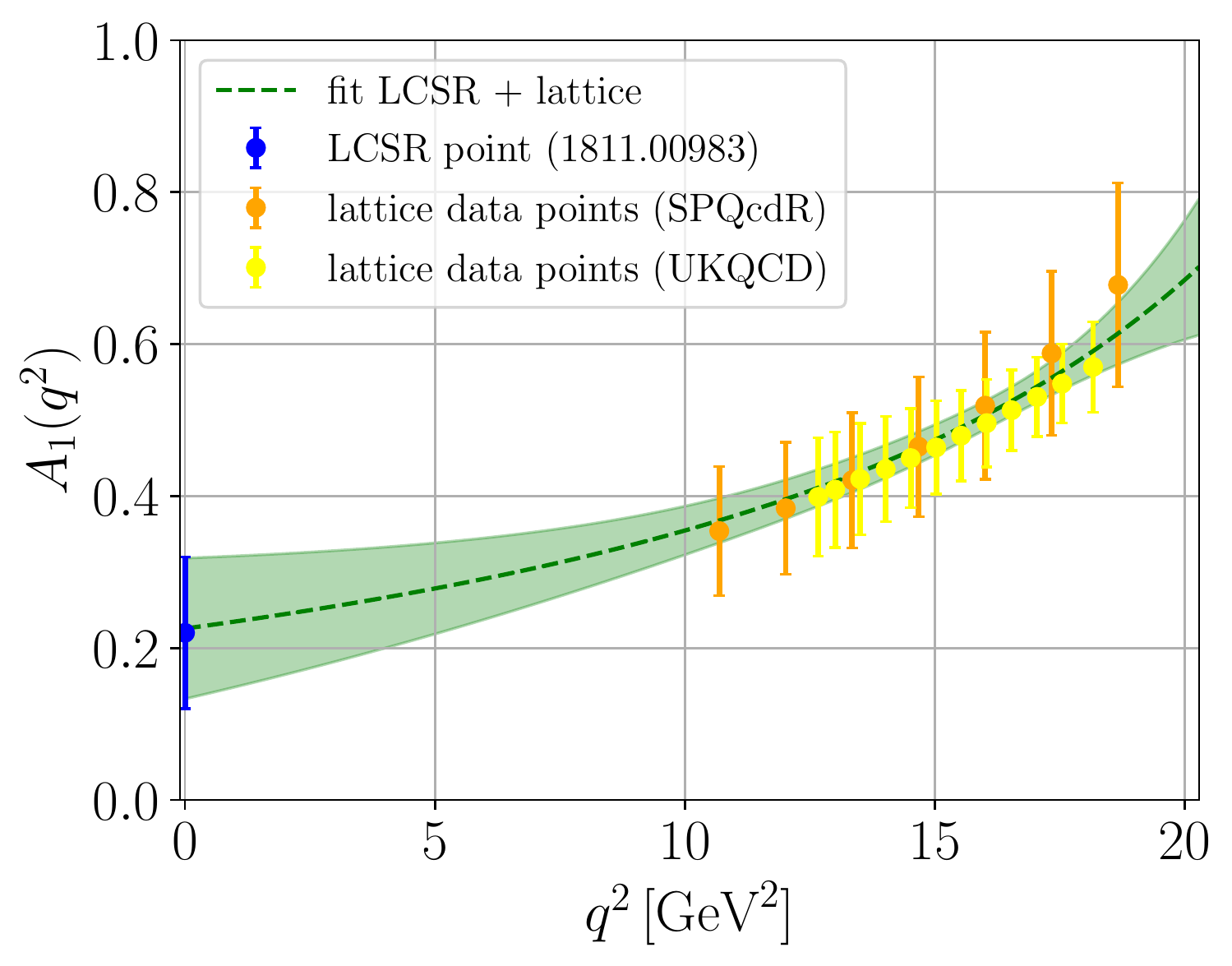} 
    \includegraphics[width=7cm,height=5.5cm]{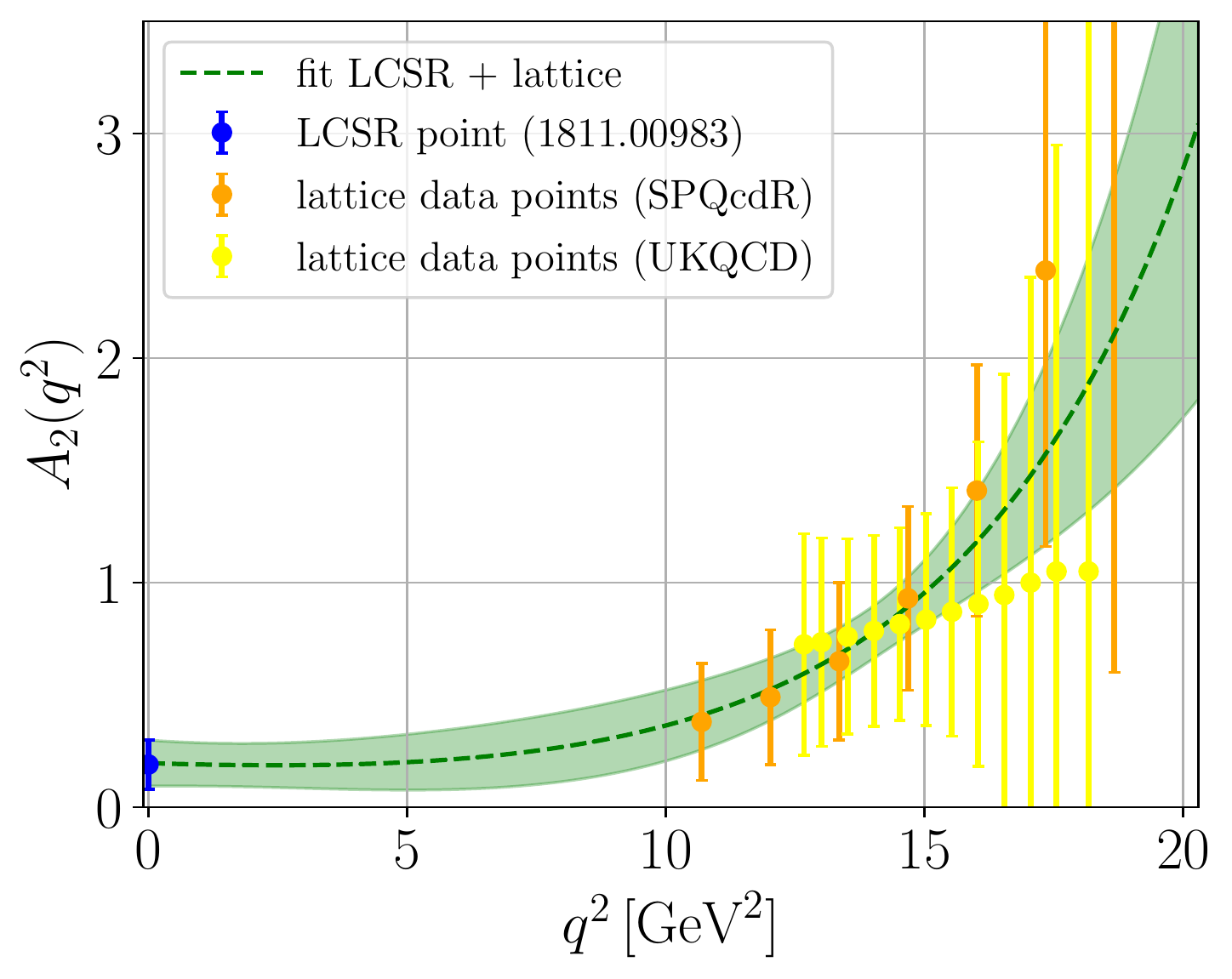}
    \caption{$B\to \rho$ form factors $V$, $A_1$ and $A_{2}$.  LCSR data~\cite{Gubernari:2018wyi} (blue points), lattice data~\cite{Flynn:2008zr} (orange and yellow points) and our fit results (green dashed line) and their  $1\,\sigma$ uncertainties (green band). 
    }
    \label{fig:FFfit_Brho_LCSR_lattice}
\end{figure}

Fig.~\ref{fig:FFfit_Brho_LCSR_lattice} shows the $q^2$--distribution and its uncertainties for the form factors $V,\,A_1$ and $A_2$ in $B \to \rho$. 
The fit results (best fit values, uncertainties, and correlations) of these form factors can be found in a supplemented file of this article on \texttt{arXiv}~\cite{rhofit}.

Assuming that experimental information of charged modes, $B\to\rho \,\ell\nu_\ell$, is saturated by SM contributions, we can use the experimental branching ratios of these modes as normalization, leading to~\cite{Aliev:1997se} 
\begin{align}\label{eq:BrhoSMratioExp}
    \frac{\mathcal{B}(B^0\to\rho^0\,\nu\bar\nu)_\text{SM}}{\mathcal{B}(B^0\to\rho^\pm\,\ell^\mp\nu_\ell)_\text{exp}}=\frac{3}{2}\,\left|\frac{V_{td}}{V_{ub}}\right|^2\left(\frac{\alphae}{4\,\pi}\right)^2\,\left| X_{\rm SM}\right|^2~,
\end{align}
for neutral modes, and
\begin{align}
    \frac{\mathcal{B}(B^\pm\to\rho^\pm\,\nu\bar\nu)_\text{SM}}{\mathcal{B}(B^\pm\to\rho^0\,\ell^\pm\nu_\ell)_\text{exp}}=6\,\left|\frac{V_{td}}{V_{ub}}\right|^2\left(\frac{\alphae}{4\,\pi}\right)^2\,\left| X_{\rm SM}\right|^2~,
\end{align}
for charged ones. Since lepton-flavor is conserved and universal in the SM \eqref{eq:SM}, any leptonic mode,  $\ell=e$, $\ell=\mu$ or $\ell=\tau$, can be used as normalization. In particular, we can use $\mathcal{B}(B^0\to\rho^\pm\,\ell^\mp\nu_\ell)_{\rm exp}=(2.94\pm0.21)\cdot 10^{-4}$~\cite{Zyla:2020zbs} and $\mathcal{B}(B^\pm\to\rho^0\,\ell^\pm\nu_\ell)_{\rm exp}=(1.58\pm0.11)\cdot 10^{-4}$~\cite{Zyla:2020zbs} where $\ell=e$ or $\ell=\mu$, not a sum over $e$ and $\mu$ modes. 
Corresponding SM branching ratios are in agreement with those based on our fit to LCSR and lattice data, see Tab.~\ref{tab:smPred}. 

When computing the ratios in Eq.~\eqref{eq:BR_ratio_rhonorm}  we only consider the leading term in Eq.~\eqref{eq:LUratio_series} since we can only extract the value of the sum ${A_+^{B^0\rho^0} + A_-^{B^0\rho^0}}$. 
Hence, the $1\,\sigma$($2\,\sigma$) intervals are obtained by varying $\mathcal{B}(B^0\to\rho^\pm\,\ell^\mp\nu_\ell)_{\rm exp}$ in Eq.~\eqref{eq:BrhoSMratioExp} within their $1\,\sigma$($2\,\sigma$) uncertainties.

\section{Global  $\boldsymbol{b \to s}$ fits}\label{app:globalfit}

Here we provide the results of our global fits to $b\to s\,\mu^+\mu^-$ data using the python package \textit{flavio}~\cite{Straub:2018kue}. 
We consider two cases: global fits including only $b\to s\,\mu^+\mu^-$ data, and others where in addition we include information from observables such as $R_{K^\ast}$ and $B^0\to K^{\ast 0}\,e^+e^-$ observables.
We follow a similar approach as Ref.~\cite{Hati:2020cyn,Kriewald:2021hfc}, where also the \textit{flavio} package is used and refer to this reference for details.
In particular we employ observables from $b\to s \,\ell \ell$ transitions listed in Tabs.~B.1-B.3 in Ref.~\cite{Hati:2020cyn}, while using the updated 2021 measurement of $R_K$ from LHCb~\cite{Aaij:2021vac}. 

However, we \emph{do not} include the observables listed in Tabs.~B.4-B.9 of Ref.~\cite{Hati:2020cyn}, which incorporate observables from charged current $B$ decays as well as strange, charm and $\tau$-decays.

We additionally include observables of radiative modes, $B_{(s)}^0 \to \mu\mu$ and $\Lambda_b$ decays listed in Tab.~\ref{tab:addDataFlavio}, which are already implemented in \emph{flavio}.

\begin{table*}[ht!]
\centering
\setlength{\tabcolsep}{5pt} 
\resizebox{\textwidth}{!}{
\renewcommand{\arraystretch}{1.3}
\begin{tabular}{lcr}
\hline
\hline
observable & SM prediction & measurement/limit  \\
\hline
$\mathcal{B}\left( B^0_s \to \mu^+ \mu^- \right)$ & $(3.67 \pm 0.14) \cdot 10^{-9}$ & $2.7 \cdot 10^{-9}$\quad combination 2020$^\dagger$~\cite{LHCb:2020zud} \\
$\mathcal{B}\left( B^0 \to \mu^+ \mu^- \right)$ & $(1.14 \pm 0.11) \cdot 10^{-10}$ & $0.6 \cdot 10^{-10}$\quad combination 2020$^\dagger$~\cite{LHCb:2020zud} \\
& &  \\
$\mathcal{B}\left( B^0 \to K^{\ast 0}\,\gamma \right)$ & $\left( 41.8 \pm 7.4 \right) \cdot 10^{-6}$ & $\left( 43.3 \pm 1.5 \right) \cdot 10^{-6}$\quad HFAG'14~\cite{HeavyFlavorAveragingGroupHFAG:2014ebo} \\
$\mathcal{B}\left( B^+ \to K^{\ast +}\,\gamma \right)$ & $\left( 42.5 \pm 8.0 \right) \cdot 10^{-6}$ & $\left( 42.1 \pm 1.8 \right) \cdot 10^{-6}$\quad HFAG'14~\cite{HeavyFlavorAveragingGroupHFAG:2014ebo} \\
$\mathcal{B}\left( B \to X_s\,\gamma \right)$ & $\left( 329 \pm 23 \right) \cdot 10^{-6}$ & $\left( 327 \pm 14 \right) \cdot 10^{-6}$\quad Belle'14~\cite{Misiak:2017bgg} \\
$\mathcal{B}\left( B_s^0 \to \phi\,\gamma \right)$ & $\left( 4.0 \pm 0.5 \right) \cdot 10^{-5}$ & $\left( 3.6 \pm 0.5 \pm 0.3 \pm 0.6 \right) \cdot 10^{-5}$\quad Belle'14~\cite{Belle:2014sac} \\
$\frac{\mathcal{B}\left( B_s^0 \to K^{\ast 0}\,\gamma \right)}{\mathcal{B}\left( B_s^0 \to \phi\,\gamma \right)}$ & $1.04 \pm 0.19$ & $1.19 \pm 0.06 \pm 0.04 \pm 0.07$\quad LHCb'12~\cite{LHCb:2012quo} \\
& & \\
$A_{CP}\left( B^0 \to K^{\ast 0}\,\gamma \right)$ & $0.005 \pm 0.002$ & $-0.002 \pm 0.015$\quad HFAG'14 \cite{HeavyFlavorAveragingGroupHFAG:2014ebo} \\
$A_{CP}\left( B_s^0 \to \phi\,\gamma \right)$ & $0.004 \pm 0.002$ & $0.11 \pm 0.29 \pm 0.11$\quad LHCb'19~\cite{LHCb:2019vks}  \\
& & \\
$\mathcal{A}^\Delta_{CP}\left(B_s^0 \to \phi \gamma \right)$ & $0.03 \pm 0.02$ & $-0.67^{+0.37}_{-0.41} \pm 0.17$\quad LHCb'19~\cite{LHCb:2019vks} \\
$S_{\phi \gamma}$ & $(-2 \pm 2) \cdot 10^{-4}$ & $0.43 \pm 0.30 \pm 0.11$\quad LHCb'19~\cite{LHCb:2019vks} \\
$S_{K^\ast \gamma}$ & $-0.023 \pm 0.014$ & $-0.16 \pm 0.22$\quad  HFAG'14~\cite{HeavyFlavorAveragingGroupHFAG:2014ebo} \\
& & \\
\hline
observable & $q^2$-bins in $\GeV^2$ & datasets \\
\hline
$\frac{\text{d}\mathcal{B}}{\text{d}q^2}\left( \Lambda_b \to \Lambda\,\mu^+ \mu^- \right)$ & $[2,4],\,[4,6],\,[15,20]$ & LHCb'15~\cite{LHCb:2015tgy} \\
$A_\text{FB}\left( B^+ \to K^{\ast +}\,\mu^+\mu^- \right)$ &$[1.1,2.5],\,[4,6],\,[15,19]$ & LHCb'20 \cite{LHCb:2020gog} \\
& &  \\
$A_\text{FB}\left( B^+ \to K^+\,\mu^+\mu^- \right)$ & $[1.1,2],\,[2,3],\,[3,4],\,[4,5],\,[5,6],\,[19,22]$ & LHCb'14~\cite{LHCb:2014auh} \\
$F_\text{H}\left( B^+ \to K^+\,\mu^+\mu^- \right)$ & $[1.1,2],\,[2,3],\,[3,4],\,[4,5],\,[5,6],\,[19,22]$ & LHCb'14~\cite{LHCb:2014auh} \\
& &  \\
$A_\text{FB}^h\left( \Lambda_b \to \Lambda\,\mu^+ \mu^- \right)$ & $[15,\,20]$ & LHCb'18~\cite{LHCb:2018jna} \\
$A_\text{FB}^l\left( \Lambda_b \to \Lambda\,\mu^+ \mu^- \right)$ & $[15,\,20]$ & LHCb'18~\cite{LHCb:2018jna} \\
$A_\text{FB}^{lh}\left( \Lambda_b \to \Lambda\,\mu^+ \mu^- \right)$ & $[15,\,20]$ & LHCb'18~\cite{LHCb:2018jna} \\
\hline
\end{tabular}}
\caption{Additional input for the performed global fit, which is not listed in Ref.~\cite{Hati:2020cyn}. 
$^\dagger${\scriptsize Combination of ATLAS, CMS and LHCb results, where we use the given multivariate numerical distribution of the observables, which is implemented in \emph{flavio}. In this table we only provide the central value for comparison.} 
}
\label{tab:addDataFlavio}
\end{table*}

\clearpage

\begin{table*}[h!]
\centering
\setlength{\tabcolsep}{7pt} 
\renewcommand{\arraystretch}{1.3}
\begin{tabular}{lcr}
\hline\hline
$R_{K^{(\ast)}}$ observables & $q^2$-bins in $\GeV^2$ & datasets \\
\hline
$R_{K^0}$ & $[0.1,4],\,[1,6],\,[14.18,19]$ & Belle'19~\cite{Abdesselam:2019lab} \\
$R_{K^+}$ & $[0.1,4],\,[1,6],\,[14.18,19]$ & Belle'19~\cite{Abdesselam:2019lab} \\
$R_{K^+}$ & $[1.1,6]$ & LHCb'21~\cite{Aaij:2021vac} \\
& & \\
$R_{K^{\ast 0}}$ & $[0.045,1.1],\,[1.1,6],\,[15,19]$ & Belle'19~\cite{Abdesselam:2019wac} \\
$R_{K^{\ast 0}}$ & $[0.045,1.1],\,[1.1,6]$ & LHCb'17~\cite{Aaij:2017vbb} \\
& & \\
$R_{K^{\ast +}}$ & $[0.045,1.1],\,[1.1,6],\,[15,19]$ & Belle'19~\cite{Abdesselam:2019wac} \\
\hline
\hline
LFU violating observables & $q^2$-bins in $\GeV^2$ & datasets \\
\hline
$Q_{4,5} = P_{4,5}^\mu - P_{4,5}^e$ & $[0.1,\,4],\,[1,\,6],\,[14.18,\,19]$ & Belle'16~\cite{Belle:2016fev} \\
\hline
\hline
$B^0\to K^{\ast 0}\,e^+e^-$ observables & $q^2$-bins in $\GeV^2$ & datasets \\
\hline
$F_L,\, P_1,\, P_2,\,\text{Im}(A_T)$ & $[0.002,\,1.12],\,[0.0008,\,0.257]$ & LHCb'20~ \cite{Aaij:2015dea,Aaij:2020umj} \\
\hline\hline
\end{tabular}
\caption{$R_{K^{(\ast)}}$ and $B^0\to K^{\ast 0}\,e^+e^-$ observables input for the global fit including $R_{K^{(\ast)}}$ data. 
}
\label{tab:dataRK}
\end{table*}

\subsection{Global fits with only $b\to s\,\mu^+\mu^-$ data}

In the global fit with only $b\to s\,\mu^+\mu^-$ data, we exclude experimental information on the LU ratios $R_{K^*}$, and $B^0\to K^{*0}\,e^+e^-$ observables. 

Due to strong correlations it is mandatory to include $\mathcal{B}(B^0\to\mu\mu)$ in addition to $\mathcal{B}(B_s^0\to\mu\mu)$ in the global fit. Although the updated 2021 branching ratios from LHCb have been recently presented, the correlations remain unavailable, therefore, we use the 2020 combination of ATLAS, CMS, and LHCb that is implemented in \textit{flavio}. 
We expect small changes when including the new LHCb measurement.

We perform five different global fits:
\begin{itemize}
    \item 1 dimensional with only $\Cnine$~,
    \item 1 dimensional with $\Cnine=-\Cten$~,
    \item 2 dimensional with $\Cnineten$~,
    \item 4 dimensional with $\Cninetenall$~,
    \item 6 dimensional  with $\Call$~.
\end{itemize}

The best fit values of the Wilson coefficients, as well as their $1\,\sigma$ uncertainties are listed in Tab.~\ref{tab:FitValues_noRk}. 
The last two columns display the reduced $\chi^2$ of the fit ($\sim 1$), with their respective pull from the SM hypothesis ($\sim 4.5\,\sigma$). 

\begin{table*}[ht!]
\def\arraystretch{1.5}
\centering
\resizebox{\textwidth}{!}{
\begin{tabular}{c|c|c|c|c|c|c|c|c}
\hline
\hline
Dim.  & $\Cseven$ & $\Csevenpr$ &$\Cnine$ & $\Cten$ & $\Cninepr$ & $\Ctenpr$ & $\chi^2/\text{dof}$ & $\text{Pull}_ \text{SM}$ \\ 
\hline
$1$  & - & - & $-0.91 \pm 0.18$ & - & - & - & $1.00$ & $4.5 \sigma$ \\ 
$1$  & - & - & $-0.68 \pm 0.16$ &  $-\Cnine$ & - & - & $0.99$ & $4.7 \sigma$ \\ 
$2$  & - & - & $-1.02 \pm 0.19$ & $0.46 \pm 0.18$ & - & - & $0.96$ & $4.9 \sigma$ \\
$4$  & - & - & $-1.13 \pm 0.18$ & $0.31 \pm 0.21$ & $0.29 \pm 0.33$ & $-0.24 \pm 0.19$ & $0.92$ & $5.0 \sigma$ \\ 
$6$  & $0.002 \pm 0.01$ & $0.02 \pm 0.02$ & $-1.15 \pm 0.18$ & $0.30 \pm 0.20$ & $0.22 \pm 0.34$ & $-0.24 \pm 0.19$ & $0.91$ & $4.6 \sigma$ \\ 
\hline
\hline
\end{tabular}
}
\caption{Best fit values and $1\sigma$ uncertainties of the Wilson coefficients from a fit with only pure $b\to  s\,\mu^+\mu^-$ data for different new physics scenarios. We also provide the $\chi^2/\text{dof}$ value and respective pull from the SM hypothesis.
} 
\label{tab:FitValues_noRk}
\end{table*}

\subsection{Global fits including $R_{K^{(\ast)}}$ data}

Assuming that electron modes does not suffer from NP effects, we can include in addition the observables from Tab.~\ref{tab:dataRK}. 
The $B^0\to K^{*0}\,e^+e^-$ observables set strong constraints on the Wilson coefficients $\mathcal{C}_7^{(\prime)}$. 
We perform five different global fits as before. The results are displayed in Tab.~\ref{tab:FitValues_Rk}, where the pull from the SM hypothesis has increased from 
$\sim 4.5\,\sigma$  to $\sim 6\,\sigma$.

\begin{table*}[hb!]
\def\arraystretch{1.5}
\centering
\resizebox{\textwidth}{!}{
\begin{tabular}{c|c|c|c|c|c|c|c|c}
\hline
\hline
Dim. & $\Cseven$ & $\Csevenpr$ &$\Cnine$ & $\Cten$ & $\Cninepr$ & $\Ctenpr$ & $\chi^2/\text{dof}$ & $\text{Pull}_ \text{SM}$ \\ 
\hline
1 & - & - & $-0.83 \pm 0.14$ & - & - & - & $0.98$ & $6.0 \sigma$ \\ 
1 & - & - & $-0.41 \pm 0.07$ &  $-\Cnine$ & - & - & $0.99$ & $6.0 \sigma$ \\ 
2  & - & - & $-0.71 \pm 0.17$ & $0.20 \pm 0.13$ & - & - & $0.97$ & $5.9 \sigma$ \\
4  & - & - & $-1.07 \pm 0.17$ & $0.18 \pm 0.15$ & $0.27 \pm 0.32$ & $-0.28 \pm 0.19$ & $0.90$ & $6.5 \sigma$ \\
6  & $0.0005 \pm 0.01$ & $0.005 \pm 0.006$ & $-1.08 \pm 0.18$ & $0.18 \pm 0.15$ & $0.27 \pm 0.34$ & $-0.28 \pm 0.17$ & $0.89$ & $6.1 \sigma$ \\
\hline
\hline
\end{tabular}
}
\caption{Best fit values and $1\sigma$ uncertainties of the Wilson coefficients from a fit also including the observables listed in Tab.~\ref{tab:dataRK} for different new physics scenarios. We also provide the $\chi^2/\text{dof}$ value and respective pull from the SM hypothesis.
} 
\label{tab:FitValues_Rk}
\end{table*}

\section{Benchmark dineutrino distributions}\label{app:diffBR}

\begin{figure*}[h!]
    \centering
    \includegraphics[width=7cm,height=5.5cm]{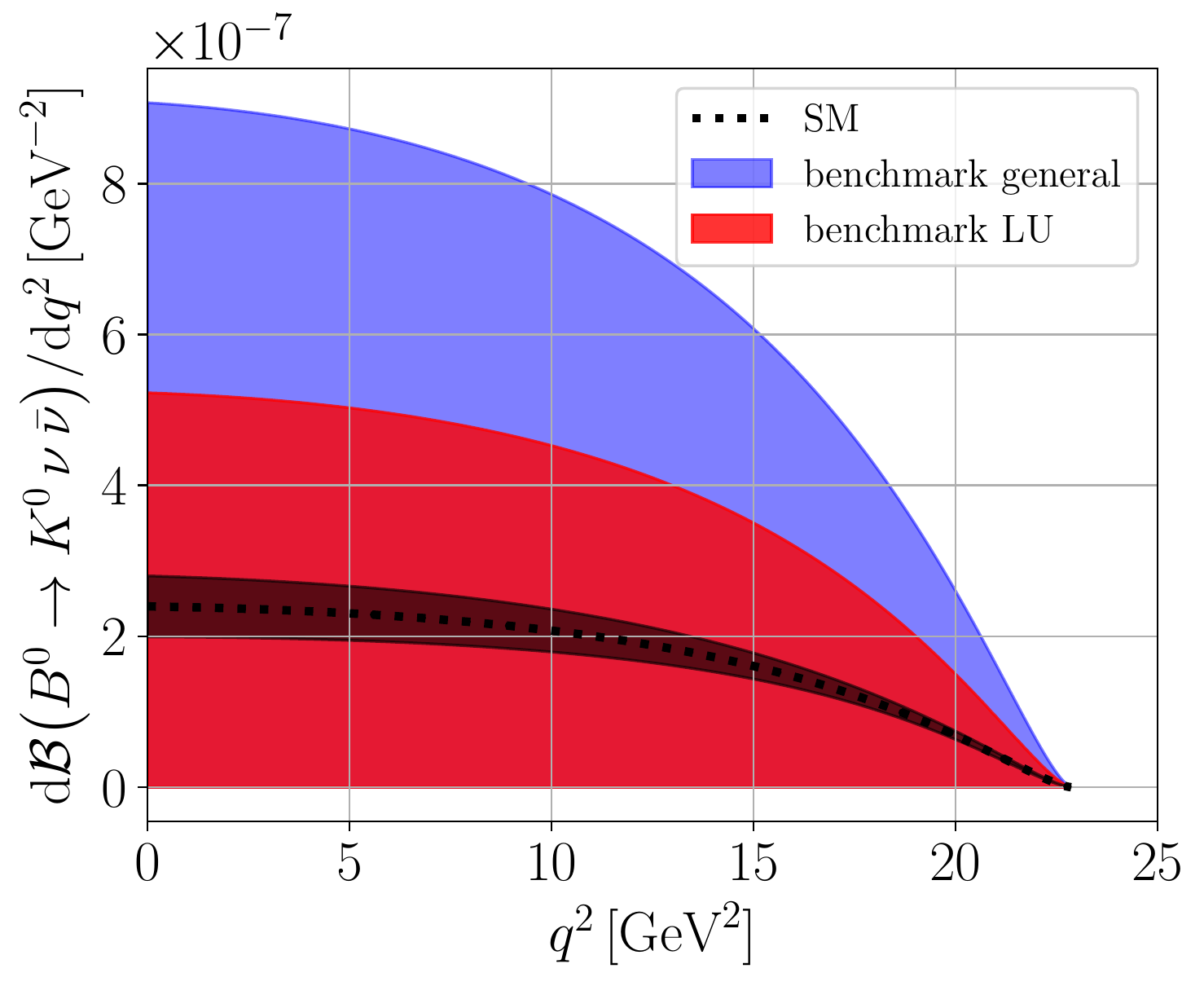}
    \includegraphics[width=7cm,height=5.5cm]{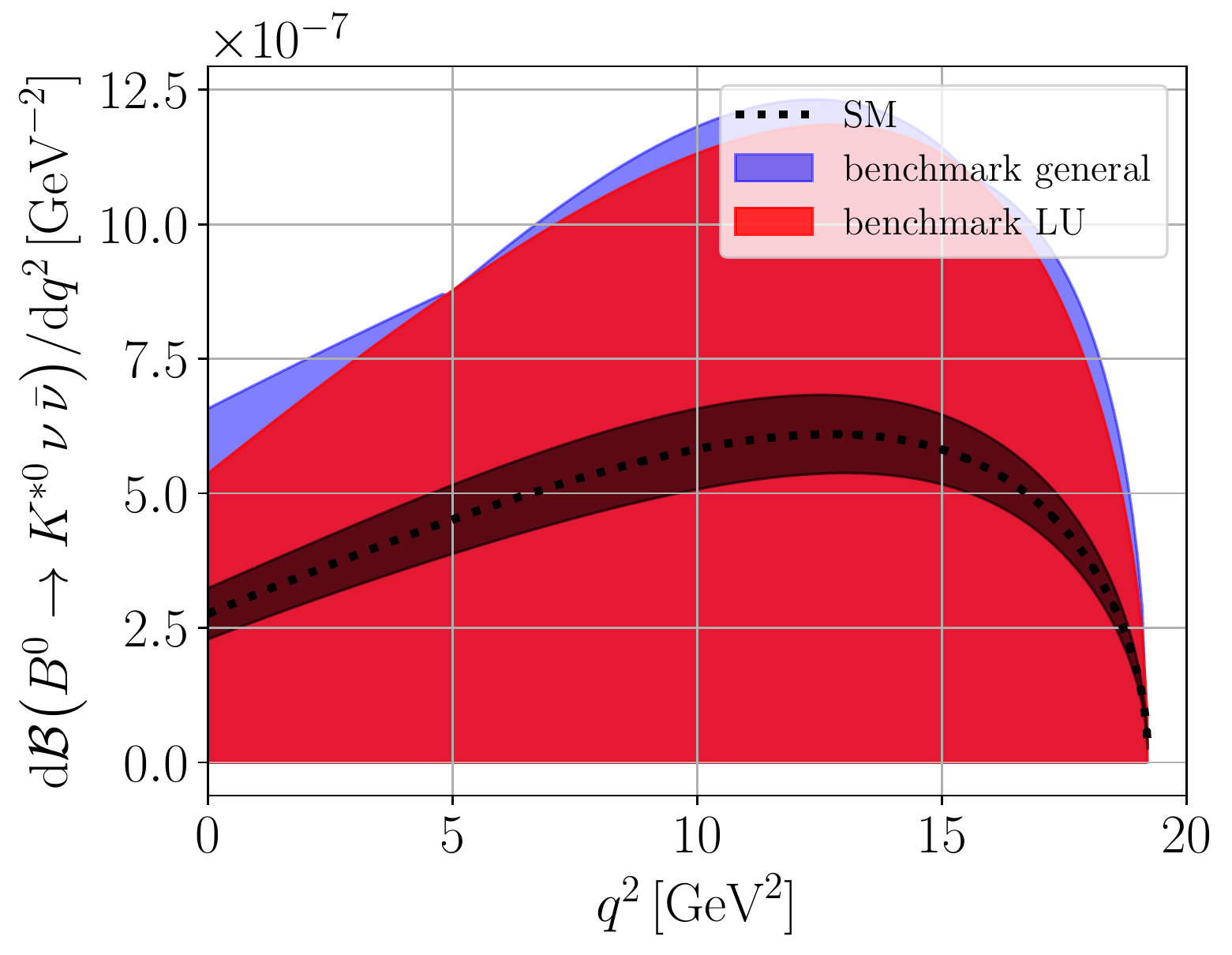}
    \includegraphics[width=7cm,height=5.5cm]{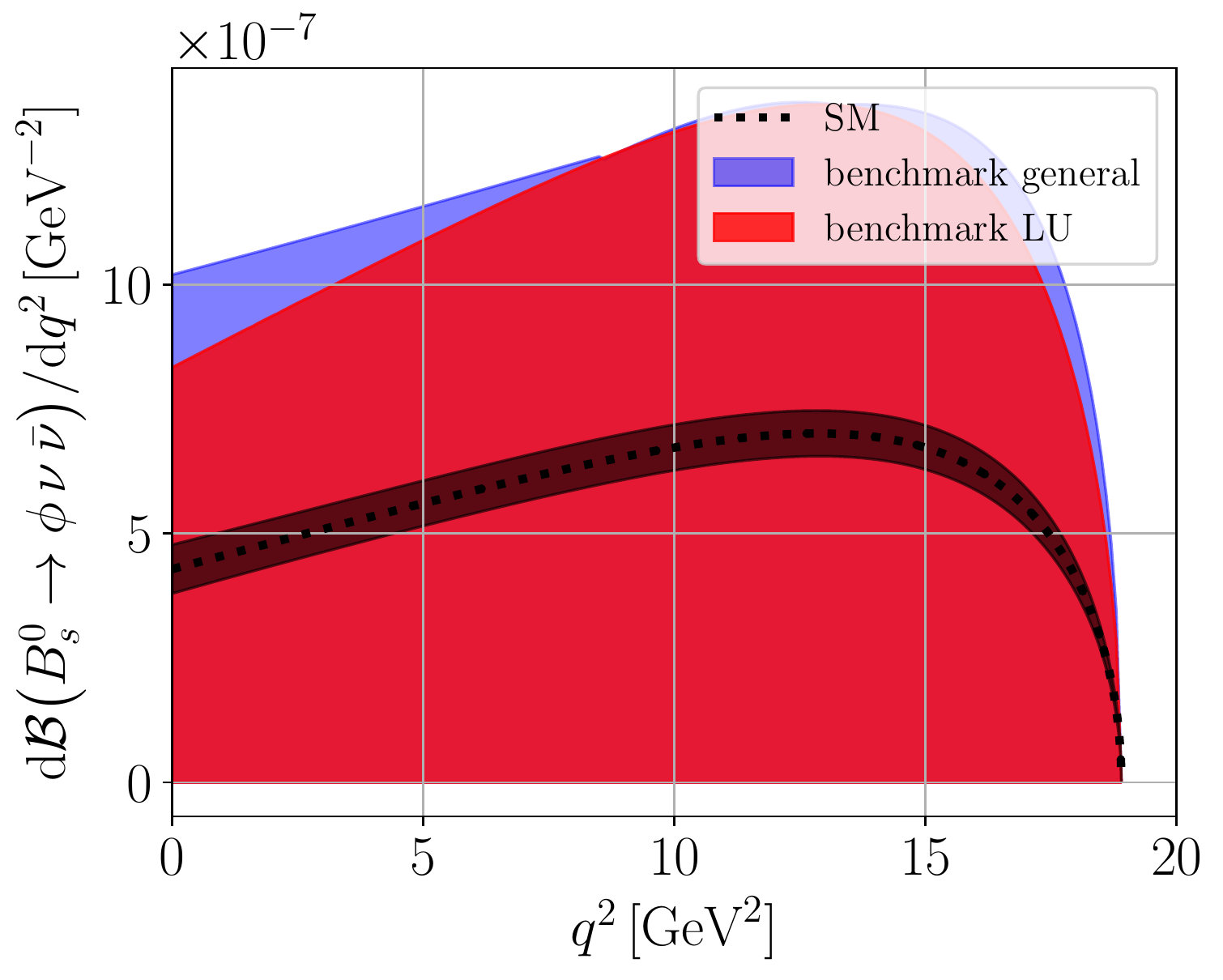}
    \includegraphics[width=7cm,height=5.5cm]{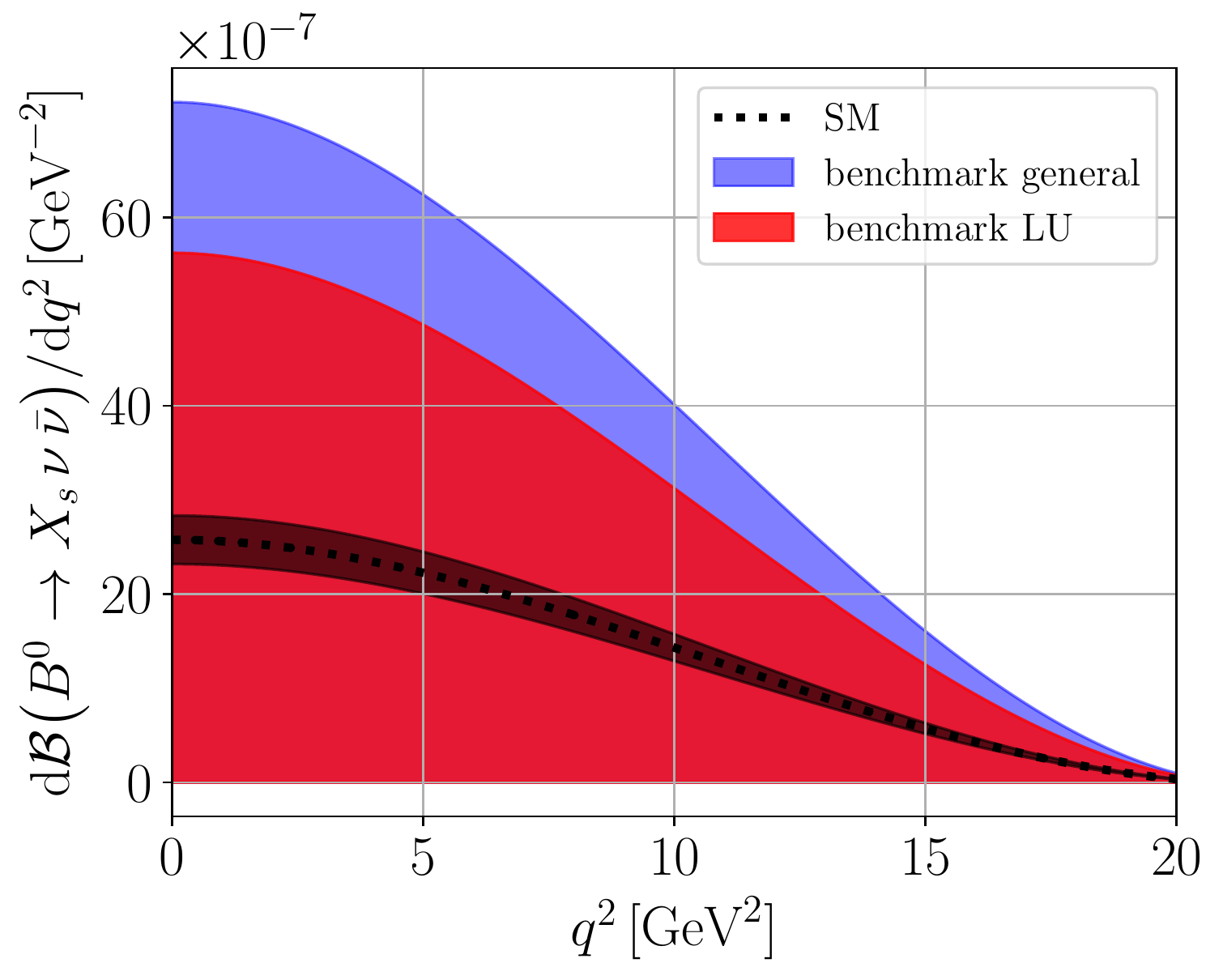}
     \includegraphics[width=7cm,height=5.5cm]{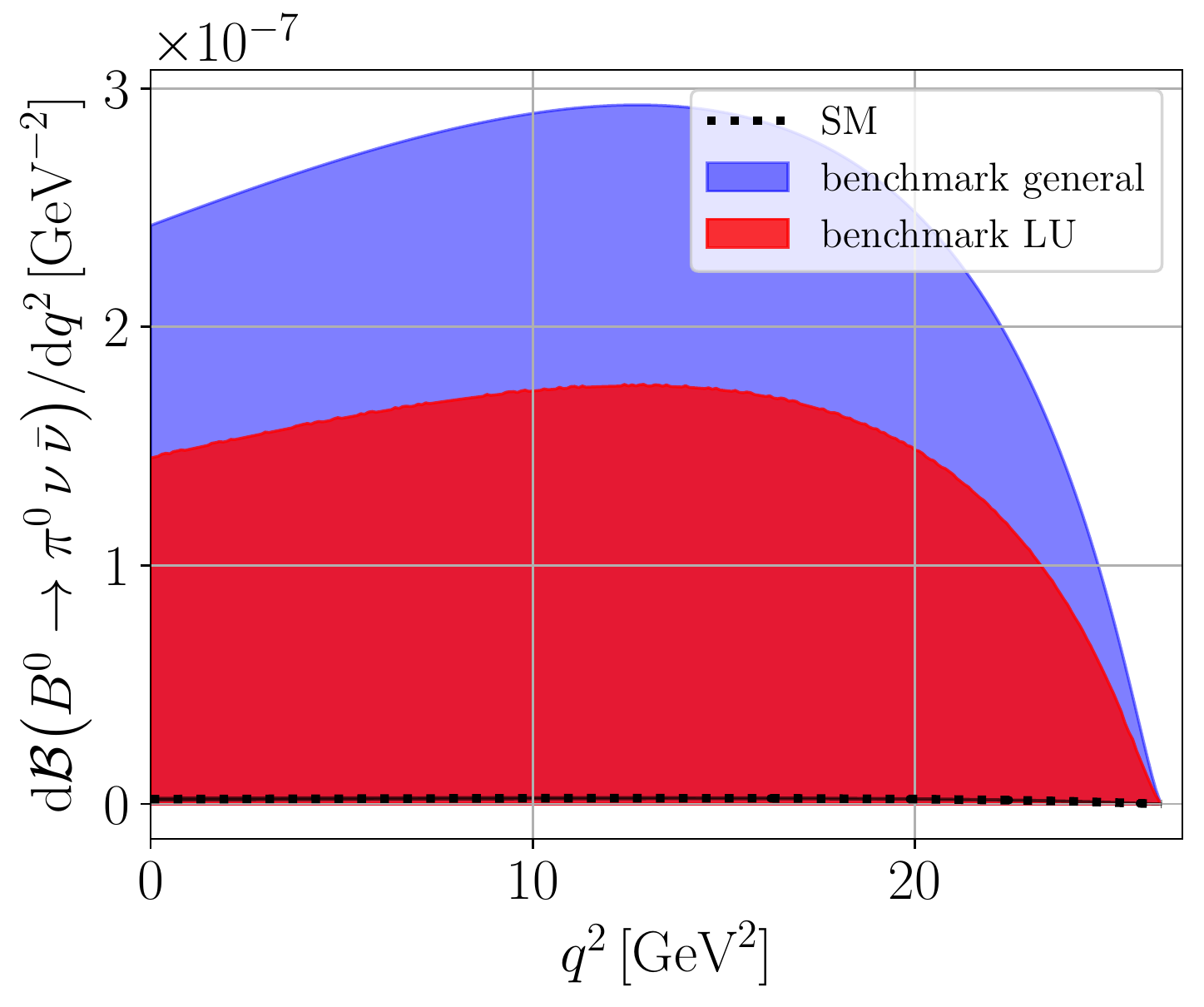}
     \includegraphics[width=7cm,height=5.5cm]{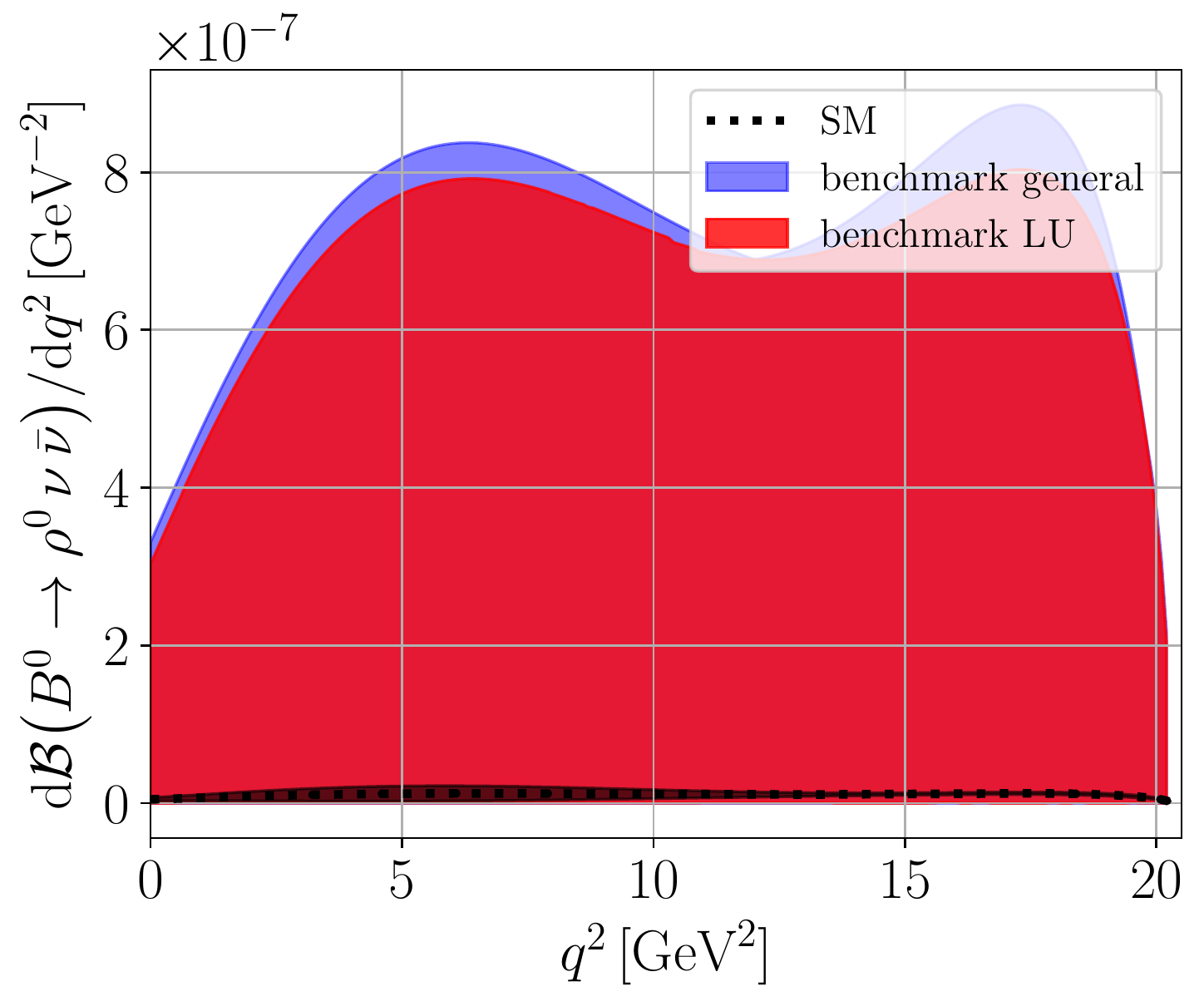}
    \caption{Differential branching ratio for $B^0\to K^0 \nu\bar\nu$, $B^0\to K^{*0} \nu\bar\nu$, $B^0_s \to \phi\, \nu\bar\nu$, $B^0 \to X_s\, \nu\bar\nu$, $B^0 \to \pi^0\, \nu\bar\nu$, and $B^0 \to \rho^0\, \nu\bar\nu$ in the SM and two NP benchmark scenarios, ``benchmark general'' using the derived EFT bounds~\eqref{eq:MIbs} and \eqref{eq:MIbd} for $b\to s\,\nu\bar\nu$ and $b\to d\,\nu\bar\nu$, respectively, and ``benchmark LU''~\eqref{eq:dBR} together with the experimental limits from Tab.~\ref{tab:smPred}. 
    See text for details.
     }
    \label{fig:plotdiffBP}
\end{figure*}

In this appendix we display the differential branching ratios of $B \to P,V$ as well as inclusive $B \to X$ dineutrino transitions for different benchmarks of $x^\pm_{{bq}}$ in Fig.~\ref{fig:plotdiffBP}. 
We show the SM distributions (black) with Wilson coefficients given by Eq.~\eqref{eq:SM}, while also including form factor uncertainties. 
The regions shown for the general benchmarks (blue) are constructed using the values of $x_{bq}^\pm$ that provide the largest (or smallest) integrated branching ratio allowed by the constraints in Eq.~\eqref{eq:MIbs} and \eqref{eq:MIbd} for $b \to s$ and $b \to d$ transitions, respectively.
For the LU benchmarks (red) we utilize Eqs.~\eqref{eq:xpmLU} and following, together with Eq.~\eqref{eq:dBR} and the experimental limits on $\mathcal{B}(B\to P\,\nu\bar\nu)$ in Tab.~\ref{tab:smPred}. Similar results are obtained for charged $B$-decay modes, which suffer from $\tau$-background contributions, see Eq.~\eqref{eq:br-long}, and are therefore not shown.

\providecommand{\href}[2]{#2}\begingroup\raggedright\endgroup

\end{document}